\pdfoutput=1
\documentclass{article}
\usepackage{arxiv_preprint,times}

\usepackage{amsmath,amsfonts,bm}

\def\eqref#1{equation~\ref{#1}}

\def\1{\bm{1}}

\DeclareMathAlphabet{\mathsfit}{\encodingdefault}{\sfdefault}{m}{sl}
\SetMathAlphabet{\mathsfit}{bold}{\encodingdefault}{\sfdefault}{bx}{n}

\def\tL{{\tens{L}}}

\def\tR{{\tens{R}}}

\newcommand{\R}{\mathbb{R}}

\PassOptionsToPackage{hyphens}{url}
\usepackage{hyperref}
\usepackage{url}
\usepackage{comment}
\usepackage{enumitem}

\usepackage[utf8]{inputenc}
\usepackage[T1]{fontenc}
\usepackage{graphicx}
\usepackage{amsmath}
\usepackage{amssymb}
\usepackage{amsfonts}
\usepackage{amsthm}
\newtheoremstyle{defnamed}%
  {\topsep}{\topsep}%
  {}%
  {}%
  {\bfseries}%
  {.}%
  {.5em}%
  {\thmname{#1}\thmnumber{ #2}\thmnote{ (\bfseries\itshape #3\normalfont)}}
\theoremstyle{defnamed}
\newtheorem{definition}{Definition}[section]
\usepackage{booktabs}
\usepackage{multirow}
\usepackage{nicefrac}
\usepackage{microtype}
\usepackage{xcolor}
\usepackage{colortbl}
\usepackage{arydshln}
\usepackage{subcaption}
\usepackage{listings}
\usepackage[most]{tcolorbox}
\usepackage{needspace}
\usepackage{pmboxdraw}
\definecolor{lstbg}{HTML}{FBFBFD}
\definecolor{lstnum}{HTML}{B7BDC4}
\definecolor{lstident}{HTML}{2B2F36}
\colorlet{lstbandcol}{black!60}
\DeclareCaptionFormat{lstdesc}{{\setlength{\fboxrule}{0.5pt}\setlength{\fboxsep}{4pt}%
  \fcolorbox{lstbandcol}{lstbandcol!7!white}{\parbox{\dimexpr\linewidth-2\fboxsep-2\fboxrule\relax}{\scriptsize\color{black!80}#3}}}}
\lstdefinestyle{prompttext}{%
  language={},
  backgroundcolor=\color{lstbg},
  basicstyle=\ttfamily\fontsize{5.0}{5.7}\selectfont\color{lstident},
  numberstyle=\tiny\color{lstnum},
  numbers=left, numbersep=8pt,
  frame=single, frameround=fttf, framerule=0.5pt, framesep=5pt, rulecolor=\color{lstbandcol},
  breaklines=true, breakatwhitespace=false, breakindent=12pt,
  columns=fullflexible, keepspaces=true, xleftmargin=14pt,
  showstringspaces=false,
  escapeinside={(*@}{@*)},
  literate={—}{{---}}1 {–}{{--}}1 {→}{{$\rightarrow$}}1 {↔}{{$\leftrightarrow$}}1 {↑}{{$\uparrow$}}1
    {ℕ}{{$\mathbb{N}$}}1 {ℤ}{{$\mathbb{Z}$}}1 {ℝ}{{$\mathbb{R}$}}1 {ℚ}{{$\mathbb{Q}$}}1
    {§}{{\S}}1 {…}{{\dots}}1 {≤}{{$\le$}}1 {≥}{{$\ge$}}1 {∃}{{$\exists$}}1 {∀}{{$\forall$}}1
    {⟨}{{$\langle$}}1 {⟩}{{$\rangle$}}1 {“}{{``}}1 {”}{{''}}1 {−}{{$-$}}1 {∈}{{$\in$}}1
    {α}{{$\alpha$}}1 {¬}{{$\lnot$}}1 {─}{{\textSFx}}1 {│}{{\textSFxi}}1 {└}{{\textSFii}}1 {├}{{\textSFviii}}1,
}
\usepackage{cleveref}
\usepackage{tikz}
\usepackage{pgfplots}
\pgfplotsset{compat=1.18}
\usetikzlibrary{pgfplots.statistics,pgfplots.groupplots,shapes.geometric}
\makeatletter
\pgfdeclareplotmark{sqstar}{%
  \pgfpathrectangle{\pgfqpoint{-\pgfplotmarksize}{-\pgfplotmarksize}}{\pgfqpoint{2\pgfplotmarksize}{2\pgfplotmarksize}}%
  \pgfusepathqfillstroke
  \pgfmathsetlength\pgf@xa{0.62\pgfplotmarksize}%
  \pgfpathmoveto{\pgfqpoint{0pt}{\pgf@xa}}\pgfpathlineto{\pgfqpoint{0pt}{-\pgf@xa}}%
  \pgfpathmoveto{\pgfqpointpolar{30}{\pgf@xa}}\pgfpathlineto{\pgfqpointpolar{210}{\pgf@xa}}%
  \pgfpathmoveto{\pgfqpointpolar{150}{\pgf@xa}}\pgfpathlineto{\pgfqpointpolar{330}{\pgf@xa}}%
  \pgfusepathqstroke
}
\pgfdeclareplotmark{circstar}{%
  \pgfpathcircle{\pgfpointorigin}{1.1\pgfplotmarksize}%
  \pgfusepathqfillstroke
  \pgfmathsetlength\pgf@xa{0.62\pgfplotmarksize}%
  \pgfpathmoveto{\pgfqpoint{0pt}{\pgf@xa}}\pgfpathlineto{\pgfqpoint{0pt}{-\pgf@xa}}%
  \pgfpathmoveto{\pgfqpointpolar{30}{\pgf@xa}}\pgfpathlineto{\pgfqpointpolar{210}{\pgf@xa}}%
  \pgfpathmoveto{\pgfqpointpolar{150}{\pgf@xa}}\pgfpathlineto{\pgfqpointpolar{330}{\pgf@xa}}%
  \pgfusepathqstroke
}
\pgfdeclareplotmark{tristar}{%
  \pgfpathmoveto{\pgfqpoint{0pt}{1.25\pgfplotmarksize}}%
  \pgfpathlineto{\pgfqpointpolar{-30}{1.25\pgfplotmarksize}}%
  \pgfpathlineto{\pgfqpointpolar{210}{1.25\pgfplotmarksize}}%
  \pgfpathclose
  \pgfusepathqfillstroke
  \pgfmathsetlength\pgf@xa{0.5\pgfplotmarksize}%
  \pgfpathmoveto{\pgfqpoint{0pt}{\pgf@xa}}\pgfpathlineto{\pgfqpoint{0pt}{-\pgf@xa}}%
  \pgfpathmoveto{\pgfqpointpolar{30}{\pgf@xa}}\pgfpathlineto{\pgfqpointpolar{210}{\pgf@xa}}%
  \pgfpathmoveto{\pgfqpointpolar{150}{\pgf@xa}}\pgfpathlineto{\pgfqpointpolar{330}{\pgf@xa}}%
  \pgfusepathqstroke
}
\pgfdeclareplotmark{circAst}{%
  \pgfpathcircle{\pgfpointorigin}{1.15\pgfplotmarksize}%
  \pgfusepathqfillstroke
  \pgfmathsetlength\pgf@xa{0.72\pgfplotmarksize}\pgfsetlinewidth{0.8pt}%
  \pgfpathmoveto{\pgfqpoint{0pt}{\pgf@xa}}\pgfpathlineto{\pgfqpoint{0pt}{-\pgf@xa}}\pgfpathmoveto{\pgfqpointpolar{30}{\pgf@xa}}\pgfpathlineto{\pgfqpointpolar{210}{\pgf@xa}}\pgfpathmoveto{\pgfqpointpolar{150}{\pgf@xa}}\pgfpathlineto{\pgfqpointpolar{330}{\pgf@xa}}
  \pgfusepathqstroke
}
\pgfdeclareplotmark{circPlus}{%
  \pgfpathcircle{\pgfpointorigin}{1.15\pgfplotmarksize}%
  \pgfusepathqfillstroke
  \pgfmathsetlength\pgf@xa{0.72\pgfplotmarksize}\pgfsetlinewidth{0.8pt}%
  \pgfpathmoveto{\pgfqpoint{0pt}{\pgf@xa}}\pgfpathlineto{\pgfqpoint{0pt}{-\pgf@xa}}\pgfpathmoveto{\pgfqpoint{-\pgf@xa}{0pt}}\pgfpathlineto{\pgfqpoint{\pgf@xa}{0pt}}
  \pgfusepathqstroke
}
\pgfdeclareplotmark{circX}{%
  \pgfpathcircle{\pgfpointorigin}{1.15\pgfplotmarksize}%
  \pgfusepathqfillstroke
  \pgfmathsetlength\pgf@xa{0.72\pgfplotmarksize}\pgfsetlinewidth{0.8pt}%
  \pgfpathmoveto{\pgfqpointpolar{45}{\pgf@xa}}\pgfpathlineto{\pgfqpointpolar{225}{\pgf@xa}}\pgfpathmoveto{\pgfqpointpolar{135}{\pgf@xa}}\pgfpathlineto{\pgfqpointpolar{315}{\pgf@xa}}
  \pgfusepathqstroke
}
\pgfdeclareplotmark{triAst}{%
  \pgfpathmoveto{\pgfqpoint{0pt}{1.45\pgfplotmarksize}}\pgfpathlineto{\pgfqpointpolar{-30}{1.45\pgfplotmarksize}}\pgfpathlineto{\pgfqpointpolar{210}{1.45\pgfplotmarksize}}\pgfpathclose
  \pgfusepathqfillstroke
  \pgfmathsetlength\pgf@xa{0.62\pgfplotmarksize}\pgfsetlinewidth{0.8pt}%
  \pgfpathmoveto{\pgfqpoint{0pt}{\pgf@xa}}\pgfpathlineto{\pgfqpoint{0pt}{-\pgf@xa}}\pgfpathmoveto{\pgfqpointpolar{30}{\pgf@xa}}\pgfpathlineto{\pgfqpointpolar{210}{\pgf@xa}}\pgfpathmoveto{\pgfqpointpolar{150}{\pgf@xa}}\pgfpathlineto{\pgfqpointpolar{330}{\pgf@xa}}
  \pgfusepathqstroke
}
\pgfdeclareplotmark{triPlus}{%
  \pgfpathmoveto{\pgfqpoint{0pt}{1.45\pgfplotmarksize}}\pgfpathlineto{\pgfqpointpolar{-30}{1.45\pgfplotmarksize}}\pgfpathlineto{\pgfqpointpolar{210}{1.45\pgfplotmarksize}}\pgfpathclose
  \pgfusepathqfillstroke
  \pgfmathsetlength\pgf@xa{0.62\pgfplotmarksize}\pgfsetlinewidth{0.8pt}%
  \pgfpathmoveto{\pgfqpoint{0pt}{\pgf@xa}}\pgfpathlineto{\pgfqpoint{0pt}{-\pgf@xa}}\pgfpathmoveto{\pgfqpoint{-\pgf@xa}{0pt}}\pgfpathlineto{\pgfqpoint{\pgf@xa}{0pt}}
  \pgfusepathqstroke
}
\pgfdeclareplotmark{triX}{%
  \pgfpathmoveto{\pgfqpoint{0pt}{1.45\pgfplotmarksize}}\pgfpathlineto{\pgfqpointpolar{-30}{1.45\pgfplotmarksize}}\pgfpathlineto{\pgfqpointpolar{210}{1.45\pgfplotmarksize}}\pgfpathclose
  \pgfusepathqfillstroke
  \pgfmathsetlength\pgf@xa{0.62\pgfplotmarksize}\pgfsetlinewidth{0.8pt}%
  \pgfpathmoveto{\pgfqpointpolar{45}{\pgf@xa}}\pgfpathlineto{\pgfqpointpolar{225}{\pgf@xa}}\pgfpathmoveto{\pgfqpointpolar{135}{\pgf@xa}}\pgfpathlineto{\pgfqpointpolar{315}{\pgf@xa}}
  \pgfusepathqstroke
}
\pgfdeclareplotmark{sqAst}{%
  \pgfpathrectangle{\pgfqpoint{-1.05\pgfplotmarksize}{-1.05\pgfplotmarksize}}{\pgfqpoint{2.1\pgfplotmarksize}{2.1\pgfplotmarksize}}%
  \pgfusepathqfillstroke
  \pgfmathsetlength\pgf@xa{0.72\pgfplotmarksize}\pgfsetlinewidth{0.8pt}%
  \pgfpathmoveto{\pgfqpoint{0pt}{\pgf@xa}}\pgfpathlineto{\pgfqpoint{0pt}{-\pgf@xa}}\pgfpathmoveto{\pgfqpointpolar{30}{\pgf@xa}}\pgfpathlineto{\pgfqpointpolar{210}{\pgf@xa}}\pgfpathmoveto{\pgfqpointpolar{150}{\pgf@xa}}\pgfpathlineto{\pgfqpointpolar{330}{\pgf@xa}}
  \pgfusepathqstroke
}
\pgfdeclareplotmark{sqPlus}{%
  \pgfpathrectangle{\pgfqpoint{-1.05\pgfplotmarksize}{-1.05\pgfplotmarksize}}{\pgfqpoint{2.1\pgfplotmarksize}{2.1\pgfplotmarksize}}%
  \pgfusepathqfillstroke
  \pgfmathsetlength\pgf@xa{0.72\pgfplotmarksize}\pgfsetlinewidth{0.8pt}%
  \pgfpathmoveto{\pgfqpoint{0pt}{\pgf@xa}}\pgfpathlineto{\pgfqpoint{0pt}{-\pgf@xa}}\pgfpathmoveto{\pgfqpoint{-\pgf@xa}{0pt}}\pgfpathlineto{\pgfqpoint{\pgf@xa}{0pt}}
  \pgfusepathqstroke
}
\pgfdeclareplotmark{sqX}{%
  \pgfpathrectangle{\pgfqpoint{-1.05\pgfplotmarksize}{-1.05\pgfplotmarksize}}{\pgfqpoint{2.1\pgfplotmarksize}{2.1\pgfplotmarksize}}%
  \pgfusepathqfillstroke
  \pgfmathsetlength\pgf@xa{0.72\pgfplotmarksize}\pgfsetlinewidth{0.8pt}%
  \pgfpathmoveto{\pgfqpointpolar{45}{\pgf@xa}}\pgfpathlineto{\pgfqpointpolar{225}{\pgf@xa}}\pgfpathmoveto{\pgfqpointpolar{135}{\pgf@xa}}\pgfpathlineto{\pgfqpointpolar{315}{\pgf@xa}}
  \pgfusepathqstroke
}
\makeatother
\definecolor{bmGray}{HTML}{8a8985}\definecolor{lcLight}{HTML}{7fb2e6}\definecolor{lcMid}{HTML}{2a78d6}\definecolor{lcDark}{HTML}{164a8a}\colorlet{numC}{lcDark}\definecolor{ogC}{HTML}{E08A1E}\colorlet{aipC}{red!70!black}\colorlet{hostNone}{black!55}\definecolor{hostCdx}{HTML}{2A78D6}\definecolor{hostCC}{HTML}{E08A1E}%
\usetikzlibrary{positioning,fit,arrows.meta,calc,backgrounds,matrix}%
\usepackage{wrapfig}
\providecommand{\rnd}[1]{\smash[t]{#1}}
\providecommand{\pred}{\texttt{[PRED]}}
\providecommand{\Mnl}{M_{\mathrm{NL}}}
\providecommand{\Tnl}{T_{\mathrm{NL}}}
\providecommand{\Pnl}{P_{\mathrm{NL}}}
\providecommand{\Mfl}{M_{\mathrm{FL}}}
\providecommand{\Tfl}{T_{\mathrm{FL}}}
\providecommand{\Pfl}{P_{\mathrm{FL}}}

\providecommand{\Mflh}{\widehat{M}_{\mathrm{FL}}}
\providecommand{\Tflh}{\widehat{T}_{\mathrm{FL}}}
\providecommand{\Pflh}{\widehat{P}_{\mathrm{FL}}}
\providecommand{\znl}{z_{\mathrm{NL}}}
\providecommand{\zfl}{z_{\mathrm{FL}}}
\tikzset{
  tok/.style={draw, rounded corners=2pt, minimum height=6.5mm, inner sep=4pt, font=\small},
  blk/.style={draw, rounded corners=3pt, minimum height=9mm, minimum width=42mm,
              align=center, font=\small, inner sep=4pt, fill=black!5},
  emb/.style={draw, rounded corners=3pt, minimum height=7mm, minimum width=16mm,
              align=center, font=\small, inner sep=3pt},
  loss/.style={draw, very thick, rounded corners=3pt, minimum height=7mm,
               minimum width=16mm, align=center, font=\small, inner sep=3pt},
  flow/.style={-{Stealth[length=2.2mm]}, semithick},
  tie/.style={densely dashed, black!55},
  lbl/.style={font=\scriptsize\itshape, text=black!55},
}

\newcommand{\ourTool}{\textsc{AIProver}}
\definecolor{nlshade}{gray}{0.92}
\definecolor{flshade}{gray}{1.00}
\definecolor{nlEdge}{HTML}{966719}
\definecolor{nlFill}{HTML}{FFF1C3}
\definecolor{flEdge}{HTML}{315A93}
\definecolor{flFill}{HTML}{DEEBFF}

\title{\ourTool{}: Agentic Auto-Formalization \\ of Mathematical Research via \\ Certificate-Driven Evolving Harness\vspace{-9pt}}

\author{%
\textbf{Prithwish Jana}$^{*1}$\textbf{,}\,\,\,
\textbf{Viet Bach Hoang}$^{\dagger1}$\textbf{,}\,\,\,
\textbf{Logan Luna}$^{\dagger1}$\textbf{,}\,\,\,
\textbf{Viresh Pati}$^{\dagger2}$ \\
\textbf{Akash Singirikonda}$^{\dagger1}$\textbf{,}\,\,\,
\textbf{Cy Xie}$^{\dagger3}$\textbf{,}\,\,\,
\textbf{Lisa Carbone}$^{\mathsection4}$\textbf{,}\,\,\,
\textbf{Wuyang Chen}$^{\mathsection5}$ \\
\textbf{Walter Moreira}$^{\mathsection6}$\textbf{,}\,\,\,
\textbf{Joe Stubbs}$^{\mathsection6}$\textbf{,}\,\,\,
\textbf{Sriram Vishwanath}$^{\mathsection1}$\textbf{,}\,\,\,
\textbf{Vijay Ganesh}$^{1}$\\[0.6mm]
\small$^{1}$Georgia Institute of Technology, USA \;\;\;$^{2}$University of Pennsylvania, USA \;\;\;$^{3}$Foothill College, USA \\
\small$^{4}$Rutgers University, USA \;\;\;$^{5}$Simon Fraser University, Canada \;\;\;$^{6}$University of Texas at Austin, USA
}

\preprintfinalcopy

\usepackage{zref-abspage,zref-savepos,zref-user}
\makeatletter
\zref@newlist{budget}
\zref@addprops{budget}{abspage,page,posy}
\newif\ifbudgetdraft \budgetdrafttrue
\definecolor{budPrithwish}{RGB}{0,114,178}
\definecolor{budAkash}{RGB}{213,94,0}
\definecolor{budLogan}{RGB}{0,158,115}
\definecolor{budViresh}{RGB}{204,121,167}
\definecolor{budBach}{RGB}{190,130,0}

\providecommand{\budgetinfo}[3]{}
\providecommand{\budgettextheight}[1]{}
\newcommand{\budgetmark}[1]{\zref@savepos\zref@labelbylist{bud:#1}{budget}}
\newcommand{\secbudget}[3]{%
  \budgetmark{#1}%
  \protected@write\@auxout{}{\string\budgetinfo{#1}{#2}{#3}}%
  \ifbudgetdraft\marginpar{\scriptsize\color{gray}\raggedright\sffamily #3\\#2~p.}\fi}
\AtBeginDocument{\immediate\write\@auxout{\string\budgettextheight{\the\textheight}}}
\makeatother

\newlength{\pnpgap}
\definecolor{pnpframe}{gray}{0.65}
\newlength{\pnprule}
\newlength{\pnppad}
\newlength{\pnpk}
\newcommand{\pnprw}{1.0}
\newlength{\pnprow}
\newcommand{\pnpbox}[2]{{\setlength{\fboxsep}{\pnppad}\setlength{\fboxrule}{\pnprule}\color{pnpframe}\fbox{\includegraphics[width=#1]{#2}}}}
\newcommand{\pnpboxz}[3]{{\setlength{\fboxsep}{0pt}\setlength{\fboxrule}{\pnprule}\color{pnpframe}\fbox{\includegraphics[width=#1,trim=#2,clip]{#3}}}}
\newcommand{\pnpaw}{0.80}
\newlength{\pnpcapskip}
\newlength{\pnpbw}
\newlength{\pnpcw}
\newlength{\pnpdw}
\newlength{\pnpew}

\makeatletter
\def\section{\@startsection{section}{1}{\z@}{-1.6ex plus -0.4ex minus -.2ex}{0.8ex plus 0.2ex minus 0.1ex}{\large\sc\raggedright}}
\def\subsection{\@startsection{subsection}{2}{\z@}{-1.4ex plus -0.4ex minus -.2ex}{0.5ex plus .1ex}{\normalsize\sc\raggedright}}
\def\subsubsection{\@startsection{subsubsection}{3}{\z@}{-1.2ex plus -0.4ex minus -.2ex}{0.3ex plus .1ex}{\normalsize\sc\raggedright}}
\makeatother
\begin{document}
\budgetmark{front}

\vspace*{-34pt}
\maketitle
\lhead{}
{\renewcommand{\thefootnote}{\fnsymbol{footnote}}%
 \footnotetext[1]{Corresponding author: \href{mailto:pjana7@gatech.edu}{pjana7@gatech.edu}. Project page: \href{https://jprithwish.github.io/AIProver/}{jprithwish.github.io/AIProver}}%
 \footnotetext[2]{Equal contribution, sorted by last name.\qquad\textsuperscript{\S}Equal senior authorship, sorted by last name.}}

\vspace{-7mm}
\begin{abstract}\fontsize{9.4pt}{10pt}\selectfont
\vspace{-1mm}
Proof auto-formalization translates natural-language (NL) theorems and proofs into a formal
language (FL) such as Lean, enabling mechanical verification. Despite rapid progress, research-level proofs often depend
on concepts missing from leading proof assistant libraries (e.g., Lean's Mathlib), and successful compilation does not guarantee that a
translation preserves the theorem's meaning or the proof's reasoning. Furthermore, aligned NL-FL training data are scarce,
and leading agents often rely on costly frontier models and manually engineered harnesses.

To address the above issues, we present \ourTool{}, an agentic framework for autonomous proof auto-formalization and proof synthesis (AFPS) that jointly
post-trains a 119B open-weight language model and evolves its agentic, tool-calling harness with HarnessEvolve. Verifiers assess type correctness, proof completeness, and semantic
correctness, returning rewards and diagnostic certificates that drive model fine-tuning and alternating
\emph{reinforcement learning via symbolic feedback} and \emph{HarnessEvolve}, a certificate-driven
evolutionary search over the whole harness control flow that re-tailors the harness to the updated model. HarnessEvolve retains and generalizes rejected
designs as negative evidence to steer later mutations away from unproductive parts of the search.
For research-level training and evaluation,
we introduce \emph{LoCoBench}, 58.9k instances from Mathlib, CSLib, Mizar Math Library, and a bounded-arithmetic textbook,
combining 18.8k labelled NL--Lean pairs and 40.1k unlabelled NL--Mizar pairs
with a
771-instance validation split whose theorem--proof pairs have no public Lean formalization. Against 39
frameworks spanning AFPS agents, frontier LLMs, and coding agents, \ourTool{} lifts pass@4 semantic correctness over its Leanstral-1.5 base from 15.7\% to 36.7\% and
outperforms every other open-weight system and Aristotle. As a Claude Code and Codex skill, it lifts
their semantic correctness from 41.9\% and 34.1\% to 79.8\% and 62.4\%, respectively, outperforming Numina-Lean-Agent (72.5\% and 44.5\%) and OpenGauss (54.3\% and 41.9\%) in that role. Further, it is also 24\% cheaper than Numina-Lean-Agent, pushing the accuracy--cost frontier of research-level AFPS.
\end{abstract}

\vspace{-3.5mm}
\section{Introduction}
\label{sec:intro}
\vspace{-1.5mm}

AI agents for mathematics are making rapid progress, with headline results such as the resolution
of the Navier--Stokes conjecture~\citep{openai2026navierstokes} and the complete formalization of
Fermat's last theorem (FLT)~\citep{anthropic2026flt}. Key to this success is \emph{auto-formalization},
the translation of natural language (NL) theorems and proofs into formal language (FL) counterparts.
Most prior work on proof auto-formalization builds on large language models (LLMs) but targets
competition and undergraduate mathematics, whose theorems and proofs are short and
self-contained~\citep{cabral2025proofflow,liu2026tomap}. Research mathematics instead involves long
chains of interdependent definitions, lemmas, and proofs often absent from libraries such
as~\citet{mathlib} and CSLib~\citep{barrett2026cslib}. A single LLM call is ill-equipped for such
intricacies. Hence, recent work, such as the FLT formalization, has turned to \emph{agentic systems}
that wrap the LLM in a \emph{harness}: in a stateful multi-turn loop with Lean and search tools,
the agent iterates until the proof checks or hits an unresolvable error. Open-source harnesses
LeanMarathon~\citep{zhang2026leanmarathon}, Numina-Lean-Agent~\citep{liu2026numinaleanagent},
and OpenGauss~\citep{mathinc2026opengauss} and the closed-source Aristotle~\citep{achim2025aristotle}
implement this loop, and general-purpose coding harnesses such as Claude Code~\citep{anthropic2025claudecode}
and Codex~\citep{openaicodex2025} are also used by mathematicians~\citep{tao2026claudecode,ilin2026expertreview}.

\vspace{-1mm}
Despite this progress, these agentic systems share five limitations.
\textbf{Cost}: Claude's FLT formalization took eleven days and six billion output
tokens~\citep{anthropic2026flt,anthropic2026fable51}, resources at frontier rates beyond typical academic budgets.
\textbf{Human-in-the-loop dependence}: mathematicians supplied Fermat's FL target statements and stayed
involved~\citep{anthropic2026flt,chen2026prove2me}, Gauss ran on an expert-built, iteratively-refined
harness~\citep{mathinc2025spnt}, and Meta's AutoformBot, with 26 textbooks formalized, calls human
involvement essential~\citep{rammal2026formalizingscale}. \textbf{Jagged intelligence}: frontier models
excel at many tasks yet falter on unfamiliar ones; expert review of Claude's formalization~\citep{ilin2026expertreview}
found strong local proof repair but poor definitions and interfaces, and no model proves over 20\% of
ArXivLean's recent research theorems~\citep{matharena2026arxivlean}. \textbf{Semantic drift}: output cannot be
trusted as-is. Lean checks the FL proof against the FL theorem, not the NL one, so compiled code may
prove the wrong theorem~\citep{han2026shadowbench} or silently patch a flawed NL proof step~\citep{cornish2026faithformbench}.
\textbf{Lack of accessibility}: headline results often rest on non-public models: FLT and
Navier--Stokes reportedly used internal ones~\citep{anthropic2026flt,openai2026navierstokes,reeve2026buckmaster},
so the auto-formalization and proof synthesis (AFPS) process cannot be cross-checked.

\vspace{-0.5mm}
These limitations set our goal: an \textit{affordable, autonomous} \textbf{AI agent for
semantically correct and faithful proof auto-formalization of research-level mathematics}. Given NL
research text interleaving theorems and proofs, it should (a)~produce a full Lean counterpart
\textit{preserving each theorem's meaning and proof's reasoning}, (b)~run \textit{unattended}, (c)~use an \textit{open-weight model} via a \textit{model-agnostic} training and
harness recipe, and (d)~be a \textit{sub-agent or skill} of frontier agents. It should serve \textit{standalone} as an
open-source platform making research-level formalization accessible to all, and \textit{in conjunction with}
frontier agents as a Lean specialist pushing their accuracy--cost frontier, since frontier models
will likely stay the most expensive for the foreseeable future~\citep{tao2025cheapexpensive}.

\vspace{-0.5mm}
Achieving this goal requires addressing four issues.
\textbf{First}, \textit{NL--FL semantic alignment}: decoder-only LLMs are fine-tuned to maximize the next-token
likelihood of a reference output~\citep{ouyang2022instructgpt,zhou2023lima}, so it never learns that an
NL theorem or proof and its FL counterpart are the same mathematical object in two syntactic forms. Prior work adds
alignment outside the generator, in a retrieval encoder (ProofBridge, \citealp{jana2025proofbridge}), a critic
(\mbox{CriticLean}, \citealp{peng2025criticlean}), or a scorer (FormalAlign, \citealp{lu2024formalalign}); we hypothesize
that an LLM internalizing this equivalence instead formalizes more accurately.
\textbf{Second}, \textit{fine-grained training signal}: most AFPS models are trained with reinforcement learning (RL)
on Lean's binary verdict~\citep{ren2025deepseekproverv2,wang2025kimina}, which never compares the output with the
NL input or reference FL, so a formalization that compiles but proves the wrong theorem counts as success.
Such drift is common: on ShadowBench~\citep{han2026shadowbench} the best agent compiles 61.8\% of research-level
statements yet only 11.2\% align semantically, and FaithformBench~\citep{cornish2026faithformbench}
finds pervasive silent correction of invalid NL steps.
\textbf{Third}, \textit{research-level training data}: learning either signal at research level needs
gold-standard NL--Lean theorem--proof pairs, which barely exist: ShadowBench~(178) and ArXivLean~(41)
are evaluation sets. Mizar~\citep{bancerek2018mizar} and Rocq MathComp~\citep{mahboubi2021mathcomp} hold decades
of verified research mathematics, but in neither NL nor Lean.
\textbf{Fourth}, \textit{AFPS-tailored harness design}: an agent's performance depends as much on its harness as
on the model~\citep{tian2026swebenchmobile,milikic2026leanflow}, yet harnesses such as LeanMarathon and
Numina-Lean-Agent retain the control flow of Claude Code and Codex that are built for
software engineering. This control flow is suboptimal for Lean proving as well as for open-weight models, whose error modes and prompt sensitivity differ~\citep{sclar2024prompt}, and hand-tailoring harnesses takes
months~\citep{cognition2026cloud}.

\textbf{Contributions.} The main contributions of this paper are summarized as follows:
\vspace{-2mm}
\begin{itemize}[leftmargin=1em,itemsep=0pt,topsep=0pt,parsep=0pt]
\item \textbf{The \ourTool{} proof auto-formalization agent with an evolving harness.} We present \ourTool{}, a
framework to post-train an open-weight LLM and to simultaneously evolve its harness. Verifiers for completeness, type- and semantic correctness return a fine-grained reward and certificate, so RL and HarnessEvolve learn
from partial success. Per round, \textit{HarnessEvolve} adapts harness control flow to the model, then RL
post-trains the model within it. To our knowledge, it is the first attempt
to discover auto-formalization harnesses aligned to the LLM (Sec.~\ref{sec:coevo}).

\item \textbf{LoCoBench, a research-level benchmark for proof auto-formalization.} We introduce \emph{LoCoBench},
58.9k instances (18.8k NL--Lean, 40.1k NL--Mizar) in \emph{algebraic structures}, \emph{foundations, logic, \&
complexity}, and \emph{number theory}. We curate 771 validation instances from Mizar theorems with the longest proofs and
a bounded-arithmetic textbook: NL theorem-proof pairs with no Lean formalizations online to retrieve or
imitate and, to our knowledge, the first Mizar-to-Lean test at scale. We build it because no existing research-level dataset offers
NL theorem--proof pairs at this scale: for instance, ShadowBench has only 178 examples and ArXivLean no proofs (Sec.~\ref{sec:LoCoBench}).

\item \textbf{State-of-the-art proof auto-formalization results.} We test \ourTool{} against 39 systems
comprising AFPS models and agents, frontier LLMs, and coding agents (Sec.~\ref{sec:results}). \textbf{\textit{As a standalone agent}} built from Leanstral-1.5
in the Vibe harness, \ourTool{} lifts pass@4 semantic correctness from 15.7\% to 36.7\%, beating the best open-weight
system OpenGauss (21.5\%) and Aristotle (18.4\%). \textbf{\textit{In conjunction with coding agents}},
\ourTool{}+Claude Code (CC) reaches 79.8\% vs.\ 41.9\% (CC alone), 72.5\% for Numina-Lean-Agent (NLA)+CC and
54.3\% for OpenGauss (OG)+CC; with Codex it reaches 62.4\% vs.\ 34.1\% (Codex alone), 44.5\% (NLA+Codex) and
41.9\% (OG+Codex). It pushes the \textit{accuracy--cost frontier}: \$0.32/attempt standalone, \$1.55 with
CC vs.\ \$2.03
(NLA+CC), \$0.85 with Codex vs.\ \$1.08 (NLA+Codex), thus higher accuracy at lower cost.

\end{itemize}

\vspace{-2mm}
\section{Related Work}
\label{sec:relatedwork}
\vspace{-2mm}

\textbf{LLMs for AFPS.}
LLMs fine-tuned for \textbf{theorem auto-formalization}~\citep{wu2022autoformalization} translate NL theorems to FL, as in
Kimina-Autoformalizer~\citep{numina2025kiminaautoformalizer}, and StepFun-Formalizer~\citep{wu2025stepfun}, while those for \textbf{proof synthesis} generate FL proofs from FL theorems, as in
DeepSeek-Prover-V2~\citep{ren2025deepseekproverv2} and Goedel-Prover-V2~\citep{lin2025goedelv2}. Chaining the two for \textbf{proof auto-formalization} has two limitations.
First, errors cascade when the first model mis-formalizes the theorem. Second, neither stage verifies or corrects the
other. Draft, Sketch, and
Prove~\citep{jiang2023dsp}, ProofFlow~\citep{cabral2025proofflow},
ToMap~\citep{liu2026tomap}, and ProofBridge~\citep{jana2025proofbridge} work end-to-end with a verifier
in the loop, but fine-tune small (1.7B--32B) models or use closed APIs. None is tested at research level or uses an agentic pipeline for \textit{long-horizon} AFPS.

\vspace{-1.5mm}
\textit{Key Differences:} \ourTool{} tackles long-horizon, research-level AFPS with one agent, post-training
a 119B open-weight LLM on a reward verifying completeness, type- and semantic correctness.

\begin{figure}[!t]
    \centering
    \setlength{\pnpcapskip}{1pt}
    \captionsetup[subfigure]{skip=\pnpcapskip,aboveskip=\pnpcapskip,belowskip=0pt}
    \captionsetup{skip=3pt}

    \setlength{\pnpgap}{3mm}
    \setlength{\pnprow}{\pnprw\textwidth}

    \begin{subfigure}[t]{\textwidth}\vspace{0pt}
        \centering
        \pnpboxz{\dimexpr\pnpaw\linewidth-2\pnprule\relax}{1.38 5.10 1.38 1.86}{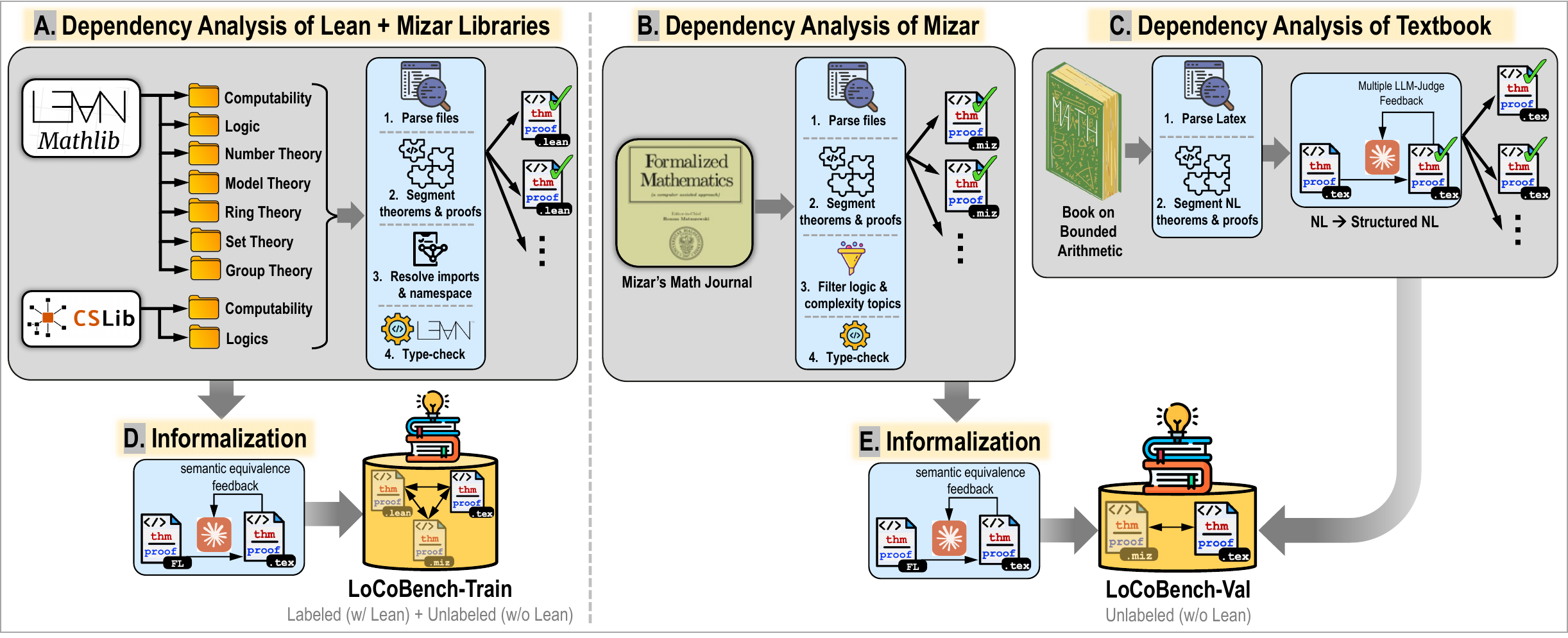}
        \caption{Data curation for \textsc{LoCoBench-Train} and \textsc{LoCoBench-Val}}
        \label{fig:AIProver_dataGenPipeline}
    \end{subfigure}

    \vspace{1mm}

    \setlength{\pnpbw}{\dimexpr0.5030\dimexpr\pnprow-\pnpgap\relax+0.1906\pnpk-0.2028\pnprule\relax}
    \setlength{\pnpcw}{\dimexpr\pnprow-\pnpgap-\pnpbw\relax}

    \makebox[\pnprow][s]{%
    \begin{subfigure}[t]{\pnpbw}\vspace{0pt}
        \centering
        \pnpboxz{\dimexpr\linewidth-2\pnprule\relax}{2.94 1.86 1.62 4.14}{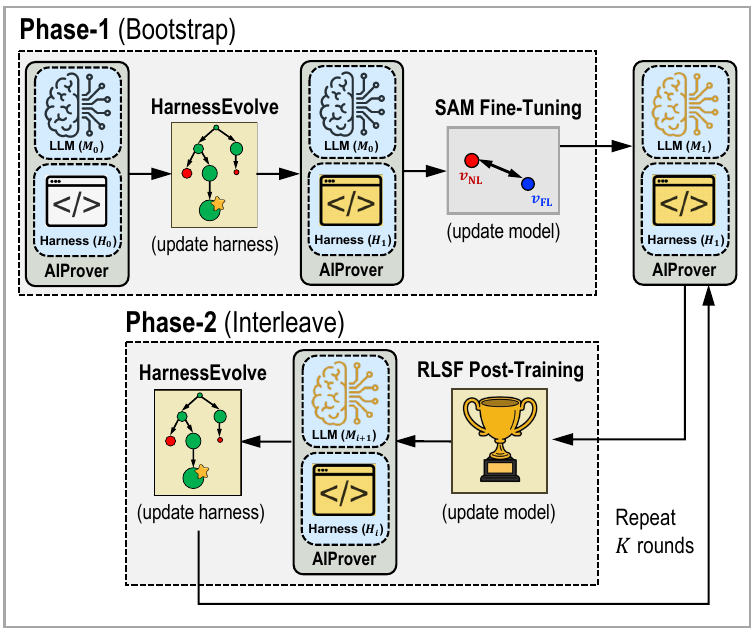}
        \caption{Overall post-training loop of \ourTool{}}
        \label{fig:AIProver_fullPipeline}
    \end{subfigure}%
    \hfill%
    \begin{subfigure}[t]{\pnpcw}\vspace{0pt}
        \centering
        \pnpbox{\dimexpr\linewidth-2\pnpk\relax}{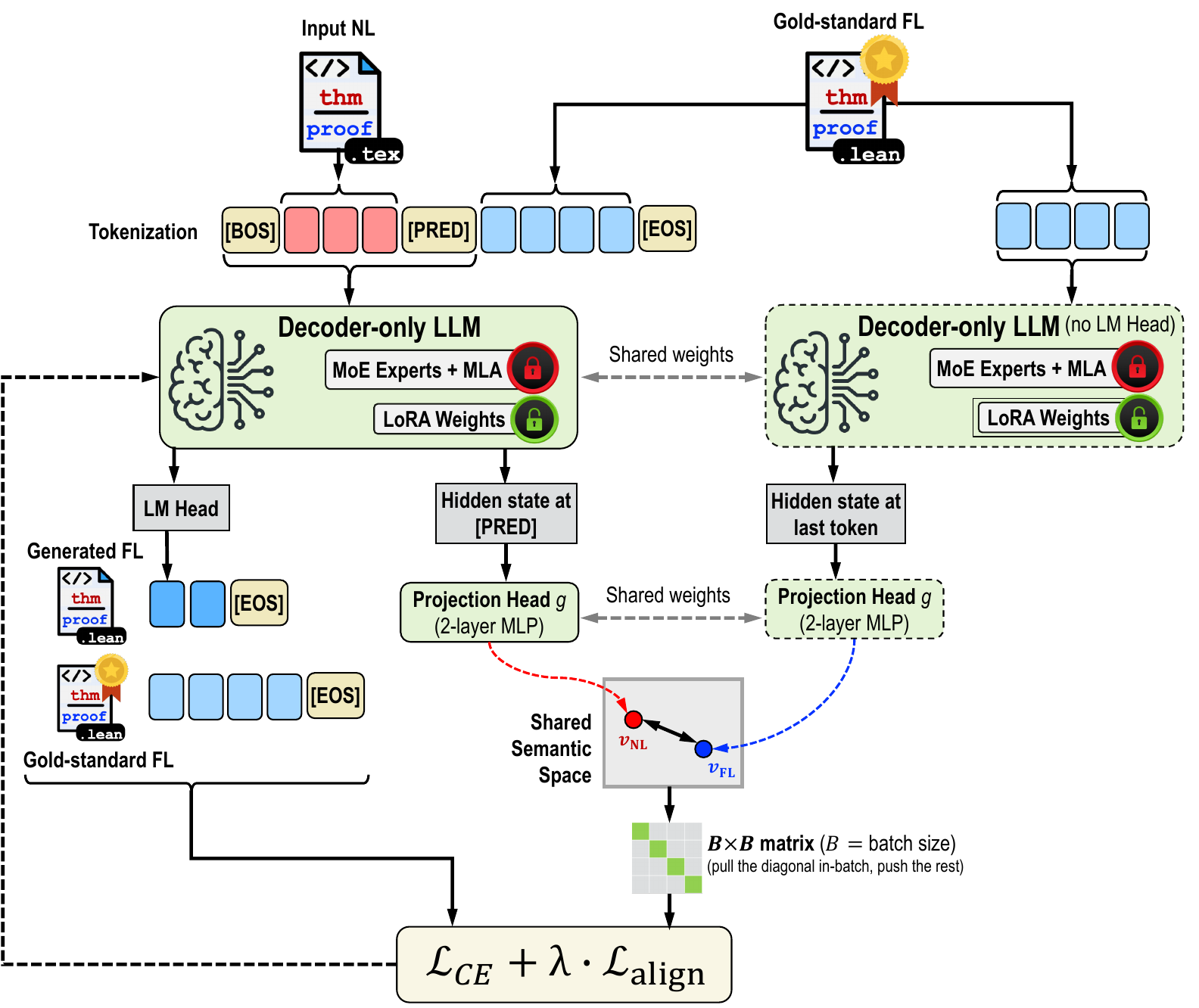}
        \caption{SAM: contrastive NL--FL alignment}
        \label{fig:AIProver_SAMpipeline}
    \end{subfigure}%
    }

    \vspace{1mm}

    \setlength{\pnpdw}{\dimexpr0.3475\dimexpr\pnprow-\pnpgap-4\pnpk\relax+2\pnpk\relax}
    \setlength{\pnpew}{\dimexpr\pnprow-\pnpgap-\pnpdw\relax}

    \makebox[\pnprow][s]{%
    \begin{subfigure}[t]{\pnpdw}\vspace{0pt}
        \centering
        \pnpbox{\dimexpr\linewidth-2\pnpk\relax}{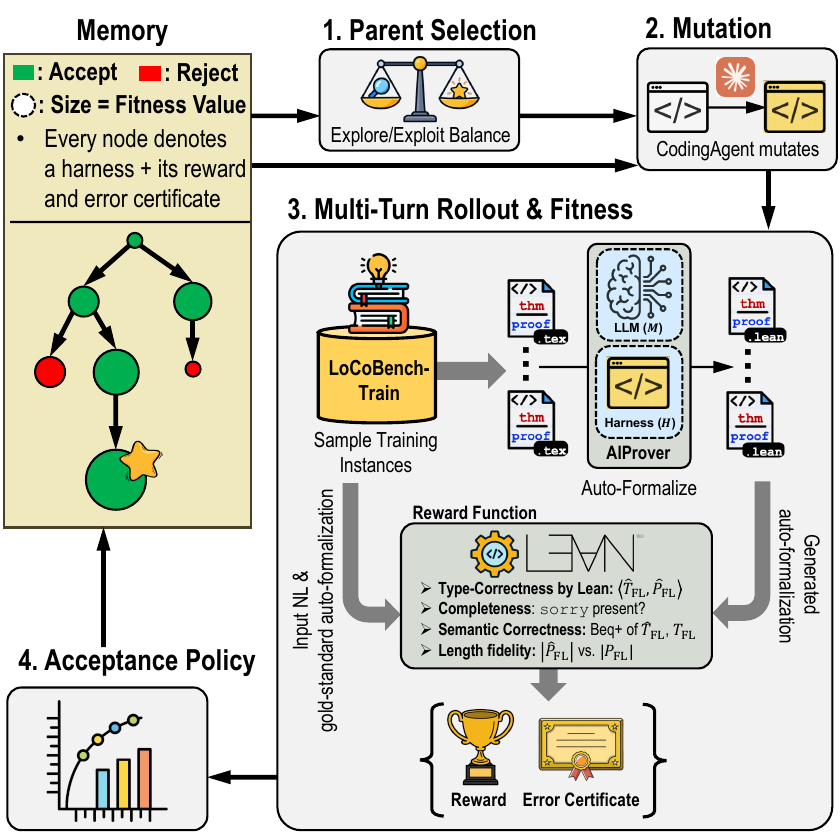}
        \caption{HarnessEvolve search}
        \label{fig:AIProver_harnessEvolution}
    \end{subfigure}%
    \hfill%
    \begin{subfigure}[t]{\pnpew}\vspace{0pt}
        \centering
        \pnpbox{\dimexpr\linewidth-2\pnpk\relax}{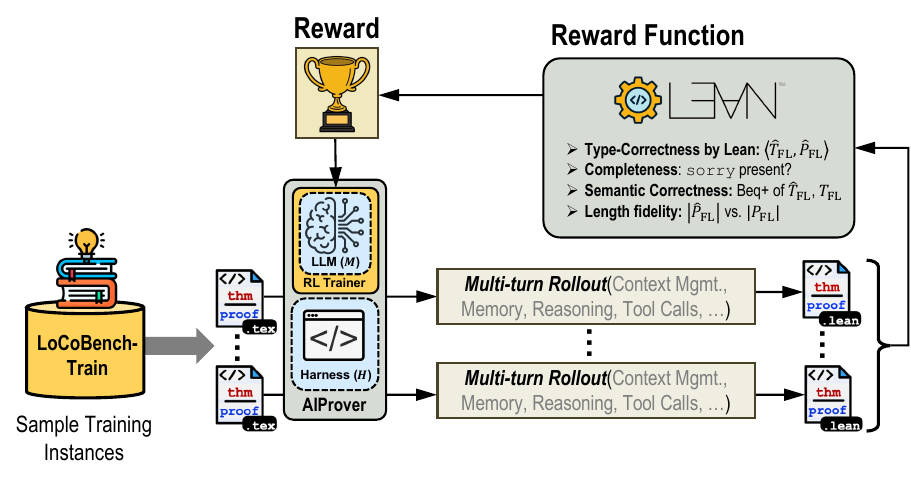}
        \caption{Agentic RLSF: post-training on verifier reward}
        \label{fig:AIProver_rl}
    \end{subfigure}%
    }

    \caption{\textbf{Pipeline of \ourTool{}.}
    (a)~LoCoBench: Mathlib, CSLib, Mizar, and textbook theorem--proof pairs as standalone NL--FL instances.
    (b)~Model and harness are optimized together: each round evolves the harness for the current model, then post-trains the model in it.
    (c)~SAM aligns NL and FL pairs in the LLM.
    (d)~HarnessEvolve optimizes the whole control flow of the harness in an evolutionary search.
    (e)~RLSF post-trains on graded verifier reward, learning from partial successes.  }
    \label{fig:pipeline}
    \vspace{-5mm}%
\end{figure}

\vspace{-1mm}
\textbf{Agentic AFPS and harness design.}
AFPS-tailored agentic harnesses such as
Numina-Lean-Agent~\citep{liu2026numinaleanagent}, OpenGauss~\citep{mathinc2026opengauss},
LeanMarathon~\citep{zhang2026leanmarathon}, and Aristotle~\citep{achim2025aristotle} use
Lean tools like lean-lsp-mcp~\citep{dressler2025leanlspmcp}. However, they optimize only type correctness, not faithfulness to the NL input. Automated harness optimization is emerging, but most methods
keep the model fixed and mutate only a small surface: prompts in
GEPA~\citep{agrawal2025gepa}, or skills and memory in A-Evolve~\citep{lin2026position}, not control flow.

\vspace{-1.5mm}
\textit{Key Differences:} \ourTool{}'s \emph{HarnessEvolve} evolves the harness's full control flow for semantic
correctness, not just type correctness, alternating with RL to tailor harness and model to each other.

\vspace{-1mm}
\textbf{Semantic alignment and auto-formalization verification.}
Checks of NL--FL semantic alignment exist at three levels. \textit{FL-to-FL checks} compare a formalization with a reference e.g.,
BEq$^{+}$~\citep{poiroux2025beq}. \textit{NL-to-FL checks} judge
a formal statement against its informal counterpart e.g., FormalAlign~\citep{lu2024formalalign},
CriticLean~\citep{peng2025criticlean}, LeanScorer~\citep{xuejun2025mathesis}, and FormalRx~\citep{wang2026formalrx}. \textit{Proof-level checks}
are rarer: ProofScore~\citep{cabral2025proofflow} scores structural fidelity to the NL proof,
and FaithformBench~\citep{cornish2026faithformbench} tests whether flawed NL steps are silently repaired.
Crucially, most serve only post-hoc evaluation or filtering.

\vspace{-1.5mm}
\textit{Key Differences:} \ourTool{} makes semantic alignment a training signal, not a post-hoc filter: SAM builds
NL--FL equivalence into the model, RL rewards semantic correctness, and HarnessEvolve selects harnesses
by that reward, so the agent learns to preserve meaning, not just to compile.

\vspace{-2mm}
\section{Preliminaries and Problem Statement}
\label{sec:prelim}
\vspace{-2mm}

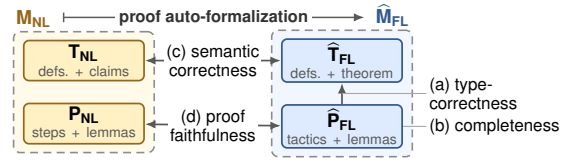
\begin{wrapfigure}[8]{r}{0.45\linewidth}
\vspace{-1.05\baselineskip}
\centering
\begin{tikzpicture}[
  font=\sffamily\fontsize{7}{8}\selectfont, line width=0.65pt,
  >={Latex[length=4pt,width=3.2pt]},
  block/.style={draw, rounded corners=2pt, align=center, inner sep=1.3pt, text width=1.5cm, minimum height=0.53cm},
  nl/.style={block, draw=nlEdge, fill=nlFill},
  fl/.style={block, draw=flEdge, fill=flFill},
  grp/.style={draw=black!45, densely dashed, rounded corners=2.5pt, inner xsep=3.5pt, inner ysep=3pt, line width=0.6pt},
  note/.style={font=\sffamily\fontsize{6.4}{7.2}\selectfont, inner sep=0.8pt, align=center},
  head/.style={font=\sffamily\bfseries\fontsize{7.2}{8}\selectfont, inner sep=1pt},
  rel/.style={<->, draw=black!65},
  chk/.style={->, draw=black!65},
  leader/.style={draw=black!35, line width=0.4pt},
]
\newcommand{\subl}[1]{\\[-1.5pt]{\fontsize{5.2}{5.8}\selectfont\color{black!65}#1}}
\node[nl] (Tnl) {$\textbf{T}_{\textbf{NL}}$\subl{defs.\ + claims}};
\node[nl, below=0.28cm of Tnl] (Pnl) {$\textbf{P}_{\textbf{NL}}$\subl{steps + lemmas}};
\node[fl, right=1.8cm of Tnl] (Tfl) {$\widehat{\textbf{T}}_{\textbf{FL}}$\subl{defs.\ + theorem}};
\node[fl, below=0.28cm of Tfl] (Pfl) {$\widehat{\textbf{P}}_{\textbf{FL}}$\subl{tactics + lemmas}};
\begin{scope}[on background layer]
\node[grp, fill=nlFill!35, fit=(Tnl)(Pnl)] (NLg) {};
\node[grp, fill=flFill!40, fit=(Tfl)(Pfl)] (FLg) {};
\end{scope}
\node[head, text=nlEdge, anchor=south west, inner xsep=0pt] at ($(NLg.north west)+(1pt,0)$) (Hnl) {$\textbf{M}_{\textbf{NL}}$};
\node[head, text=flEdge, anchor=south east, inner xsep=0pt] at ($(FLg.north east)+(-1pt,0)$) (Hfl) {$\widehat{\textbf{M}}_{\textbf{FL}}$};
\coordinate (yref) at ($(Hnl.base)+(0,2.6pt)$);
\draw[{Bar[width=3.6pt]}-{Latex[length=4pt,width=3.2pt]}, draw=black!65, line width=0.55pt]
  ($(Hnl.east |- yref)+(4pt,0)$) -- node[fill=white, inner xsep=1.5pt, inner ysep=0.6pt, text=black!80, font=\sffamily\bfseries\fontsize{6.2}{7}\selectfont] {proof auto-formalization} ($(Hfl.west |- yref)+(-4pt,0)$);
\draw[rel] (Tnl) -- node[note, fill=white] {(c) semantic\\correctness} (Tfl);
\draw[rel] (Pnl) -- node[note, fill=white] {(d) proof\\faithfulness} (Pfl);
\draw[chk] (Pfl) -- coordinate[pos=0.5] (amid) (Tfl);
\node[note, anchor=west, align=left] at ($(FLg.east |- amid)+(5pt,0)$) (La) {(a) type-\\correctness};
\draw[leader] (amid) -- (La.west);
\node[note, anchor=west] at ($(FLg.east |- Pfl)+(5pt,0)$) (Lb) {(b) completeness};
\draw[leader] (Pfl.east) -- (Lb.west);
\end{tikzpicture}
\vspace{-8pt}
\captionsetup{font=small, skip=1pt}
\caption{\textbf{Proof auto-formalization.} Lean checks (a) and (b); a judge checks (c) and (d).}
\vspace{3pt}
\vspace{-0.95\baselineskip}
\end{wrapfigure}
We focus on auto-formalization from NL to Lean~4~\citep{moura2021lean}, a functional programming language and interactive theorem prover. Lean represents theorems as types and proofs as terms of those types, mechanically checked by a small trusted kernel. Other proof assistants include Isabelle/HOL~\citep{nipkow2002isabelle}, Rocq~\citep{rocqprover}, and Mizar~\citep{trybulec1985computer}. Lean, however, is a rapidly growing modern proof assistant with a large and active community. Its extensive library,~\citet{mathlib}, provides a rich collection of formalized definitions and lemmas.

\begin{definition}[Agents and Harnesses]
\label{def:agents}
An agent $A=\langle\mathcal{M},\mathcal{H}\rangle$ pairs an LLM $\mathcal{M}$ with a \emph{harness} (or scaffolding) $\mathcal{H}$, the software layer that executes the actions $\mathcal{M}$ proposes in a stateful environment. It governs the agent's context, tools, memory, and control flow, e.g., the system prompt, how tool calls are handled, what state persists across turns, and when it stops. We focus on ReAct-style~\citep{yao2023react} Python harnesses. \ourTool{} optimizes the full agent, post-training $\mathcal{M}$ and evolving $\mathcal{H}$.
\end{definition}

\vspace{-1.5mm}%
\textbf{Problem Statement \textit{(Proof Auto-Formalization)}.}
The input is a self-contained NL passage $\Mnl=\langle\Tnl,\Pnl\rangle$: $\Tnl$ states a theorem, possibly in several parts, with the definitions and assumptions it relies on, and $\Pnl$ proves it, possibly through intermediate lemmas. The goal is to learn a map $\Mnl\mapsto\rnd{\Mflh}$ to an \emph{equivalent} FL pair $\rnd{\Mflh}=\langle\rnd{\Tflh},\rnd{\Pflh}\rangle$ in Lean, where $\rnd{\Tflh}$ formalizes the definitions and theorems and $\rnd{\Pflh}$ the proof with auxiliary lemmas, subject to four properties.
(a)~\textbf{Type-correctness}: Lean's kernel accepts $\rnd{\Mflh}$.
(b)~\textbf{Completeness}: $\rnd{\Pflh}$ uses no Lean placeholder such as \texttt{sorry}, \texttt{admit}, an empty \texttt{by} block, \texttt{by sorry}, or \texttt{by admit}.
(c)~\textbf{Semantic correctness}: $\rnd{\Tflh}$ states exactly the theorem $\Tnl$ under the same definitions, neither adding assumptions nor dropping conditions.
(d)~\textbf{Proof faithfulness}: $\rnd{\Pflh}$ follows the proof strategy of $\Pnl$, including its intermediate lemmas, without new assumptions.
When all four hold, we call $\rnd{\Mflh}$ equivalent to $\Mnl$. Lean mechanically certifies (a) and (b). Properties (c) and (d) compare the formalization with its NL source; this is undecidable~\citep{church1936unsolvable,turing1936computable} and thus requires a judge.

\clearpage
\section{Proposed Methodology}
\label{sec:method}
\vspace{-1mm}

We introduce \ourTool{}, which jointly post-trains an open-weight LLM $\mathcal{M}$ and evolves the harness $\mathcal{H}$ around it (Fig.~\ref{fig:AIProver_fullPipeline}). Four components drive it: \emph{SAM fine-tuning} (Sec.~\ref{sec:sam}) and \emph{Agentic RLSF} (Sec.~\ref{sec:rl}) for $\mathcal{M}$, \emph{HarnessEvolve} (Sec.~\ref{sec:harness}) for $\mathcal{H}$, and a shared \emph{verifier-based reward} (Sec.~\ref{sec:reward}), averaged over tasks as \emph{fitness}. From an initial agent $A_0=\langle\mathcal{M}_0,\mathcal{H}_0\rangle$, we bootstrap by evolving $\mathcal{H}_0\!\to\!\mathcal{H}_1$ under $\mathcal{M}_0$ and SAM-tuning $\mathcal{M}_0\!\to\!\mathcal{M}_1$. HarnessEvolve and RLSF then alternate for $K$ rounds: round $i$ evolves $\mathcal{H}_i\!\to\!\mathcal{H}_{i+1}$ with $\mathcal{M}_i$ frozen, then post-trains $\mathcal{M}_i\!\to\!\mathcal{M}_{i+1}$ with $\mathcal{H}_{i+1}$ frozen, so each update builds on the other's latest gain. $\mathcal{H}$ and $\mathcal{M}$ use disjoint training subsets, and each round is \emph{gated}: $\langle\mathcal{M}_{i+1},\mathcal{H}_{i+1}\rangle$ is kept only if its held-out fitness improves, else rolled back.

\vspace{-1mm}
\subsection{Semantic Alignment Model (SAM) Fine-Tuning}
\label{sec:sam}

We propose \textit{Semantic Alignment Model (SAM)} (Fig.~\ref{fig:AIProver_SAMpipeline}), a decoder-only LLM fine-tuned to represent $\Mnl$ and $\Mfl$ as the same mathematical object, which language modeling alone does not induce (Sec.~\ref{sec:intro}). We do so by jointly optimizing a contrastive term with the language-modeling loss, aligning the representations of $\Mnl$ and $\Mfl$ in an \textit{NL/FL shared semantic space}. SAM thus remains a generative LLM, but conditions generation on semantic meaning rather than structural form alone.

\textbf{Two views.} Let $\mathcal{M}_\theta$ be the frozen decoder-only base LLM with parameters $\theta$, and $\mathcal{M}_{\theta,\Delta}$ the same model with trainable low-rank adapters $\Delta$ on attention projections~\citep{hu2021lora}. Given an instance $\langle\Mnl,\Mfl\rangle$ with gold formalization $\Mfl$, both are tokenized and joined via $\pred$, a reserved tokenizer token. This gives $\rnd{u=\Mnl\,\Vert\,\pred\,\Vert\,\Mfl}$, and two forward passes yield two views. The first pass runs $\mathcal{M}_{\theta,\Delta}$ on $u$ and takes the state $h_{\mathrm{NL}}\in\mathbb{R}^{d}$ at $\pred$ as the \textbf{NL view}. By causal attention, $h_{\mathrm{NL}}$ depends only on $\Mnl$ and is where generation of $\rnd{\Mflh}$ begins. The second pass runs on $\Mfl$ alone and takes its final state $h_{\mathrm{FL}}\in\mathbb{R}^{d}$ as the \textbf{FL view}. This bars the FL view from $\Mnl$ and from matching the NL view via shared context, so they align on meaning alone.

\textbf{Training losses.} The two passes support two training objectives. In the pass over $u$, the \textit{language-modeling head} predicts the next token at every position, and we compute the cross-entropy loss $\mathcal{L}_{\mathrm{CE}}$ only on the gold tokens $F=\{\,t : u_t \in \Mfl\,\}$ of $\Mfl$, with $\Mnl,\pred$ as context. The trainable \textit{projection head} $g_\phi:\mathbb{R}^{d}\!\to\!\mathbb{R}^{m}$ is a two-layer MLP shared by both views. It maps each view to the unit sphere, $\rnd{\znl=g_\phi(h_{\mathrm{NL}})/\lVert g_\phi(h_{\mathrm{NL}})\rVert\in\mathbb{S}^{m-1}}$, and likewise $h_{\mathrm{FL}}$ to $\zfl$. Over a batch of $B$ NL--FL pairs, let $S_{ij}={z_{\mathrm{NL},i}}^{\!\top}\, z_{\mathrm{FL},j}/\tau$ be the cosine similarity of the $i$-th NL view and $j$-th FL view, both unit vectors, scaled by temperature $\tau$. The symmetric InfoNCE loss $\mathcal{L}_{\mathrm{align}}$~\citep{oord2018cpc} pulls each pair's two views together and contrasts them with the other $B-1$ pairs:
{\setlength{\abovedisplayskip}{3pt}\setlength{\belowdisplayskip}{3pt}\setlength{\abovedisplayshortskip}{1pt}\setlength{\belowdisplayshortskip}{3pt}%
\begin{equation}
  \mathcal{L}_{\mathrm{CE}} = -\frac{1}{|F|}\sum_{t\in F}\log P_{\theta,\Delta}\!\big(u_t \mid u_{<t}\big),
  \qquad
  \mathcal{L}_{\mathrm{align}} = -\frac{1}{2B}\sum_{i=1}^{B}\bigg[\log\frac{e^{S_{ii}}}{\sum_{j} e^{S_{ij}}}+\log\frac{e^{S_{ii}}}{\sum_{j} e^{S_{ji}}}\bigg].
  \label{eq:losses}
\end{equation}}
\textbf{Full objective.} We sum the two losses with weight $\lambda>0$, $\rnd{\mathcal{L}=\mathcal{L}_{\mathrm{CE}}+\lambda\,\mathcal{L}_{\mathrm{align}}}$, and train only $\Delta$, $g_\phi$, and the $\pred$ embedding $p$. At $\lambda{=}0$ this reduces to supervised fine-tuning (SFT), and $\lambda{=}1$ gives full SAM. Since $g_\phi$ is discarded at generation, alignment shapes generation only through the shared $\Delta$, pushing $h_{\mathrm{NL}}$ to encode the formalization's meaning, not only next-token statistics.

\vspace{-1mm}
\subsection{Model-Harness Co-Evolution}
\label{sec:coevo}

\vspace{-2mm}
\subsubsection{Reward Design}
\label{sec:reward}

\begin{wraptable}[10]{r}{0.51\linewidth}
\vspace{-1.7\baselineskip}%
\centering\small
\setlength{\tabcolsep}{2.5pt}
\captionsetup{font=small, skip=3pt}
\caption{\textbf{Reward Ladder.} Reward $r$ of a candidate $\rnd{\Mflh}=\langle\rnd{\Tflh},\rnd{\Pflh}\rangle$ per outcome of the four checks; $v_{\mathrm{TC}},v_{\mathrm{CP}}\in\{0,1\}$, $v_{\mathrm{SC}}\in\{0,0.5,1\}$, $v_{\mathrm{LF}}\in(0,1]$; \checkmark/$\times$ denote $1$/$0$.}
\label{tab:reward}
\begin{tabular}{@{}lccc l@{}}
\toprule
\textbf{Outcome} & TC & CP & SC & \textbf{Reward} ($r$) \\
\midrule
No answer        & --          & --          & --                & $0$ \\
Ill-typed        & $\times$    & --          & --                & $0.05$ \\
Incomplete proof & \checkmark  & $\times$    & $v_{\mathrm{SC}}$ & $0.15+0.35\,v_{\mathrm{SC}}$ \\
Complete proof   & \checkmark  & \checkmark  & $v_{\mathrm{SC}}$ & $0.30+0.60\,v_{\mathrm{SC}}$ \\
Solved           & \checkmark  & \checkmark  & \checkmark        & $1-0.10\,(1-v_{\mathrm{LF}})$ \\
\bottomrule
\end{tabular}
\end{wraptable}
Given $\Mnl$, the agent $A$ returns a candidate $\rnd{\Mflh}=\langle\rnd{\Tflh},\rnd{\Pflh}\rangle$; during the co-evolution, a gold formalization $\rnd{\Mfl=\langle\Tfl,\Pfl\rangle}$ is also available. We score the candidate with four \emph{checks}\footnote{Currently, no reliable check exists for NL--FL proof faithfulness: heuristics such as ProofScore~\citep{cabral2025proofflow} and FaithformBench~\citep{cornish2026faithformbench} rely on LLM judges, which are stochastic and slow when scoring many rollouts concurrently. We thus design our reward around sound formal verifiers.}, ordered so that each presupposes the preceding one. Every check $k$ returns a value $v_k\in[0,1]$ and a \emph{certificate} $\xi_k$ from the verifier.
(1)~\textbf{Type-correctness (TC)}: whether Lean accepts $\rnd{\Mflh}$: every definition and tactic step resolves and its trusted kernel type-checks the proof term, so $v_{\mathrm{TC}}\in\{0,1\}$. $\xi_{\mathrm{TC}}$ contains Lean's diagnostics verbatim: each error's position, the failing tactic with its open goals, or the expected versus actual type.
(2)~\textbf{Completeness (CP)}: whether the proof closes every goal without a placeholder (\texttt{sorry}) or axioms beyond Lean's standard ones. Given a type-correct proof, CP checks for \texttt{sorry} and audits its axioms (\texttt{\#print axioms}), failing on a kernel bypass (\texttt{native\_decide}) or an agent-introduced axiom, so $v_{\mathrm{CP}}\in\{0,1\}$. $\xi_{\mathrm{CP}}$ identifies the placeholder's position or violating axiom.
(3)~\textbf{Semantic correctness (SC)}: whether $\rnd{\Tflh}$ states the same theorem as $\Tfl$.
Since a correct formalization may differ from the gold in variable names, hypothesis order, or Mathlib lemmas, we test logical equivalence with an extended BEq$^{+}$~\citep{poiroux2025beq} (App.~\ref{app:beq}), which attempts to prove $\Tfl\Rightarrow\rnd{\Tflh}$ and $\rnd{\Tflh}\Rightarrow\Tfl$ in Lean with automated tactics. One direction alone shows $\rnd{\Tflh}$ weaker or stronger than $\Tfl$ ($v_{\mathrm{SC}}=0.5$); both certify $\Tfl\Leftrightarrow\rnd{\Tflh}$ ($v_{\mathrm{SC}}=1$). $\xi_{\mathrm{SC}}$ records both verdicts, and, on a miss, their asymmetry.
(4)~\textbf{Length fidelity (LF)}: how close the length of $\rnd{\Pflh}$ is that of $\Pfl$, $v_{\mathrm{LF}}=\exp\!\big(-\big|\ln\tfrac{|\rnd{\Pflh}|+c}{|\Pfl|+c}\big|\big)\in(0,1]$ with $c=60$ characters of slack. This is because, a far shorter proof typically relies on heavy automation like \texttt{grind}, and a longer one meanders; both less \textit{human-readable}. $\xi_{\mathrm{LF}}$ reports length ratio.

\vspace{-1mm}
We compose the checks into one consolidated \textbf{reward} $r\in[0,1]$ and one consolidated \textbf{certificate} $\xi$ per instance. The reward follows the ladder in Table~\ref{tab:reward}:
no Lean file, crash, or timeout earns $r=0$, and an ill-typed file $0.05$. Once it type-checks, reward grows with semantic match: an incomplete proof earns $0.15$, $0.325$, or $0.50$ for an unmatched, half-matched, or matched theorem, and a complete proof $0.30$, $0.60$, or $0.90$. A matched theorem thus outweighs completeness, as a flawless proof of the wrong theorem is not useful. Finally, a solved instance loses at most $0.10$ for length. The certificate $\xi$ states the verdict and $r$, then concatenates $\xi_{\mathrm{TC}},\dots,\xi_{\mathrm{LF}}$ in ladder order, so the first failing check is the first to fix. We use $r$ as the per-rollout reward for RLSF (Sec.~\ref{sec:rl}) and, averaged over tasks, as a harness's fitness, while $\xi$ is read by HarnessEvolve's mutator agent as feedback (Sec.~\ref{sec:harness}).

\vspace{-2mm}
\subsubsection{HarnessEvolve: Certificate-Driven Evolutionary Search}
\label{sec:harness}
\vspace{-1.5mm}
We propose \textit{HarnessEvolve} (Fig.~\ref{fig:AIProver_harnessEvolution}) to evolve the current harness $\mathcal{H}_i$ into $\mathcal{H}_{i+1}$ while the model $\mathcal{M}_i$ stays frozen. A harness is a finite but arbitrarily long program, so we search a countably infinite space for the harness maximizing the \emph{fitness} $f(\cdot)$ on the training set $\rnd{\mathcal{D}=\{\langle\Mnl,\Mfl\rangle\}}$ of NL passages and gold formalizations. Each call runs $R$ rounds. In round $r\in[1,R]$, a parent harness is selected from the \emph{search memory} and a coding agent mutates it into a child, which is evaluated and recorded.

\vspace{-1mm}
\textbf{Search memory.} The memory is a tree rooted at the current harness $\mathcal{H}_i$, which seeds the search. A node stores a harness $h$ with its fitness $f(h)$, its per-instance rewards $r$ and certificates $\xi$, and its verdict, \emph{accepted} or \emph{rejected}. An edge from parent to child stores the source diff between the two harnesses and their fitness difference. Crucially, the memory keeps every evaluated harness regardless of its verdict. The mutator thus learns which changes raised fitness from the accepted harnesses and their certificates, and which changes to avoid from the rejected ones. This is inspired by conflict-driven clause learning in SAT solvers~\citep{buss2026ercl}, which learn a clause from every refuted assignment: here, later mutations are conditioned on every refuted design rather than on the selected parent alone. Each round grows the tree by one node through the following four stages:

\vspace{-1mm}
\textbf{1.~Parent selection.} We select a parent from one of the accepted nodes of earlier rounds by an upper-confidence rule as in UCB1 and UCT~\citep{auer2002ucb,kocsis2006uct}, which balances \textit{exploiting} high-fitness nodes against \textit{exploring} rarely expanded ones. The parent $h_{\mathrm{par}}$ is the accepted $h$ that maximizes $\rnd{Q(h)+c\sqrt{\ln(1+t)/(1+k(h))}}$ with $\rnd{Q(h)=1-\min(1,(f^{*}-f(h))/3\sigma)}$, where $t$ is the round index, $f^{*}$ the best fitness in memory, $\sigma$ the fitness standard error, and $k(h)$ the number of children of $h$. The first term $Q$ exploits, measuring the gap to the best fitness in standard errors so that nodes within noise of one another score alike. The second term explores, favoring nodes with few children and growing with $t$, and its weight $c$ rises with each consecutive non-accepting round.

\vspace{-1.5mm}
\textbf{2.~Mutation.} Given the parent $h_{\mathrm{par}}$ and the memory tree, we prompt a frontier coding agent to rewrite $h_{\mathrm{par}}$ into a child $h_{\mathrm{child}}$. The tree, with every node's fitness and per-instance certificates, shows the mutator which edits raised fitness and which did not, so it avoids refuted hypotheses. Using read, write, search, and bash tools, the mutator infers recurring failures from the certificates of $h_{\mathrm{par}}$, e.g., low type-correctness or repeated \texttt{sorry}, and changes the control flow expecting to raise fitness.

\vspace{-1.5mm}
\textbf{3.~Rollout and fitness.} We run $A_{\mathrm{child}}=\langle\mathcal{M}_i,h_{\mathrm{child}}\rangle$ on every instance $d=\langle\Mnl,\Mfl\rangle\in\mathcal{D}$, $k$ times each to average out stochasticity. Rollout $j$ on $d$ earns reward $r_j(d)$ (Sec.~\ref{sec:reward}), and the fitness is their mean, $\rnd{f(h_{\mathrm{child}})=\tfrac{1}{k|\mathcal{D}|}\sum_{d\in\mathcal{D}}\sum_{j\le k} r_j(d)}$. The certificates $\xi$ are stored in memory.

\vspace{-1.5mm}
\textbf{4.~Child acceptance.} The child is labeled \emph{accepted} if $f(h_{\mathrm{child}})>f^{*}$, else \emph{rejected}, and is added to the memory in either case. After $R$ rounds, the best accepted harness goes to RLSF (Sec.~\ref{sec:rl}).

\vspace{-1mm}
\begin{table}[t]
\centering
\captionsetup{font=small, skip=4pt}
\begin{minipage}[t]{0.49\textwidth}\vspace{0pt}
\caption{\textbf{LoCoBench composition.} Mathlib+CSLib entries carry gold Lean, MML (Mizar Math.\ Lib.) ones are NL-only, and Bounded Arith.\ is from~\citet{Li25-91}.}
\label{tab:expert_grouping}
\scriptsize
\renewcommand{\arraystretch}{1.12}
\setlength{\tabcolsep}{2.6pt}
\setlength{\aboverulesep}{0pt}\setlength{\belowrulesep}{0pt}
\begin{tabular}{@{}l r r r r@{}}
\toprule
\textbf{Domain} & \multicolumn{2}{c}{\textbf{Train}} & \textbf{Val} & \textbf{Total} \\
\cmidrule(lr){2-3}\cmidrule(lr){4-4}
\addlinespace[2.5pt]
 & \shortstack[r]{Mathlib+CSLib\\{\color{black!55}\itshape NL+Lean}} & \shortstack[r]{MML\\{\color{black!55}\itshape NL only}} & \shortstack[r]{MML/\\textbook\\{\color{black!55}\itshape NL only}} & \\
\midrule
\rowcolor{black!6}\multicolumn{4}{@{}l}{\textit{Algebraic Structures}} & $28{,}432$ \\
\enspace Ring Theory        & $6{,}841$  & $3{,}303$   & $100$ & \\
\enspace Group Theory       & $3{,}310$  & $14{,}778$  & $100$ & \\
\addlinespace[2pt]
\rowcolor{black!6}\multicolumn{4}{@{}l}{\textit{Foundations, Logic \& Complexity}} & $14{,}010$ \\
\enspace Set Theory         & $2{,}871$  & $555$       & $100$ & \\
\enspace Logic              & $1{,}283$  & $2{,}637$   & $100$ & \\
\enspace Computability      & $938$      & $3{,}579$   & $100$ & \\
\enspace Model Theory       & $819$      & $857$       & $100$ & \\
\enspace Bounded Arith.\ & $-$        & $-$         & $71$  & \\
\addlinespace[2pt]
\rowcolor{black!6}\multicolumn{4}{@{}l}{\textit{Number Theory}} & $16{,}471$ \\
\enspace Number Theory      & $2{,}695$  & $13{,}676$  & $100$ & \\
\midrule
\textbf{Total} & $\mathbf{18{,}757}$ & $\mathbf{39{,}385}$ & $\mathbf{771}$ & $\mathbf{58{,}913}$ \\
\bottomrule
\end{tabular}
\end{minipage}\hfill
\begin{minipage}[t]{6.62cm}\vspace{0pt}

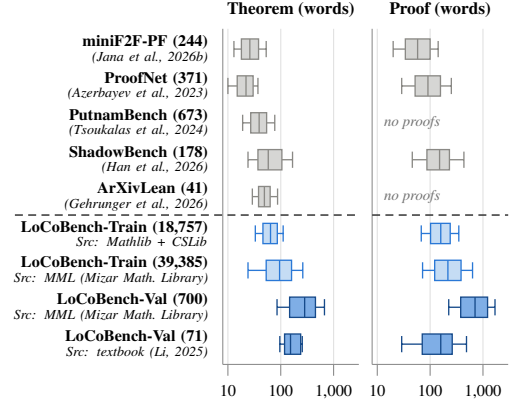
\captionof{figure}{\textbf{NL length of LoCoBench vs.\ prior benchmarks.} Boxes span the quartiles and whiskers the 10th--90th \%iles. Our Val proofs are ${\sim}8\times$ ProofNet's.}
\label{fig:locobench_lengths}
\raggedleft\par\vspace{-4pt}
\begin{tikzpicture}[font=\scriptsize]
\begin{groupplot}[
  group style={group size=2 by 1, horizontal sep=5pt, yticklabels at=edge left},
  width=1.8cm, height=4.44cm, scale only axis, xmode=log, log basis x=10, clip=false,
  ytick={1,...,9}, ytick style={draw=none}, y dir=reverse,
  yticklabels={
    {\textbf{miniF2F-PF (244)}\\[-0.6pt]{\fontsize{5.2}{5.4}\selectfont\itshape \citep{jana2025proofbridge}}},
    {\textbf{ProofNet (371)}\\[-0.6pt]{\fontsize{5.2}{5.4}\selectfont\itshape \citep{azerbayev2023proofnet}}},
    {\textbf{PutnamBench (673)}\\[-0.6pt]{\fontsize{5.2}{5.4}\selectfont\itshape \citep{tsoukalas2024putnambench}}},
    {\textbf{ShadowBench (178)}\\[-0.6pt]{\fontsize{5.2}{5.4}\selectfont\itshape \citep{han2026shadowbench}}},
    {\textbf{ArXivLean (41)}\\[-0.6pt]{\fontsize{5.2}{5.4}\selectfont\itshape \citep{matharena2026arxivlean}}},
    {\textbf{LoCoBench-Train (18,757)}\\[-0.6pt]{\fontsize{5.2}{5.4}\selectfont\itshape Src: Mathlib + CSLib}},
    {\textbf{LoCoBench-Train (39,385)}\\[-0.6pt]{\fontsize{5.2}{5.4}\selectfont\itshape Src: MML (Mizar Math.\ Library)}},
    {\textbf{LoCoBench-Val (700)}\\[-0.6pt]{\fontsize{5.2}{5.4}\selectfont\itshape Src: MML (Mizar Math.\ Library)}},
    {\textbf{LoCoBench-Val (71)}\\[-0.6pt]{\fontsize{5.2}{5.4}\selectfont\itshape Src: textbook~\citep{Li25-91}}}},
  yticklabel style={align=right, text width=2.55cm, font=\fontsize{6}{6.1}\selectfont, inner xsep=1.5pt, inner ysep=0pt},
  xtick={10,100,1000}, xticklabels={10,100,1{,}000},
  xtick pos=bottom, axis x line*=bottom, axis y line*=left, axis line style={black!45},
  xmajorgrids, grid style={black!12},
  ymin=0.45, ymax=9.55, xmin=8, xmax=3000,
  boxplot/draw direction=x, boxplot/box extend=0.55,
  every boxplot/.style={line width=0.5pt, mark=none},
  title style={font=\fontsize{6.5}{7.5}\selectfont\bfseries, yshift=-5pt},
  tick label style={font=\fontsize{6.2}{7}\selectfont}, tick align=outside,
]
\nextgroupplot[title={Theorem (words)}]
\addplot[boxplot prepared={draw position=1, lower whisker=13, lower quartile=18, median=26, upper quartile=38, upper whisker=53}, fill=bmGray!35, draw=bmGray] coordinates {};
\addplot[boxplot prepared={draw position=2, lower whisker=10, lower quartile=15, median=22, upper quartile=30, upper whisker=37}, fill=bmGray!35, draw=bmGray] coordinates {};
\addplot[boxplot prepared={draw position=3, lower whisker=19, lower quartile=27, median=39, upper quartile=54, upper whisker=77}, fill=bmGray!35, draw=bmGray] coordinates {};
\addplot[boxplot prepared={draw position=4, lower whisker=24, lower quartile=37, median=58, upper quartile=105, upper whisker=167}, fill=bmGray!35, draw=bmGray] coordinates {};
\addplot[boxplot prepared={draw position=5, lower whisker=29, lower quartile=38, median=49, upper quartile=63, upper whisker=87}, fill=bmGray!35, draw=bmGray] coordinates {};
\addplot[boxplot prepared={draw position=6, lower whisker=33, lower quartile=46, median=64, upper quartile=86, upper whisker=111}, fill=lcLight!45, draw=lcMid] coordinates {};
\addplot[boxplot prepared={draw position=7, lower whisker=24, lower quartile=53, median=95, upper quartile=160, upper whisker=261}, fill=lcLight!45, draw=lcMid] coordinates {};
\addplot[boxplot prepared={draw position=8, lower whisker=85, lower quartile=147, median=283, upper quartile=453, upper whisker=671}, fill=lcMid!60, draw=lcDark] coordinates {};
\addplot[boxplot prepared={draw position=9, lower whisker=96, lower quartile=118, median=153, upper quartile=238, upper whisker=255}, fill=lcMid!60, draw=lcDark] coordinates {};
\draw[densely dashed, line width=0.6pt, black!75] ([xshift=-2.75cm]axis cs:8,5.5) -- (axis cs:3000,5.5);
\nextgroupplot[title={Proof (words)}]
\addplot[boxplot prepared={draw position=1, lower whisker=20, lower quartile=34, median=58, upper quartile=100, upper whisker=142}, fill=bmGray!35, draw=bmGray] coordinates {};
\addplot[boxplot prepared={draw position=2, lower whisker=29, lower quartile=52, median=91, upper quartile=157, upper whisker=252}, fill=bmGray!35, draw=bmGray] coordinates {};
\node[font=\fontsize{5.6}{6.4}\selectfont\itshape, text=black!55, anchor=west] at (axis cs:9,3) {no proofs};
\addplot[boxplot prepared={draw position=4, lower whisker=46, lower quartile=87, median=151, upper quartile=232, upper whisker=436}, fill=bmGray!35, draw=bmGray] coordinates {};
\node[font=\fontsize{5.6}{6.4}\selectfont\itshape, text=black!55, anchor=west] at (axis cs:9,5) {no proofs};
\addplot[boxplot prepared={draw position=6, lower whisker=68, lower quartile=102, median=159, upper quartile=240, upper whisker=349}, fill=lcLight!45, draw=lcMid] coordinates {};
\addplot[boxplot prepared={draw position=7, lower whisker=72, lower quartile=122, median=217, upper quartile=385, upper whisker=635}, fill=lcLight!45, draw=lcMid] coordinates {};
\addplot[boxplot prepared={draw position=8, lower whisker=225, lower quartile=379, median=708, upper quartile=1222, upper whisker=1689}, fill=lcMid!60, draw=lcDark] coordinates {};
\addplot[boxplot prepared={draw position=9, lower whisker=29, lower quartile=71, median=159, upper quartile=262, upper whisker=488}, fill=lcMid!60, draw=lcDark] coordinates {};
\draw[densely dashed, line width=0.6pt, black!75] (axis cs:8,5.5) -- (axis cs:3000,5.5);
\end{groupplot}
\end{tikzpicture}
\end{minipage}
\vspace{-20pt}
\end{table}

\subsubsection{Agentic RLSF: Multi-Turn Post-Training via Symbolic Feedback}
\label{sec:rl}
\vspace{-1mm}
We propose \textit{Agentic Reinforcement Learning via Symbolic Feedback (RLSF)} for multi-turn agents, to post-train $\mathcal{M}_i$ into $\mathcal{M}_{i+1}$ inside the frozen harness $\mathcal{H}_{i+1}$ (Fig.~\ref{fig:AIProver_rl}). Given an instance $\langle\Mnl,\Mfl\rangle$, an episode is one run of the agent $\langle\mathcal{M}_i,\mathcal{H}_{i+1}\rangle$ on $\Mnl$: within the control flow of $\mathcal{H}_{i+1}$, the model attempts auto-formalization by reasoning, calling tools, and observing their results turn after turn. Only the Lean file it finally writes is rewarded, which favors reaching a good final state over a long horizon of actions. The reward itself (Sec.~\ref{sec:reward}) follows the paradigm of rewarding with sound symbolic tools~\citep{jha2025rlsf} instead of unreliable learned reward models. Our tools are the Lean kernel and BEq$^{+}$, and the reward further scores semantic correctness and proof length (Table~\ref{tab:reward}). 

\vspace{-1mm}
We optimize with Group-Relative Policy Optimization (GRPO)~\citep{shao2024deepseekmath}, which is outcome-supervised. For each training instance $d=\langle\Mnl,\Mfl\rangle$, we run $K$ episodes of the agent on $\Mnl$, obtaining $K$ trajectories (\emph{rollouts}) whose final Lean files are scored against $\Mfl$ using the reward of Sec.~\ref{sec:reward}, yielding $r_1,\dots,r_K$. GRPO maximizes the clipped PPO surrogate objective~\citep{schulman2017ppo} with a small KL penalty to $\mathcal{M}_i$, but replaces the critic baseline with the group mean $\bar{r}$, giving rollout $k$ the advantage $A_k=r_k-\bar{r}$. We do not divide by the group standard deviation~\citep{liu2025drgrpo}, since most groups have little spread and normalization would amplify the noisiest ones. The objective applies \textit{only to tokens} the model generated. Lean output and tool results condition the policy but receive no gradient. Outcome supervision makes the \textit{fine-grained ladder} in Table~\ref{tab:reward} essential. Under the binary verification reward of RL-trained provers~\citep{xin2024deepseekproverv15,wang2025kimina}, a group whose rollouts all fail or all succeed has $A_k=0$ for every $k$ and yields no gradient. Our ladder instead separates ill-typed, incomplete, and complete proofs even within a failing group. We train low-rank adapters~\citep{hu2021lora} rather than full-parameter fine-tuning (see App.~\ref{app:impl}).

\vspace{-4mm}
\section{Experimental Evaluation}
\label{sec:results}
\begin{table*}[t]
\centering
\captionsetup{font=small, skip=3pt}
\caption{\textbf{Proof auto-formalization on LoCoBench-Val: \ourTool{} vs.\ 39 SoTA systems.} Pass@4 (\%) per field and overall (shaded columns) under the metrics of Sec.~\ref{sec:metrics}: TC (compiles), TC+SC w/ or w/o \texttt{sorry} (also semantically correct), and TC+SC full proofs (complete). \textbf{Bold}/\underline{underline} = best/second best per column.  }
\label{tab:main_results}
\renewcommand{\arraystretch}{0.78}
\setlength{\tabcolsep}{2.5pt}
\setlength{\aboverulesep}{0pt}
\setlength{\belowrulesep}{0pt}
\resizebox{\textwidth}{!}{%
\begin{tabular}{@{} l ccc !{\hspace{2pt}{\color{black!45}\vrule width 0.5pt}\hspace{2pt}} ccc !{\hspace{2pt}{\color{black!45}\vrule width 0.5pt}\hspace{2pt}} ccc !{\hspace{2pt}{\color{black!45}\vrule width 0.5pt}\hspace{2pt}} >{\columncolor{gray!15}}c >{\columncolor{gray!15}}c >{\columncolor{gray!15}}c @{}}
\toprule
&
\multicolumn{3}{c}{\shortstack{\textbf{Algebraic Structures}\\{\scriptsize($n$=200)}}} &
\multicolumn{3}{c}{\shortstack{\textbf{Foundations, Logic}\\\textbf{\& Complexity} {\scriptsize($n$=471)}}} &
\multicolumn{3}{c}{\shortstack{\textbf{Number Theory}\\{\scriptsize($n$=100)}}} &
\multicolumn{3}{c}{\cellcolor{gray!15}\shortstack{\textbf{Overall}\\{\scriptsize($n$=771)}}} \\
\cmidrule(lr){2-4} \cmidrule(lr){5-7} \cmidrule(lr){8-10} \cmidrule(lr){11-13}
\addlinespace[2.5pt]
\textbf{Model Name} &
TC & TC+SC & TC+SC & TC & TC+SC & TC+SC & TC & TC+SC & TC+SC & TC & TC+SC & TC+SC \\
& & {\scriptsize\color{black!55}(w/ or w/o sorry)} & {\scriptsize\color{black!55}(full proofs)}
& & {\scriptsize\color{black!55}(w/ or w/o sorry)} & {\scriptsize\color{black!55}(full proofs)}
& & {\scriptsize\color{black!55}(w/ or w/o sorry)} & {\scriptsize\color{black!55}(full proofs)}
& & {\scriptsize\color{black!55}(w/ or w/o sorry)} & {\scriptsize\color{black!55}(full proofs)} \\
\midrule
\addlinespace[1.2pt]
\rowcolor{black!9}\multicolumn{13}{@{}l}{\textbf{AFPS Models (single-turn)}} \\
Kimina-Autoformalizer-7B
  & 52.0 & 3.5 & 0.0 & 57.3 & 1.1 & 0.0 & 80.0 & 26.0 & 0.0 & 58.9 & 4.9 & 0.0 \\
StepFun-Formalizer-7B
  & 16.0 & 5.5 & 0.0 & 9.8 & 1.5 & 0.0 & 51.0 & 26.0 & 0.0 & 16.7 & 5.7 & 0.0 \\
Goedel-Prover-V2-32B
  & 29.5 & 0.0 & 0.0 & 56.5 & 0.6 & 0.6 & 22.0 & 0.0 & 0.0 & 45.0 & 0.4 & 0.4 \\
DeepSeek-Prover-V2-7B
  & 10.0 & 0.0 & 0.0 & 12.7 & 0.2 & 0.2 & 7.0 & 0.0 & 0.0 & 11.3 & 0.1 & 0.1 \\
Kimina-Prover-Distill-8B
  & 13.0 & 1.0 & 1.0 & 15.1 & 0.6 & 0.6 & 15.0 & 10.0 & 10.0 & 14.5 & 1.9 & 1.9 \\
Kimina-Prover-RL-1.7B
  & 1.5 & 0.5 & 0.5 & 1.5 & 0.0 & 0.0 & 8.0 & 7.0 & 7.0 & 2.3 & 1.0 & 1.0 \\
Kimina-Autoformalizer-7B $\to$ Goedel-Prover-V2-32B
  & 14.5 & 1.0 & 1.0 & 24.4 & 0.2 & 0.2 & 32.0 & 10.0 & 10.0 & 22.8 & 1.7 & 1.7 \\
Kimina-Autoformalizer-7B $\to$ DeepSeek-Prover-V2-7B
  & 16.0 & 1.5 & 1.5 & 22.7 & 0.2 & 0.2 & 34.0 & 11.0 & 10.0 & 22.4 & 1.9 & 1.8 \\
Kimina-Autoformalizer-7B $\to$ Kimina-Prover-Distill-8B
  & 12.0 & 1.0 & 1.0 & 23.4 & 0.8 & 0.8 & 28.0 & 12.0 & 12.0 & 21.0 & 2.3 & 2.3 \\
StepFun-Formalizer-7B $\to$ Goedel-Prover-V2-32B 
  & 10.0 & 3.0 & 3.0 & 6.2 & 0.8 & 0.8 & 18.0 & 12.0 & 8.0 & 8.7 & 2.9 & 2.3 \\
StepFun-Formalizer-7B $\to$ DeepSeek-Prover-V2-7B 
  & 18.5 & 5.5 & 0.0 & 10.6 & 1.5 & 0.0 & 56.0 & 26.0 & 1.0 & 18.5 & 5.7 & 0.1 \\
StepFun-Formalizer-7B $\to$ Kimina-Prover-Distill-8B 
  & 18.5 & 5.5 & 2.0 & 10.6 & 1.5 & 0.4 & 56.0 & 26.0 & 3.0 & 18.5 & 5.7 & 1.2 \\
\midrule
\addlinespace[1.2pt]
\rowcolor{black!9}\multicolumn{13}{@{}l}{\textbf{AFPS Agents (multi-turn)}} \\
StepFun-Formlzr.-7B $\to$ Axiom AXLE (w/ GPT-OSS-120B) & 18.5 & 5.5 & 4.5 & 13.8 & 1.3 & 0.8 & 61.0 & 29.0 & 12.0 & 21.1 & 6.0 & 3.2 \\
StepFun-Formlzr.-7B $\to$ Hilbert
  & 17.5 & 6.0 & 5.0 & 10.6 & 1.1 & 0.6 & 56.0 & 27.0 & 17.0 & 18.3 & 5.7 & 3.9 \\
Leanstral-1.5-119B-A6B (no tools)
  & 16.5 & 5.5 & 4.5 & 20.2 & 2.3 & 2.1 & 14.0 & 5.0 & 4.0 & 18.4 & 3.5 & 3.0 \\
Leanstral-1.5-119B-A6B (w/ tools)
  & 21.0 & 14.5 & 14.5 & 29.3 & 10.2 & 10.2 & 31.0 & 20.0 & 20.0 & 27.4 & 12.6 & 12.6 \\
Aristotle
  & 43.0 & 25.0 & 25.0 & 42.9 & 13.6 & 13.4 & 45.0 & 29.0 & 29.0 & 43.2 & 18.5 & 18.4 \rule[-0.8ex]{0pt}{0pt} \\
\hdashline
\rule{0pt}{2.5ex}Math-Inc OpenGauss (w/ Leanstral-1.5-119B-A6B)
  & 62.0 & 29.0 & 28.5 & 54.8 & 15.1 & 14.9 & 60.0 & 42.0 & 39.0 & 57.3 & 22.2 & 21.5 \\
\phantom{Math-Inc OpenGauss }+ Codex (w/ GPT-5.5)
  & 92.5 & 23.5 & 23.5 & 96.6 & 16.8 & 16.6 & 90.0 & 32.0 & 32.0 & 94.7 & 20.5 & 20.4 \\
\phantom{Math-Inc OpenGauss }+ Codex (w/ GPT-5.6-Sol)
  & \underline{99.5} & 59.0 & 58.5 & 98.5 & 29.9 & 29.9 & \underline{99.0} & 66.0 & 65.0 & 98.8 & 42.2 & 41.9 \\
\phantom{Math-Inc OpenGauss }+ Claude Code (w/ Claude-Opus-4.7)
  & \underline{99.5} & 46.0 & 43.0 & 99.2 & 16.6 & 15.9 & \underline{99.0} & 55.0 & 50.0 & \underline{99.2} & 29.2 & 27.4 \\
\phantom{Math-Inc OpenGauss }+ Claude Code (w/ Claude-Opus-5)
  & 98.0 & 67.5 & 67.5 & \underline{99.4} & 45.0 & 45.0 & \underline{99.0} & 73.0 & 72.0 & 99.0 & 54.5 & 54.3 \rule[-0.8ex]{0pt}{0pt} \\
\hdashline
\rule{0pt}{2.5ex}Numina-Lean-Agent + Codex (w/ GPT-5.6-Sol)%
  & \textbf{100.0} & 56.0 & 56.0 & \textbf{100.0} & 36.1 & 35.9 & \textbf{100.0} & 62.0 & 62.0 & \textbf{100.0} & 44.6 & 44.5 \\
\phantom{Numina-Lean-Agent }+ Claude Code (w/ Claude-Opus-5)
  & \textbf{100.0} & 77.0 & 77.0 & \textbf{100.0} & \underline{70.1} & \underline{70.1} & \textbf{100.0} & 76.0 & 75.0 & \textbf{100.0} & \underline{72.6} & \underline{72.5} \\
\midrule
\addlinespace[1.2pt]
\rowcolor{black!9}\multicolumn{13}{@{}l}{\textbf{Foundation Models (single-turn)}} \\
Qwen3-Coder-30B-A3B
  & 0.0 & 0.0 & 0.0 & 1.5 & 0.2 & 0.2 & 1.0 & 1.0 & 0.0 & 1.0 & 0.3 & 0.1 \\
Gemma-3-27B
  & 0.0 & 0.0 & 0.0 & 1.9 & 0.2 & 0.2 & 1.0 & 0.0 & 0.0 & 1.3 & 0.1 & 0.1 \\
Qwen3-Coder-Next (80B-A3B)
  & 3.0 & 0.5 & 0.5 & 1.3 & 0.2 & 0.2 & 0.0 & 0.0 & 0.0 & 1.6 & 0.3 & 0.3 \\
Llama-4-Maverick-17B-128E
  & 1.0 & 1.0 & 1.0 & 4.7 & 1.3 & 1.3 & 3.0 & 2.0 & 2.0 & 3.5 & 1.3 & 1.3 \\
DeepSeek-V3.2 (685B)
  & 2.5 & 2.0 & 2.0 & 5.9 & 2.3 & 2.3 & 4.0 & 2.0 & 2.0 & 4.8 & 2.2 & 2.2 \\
GPT-OSS-20B
  & 33.0 & 3.5 & 3.5 & 61.4 & 2.1 & 2.1 & 30.0 & 7.0 & 6.0 & 49.9 & 3.1 & 3.0 \\
GPT-OSS-120B
  & 20.5 & 5.0 & 5.0 & 42.5 & 3.2 & 3.0 & 21.0 & 3.0 & 3.0 & 34.0 & 3.6 & 3.5 \\
GPT-5.5
  & 70.5 & 29.5 & 29.5 & 76.9 & 20.0 & 20.0 & 66.0 & 40.0 & 40.0 & 73.8 & 25.0 & 25.0 \\
GPT-5.6-Sol & 55.5 & 27.5 & 27.5 & 67.5 & 21.4 & 21.4 & 62.0 & 38.0 & 38.0 & 63.7 & 25.2 & 25.2 \\
Claude-Opus-4.7
  & 80.0 & 17.0 & 14.5 & 93.8 & 8.7 & 7.6 & 83.0 & 35.0 & 22.0 & 88.8 & 14.3 & 11.3 \\
Claude-Opus-5
  & 85.5 & 29.0 & 29.0 & 92.8 & 17.0 & 16.8 & 81.0 & 40.0 & 40.0 & 89.4 & 23.1 & 23.0 \\
\midrule
\addlinespace[1.2pt]
\rowcolor{black!9}\multicolumn{13}{@{}l}{\textbf{Coding Agents (multi-turn)}} \\
Codex (w/ GPT-5.5)
  & 72.5 & 33.5 & 32.5 & 68.6 & 14.4 & 14.4 & 74.0 & 43.0 & 43.0 & 70.3 & 23.1 & 22.8 \\
Codex (w/ GPT-5.6-Sol)
 & 95.0 & 45.5 & 45.0 & 98.5 & 24.0 & 23.6 & 94.0 & 63.0 & 62.0 & 97.0 & 34.6 & 34.1 \\
Claude Code (w/ Claude-Opus-4.7)
  & 90.0 & 22.5 & 19.5 & 94.9 & 10.6 & 9.3 & 92.0 & 44.0 & 31.0 & 93.3 & 18.0 & 14.8 \\
Claude Code (w/ Claude-Opus-5)
& 89.5 & 54.5 & 54.5 & 94.3 & 33.3 & 33.3 & 88.0 & 57.0 & 57.0 & 92.2 & 41.9 & 41.9 \\
\midrule
\addlinespace[1.2pt]
\rowcolor{black!9}\multicolumn{13}{@{}l}{\textbf{Proposed (ours)}} \\
\textbf{\ourTool{}-Baseline} {(Model = Leanstral-1.5-119B-A6B)} &  &  &  &  &  &  &  &  &  &  &  &  \\
\hspace{1.5mm} w/ Seed Harness
  & 59.5 & 24.5 & 20.0 & 54.6 & 11.7 & 10.4 & 68.0 & 49.0 & 32.0 & 57.6 & 19.8 & 15.7 \\
\hspace{1.5mm} w/ HarnessEvolve
  & 90.0 & 37.5 & 28.0 & 80.9 & 18.5 & 15.7 & 90.0 & 59.0 & 44.0 & 84.4 & 28.7 & 22.6 \\
\hdashline
\rule{0pt}{2.5ex}\textbf{\ourTool{}} {(SAM + Interleaved RL/HarnessEvolve)}   & 97.5 & 50.0 & 50.0 & 93.8 & 24.6 & 24.6 & 94.0 & 85.0 & 67.0 & 94.8 & 39.0 & 36.7 \\
\phantom{\textbf{\ourTool{}} }+ Codex (w/ GPT-5.6-Sol) & 95.0 & \underline{85.5} & \underline{85.5} & \textbf{100.0} & 46.1 & 46.1 & \textbf{100.0} & \textbf{93.0} & \textbf{93.0} & 98.7 & 62.4 & 62.4 \\
\phantom{\textbf{\ourTool{}} }+ Claude Code (w/ Claude-Opus-5) & \textbf{100.0} & \textbf{95.0} & \textbf{95.0} & \textbf{100.0} & \textbf{70.7} & \textbf{70.7} & \textbf{100.0} & \underline{92.0} & \underline{92.0} & \textbf{100.0} & \textbf{79.8} & \textbf{79.8} \\
\bottomrule
\end{tabular}%
}
\vspace{-7mm}%
\end{table*}

\newcommand{\costGPU}{0.32}%
\newcommand{\costNumina}{2.03}\newcommand{\costNumCodex}{1.08}%
\newcommand{\costOGCC}{1.80}\newcommand{\costOGCodex}{1.05}%
\newcommand{\costOGAlone}{\costGPU}\newcommand{\accOGAlone}{21.5}%
\newcommand{\costAIPCC}{1.55}\newcommand{\accAIPCC}{79.8}%
\newcommand{\costAIPCodex}{0.85}\newcommand{\accAIPCodex}{62.4}%
\newcommand{\costAIPAlone}{\costGPU}\newcommand{\accAIPAlone}{36.7}%
\newcommand{\costCCAlone}{0.81}\newcommand{\accCCAlone}{41.9}%
\newcommand{\costCodexAlone}{0.53}\newcommand{\accCodexAlone}{34.1}%

\vspace{-2mm}

\subsection{LoCoBench: A Benchmark for Research-Level Mathematics}
\label{sec:LoCoBench}
\vspace{-1mm}

We curate a benchmark (\textbf{LoCoBench}) of 58.9k instances for proof-autoformalization of research-level mathematics,
in three fields: \textit{Algebraic Structures} (ring and group theory), \textit{Foundations, Logic \& Complexity} (set theory, logic, computability, model theory, bounded arithmetic), and \textit{Number Theory}. 

\vspace{-1mm}
\textbf{Sources and Data Curation.} From \textit{Mathlib}~\citeyearpar{mathlib} and \textit{CSLib}~\citep{barrett2026cslib}, we collect 10.2k, 5.9k, and 2.7k Lean theorem+proof pairs for the three fields, respectively. Both libraries are organized into topic folders, so we take those aligned with our fields. Each file, however, holds many theorems, and a proof may depend on lemmas and definitions in other files. We therefore run a dependency analysis that emits one standalone \texttt{.lean} file per pair, resolving its imports and unifying namespaces, and type-check each file individually. From the Mizar Mathematical Library~\citep[MML;][]{alama2011mizar}, published in the \textit{Formalized Mathematics} journal, we collect 18.3k, 8.0k, and 13.8k theorem+proof pairs from the research papers matching keywords for our fields, splitting into standalone \texttt{.miz} files, one per theorem. Lastly, we obtained the LaTeX source of a \textit{research-level bounded arithmetic textbook}~\citep{Li25-91} from its author and parsed its 71 theorem+proof pairs into standalone \texttt{.tex} files, matching each proof to its theorem across cross-references (App.~\ref{app:curation}).

\textbf{Informalization.} LLMs informalize (FL-to-NL) far more accurately~\citep{azerbayev2023proofnet,wu2022autoformalization} than they formalize. To obtain the NL theorem+proof pair per \texttt{.lean} and \texttt{.miz} file, we informalize it with a coding agent (Claude Code w/ Opus~4.8) in a feedback loop with CriticLeanGPT~\citep{peng2025criticlean}, an NL--FL faithfulness judge, revising until it passes (App.~\ref{app:curation}).

\vspace{-1mm}
\textbf{Train/Val Split.} We split LoCoBench into LoCoBench-Train and LoCoBench-Val so that Val is out of distribution: its Mizar theorems come from MML papers disjoint from those in Train, and its textbook domain (bounded arithmetic) appears in neither library. Mathlib+CSLib instances go to Train because their Lean is public. Mizar has no Lean counterpart, so its 100 longest-proof theorems per domain form Val, plus 71 theorems from a bounded-arithmetic textbook. These 771 instances have more auxiliary theorems and far longer NL than Train and prior benchmarks (Fig.~\ref{fig:locobench_lengths}, App.~\ref{app:curation}). Train also holds the remaining Mizar theorems, for which we distill Lean formalizations from a teacher model with verifier feedback: each must compile and pass a semantic check (App.~\ref{app:distillation}).

\vspace{-2mm}

\subsection{State-of-the-Art Baselines and Evaluation Metrics}
\label{sec:metrics}
\vspace{-1mm}

\textbf{SoTA Baselines.} We compare 39 systems in four families (Table~\ref{tab:main_results}, App.~\ref{app:baselines}) on two axes: \textit{specialization} (AFPS vs.\ general-purpose) and \textit{interaction} (single-turn \emph{model} vs.\ multi-turn \emph{agent}). \emph{AFPS models} are formalizers and provers, while \emph{AFPS agents} wrap a backbone in a Lean-tailored harness. \emph{Foundation models} emit Lean in one pass, while \emph{coding agents} wrap them in a general coding harness. 

\vspace{-1mm}
\textbf{Metrics.} We report \emph{pass@4} over four attempts, under three nested criteria (Sec.~\ref{sec:prelim}). \textbf{Type-correctness (TC)} requires the file to compile. \textbf{TC+SC (w/ or w/o \texttt{sorry})} adds \emph{semantic correctness (SC)} but not completeness, so \texttt{sorry} is allowed. \textbf{TC+SC (full proofs)} also requires completeness. Val has no Lean gold, but its Mizar instances carry gold in another FL, so we define SC as a \emph{4-way check}: two frontier judges each compare the candidate Lean with the gold (FL$\leftrightarrow$FL$'$) and its informalization with the NL input (NL$\leftrightarrow$NL$'$), and all four must pass (App.~\ref{app:judge}). We do not score proof faithfulness, as no reliable check exists yet and LLM judges are far less reliable on proofs than theorems (Sec.~\ref{sec:reward}).

\vspace{-2mm}
\subsection{Experimental Results}
\label{sec:mainresults}
\vspace{-1mm}

\textbf{\ourTool{} Setup.} We start from Leanstral-1.5~\citep{mistral2026leanstral} as $\mathcal{M}_0$ and construct a \textit{seed harness} $\mathcal{H}_0$ by adapting Mistral's Vibe~\citep{mistral2025vibe} with lean-lsp-mcp~\citep{dressler2025leanlspmcp} (App.~\ref{app:impl}). We evaluate in two stages: \textbf{(i)} \textbf{\ourTool{}-Baseline} runs $\mathcal{M}_0$ with either $\mathcal{H}_0$ (\textbf{w/ Seed Harness}) or a harness evolved from $\mathcal{H}_0$ by HarnessEvolve (\textbf{w/ HarnessEvolve}); and \textbf{(ii)} \ourTool{} post-trains $\mathcal{M}_0$ with SAM, then interleaved Agentic RLSF and HarnessEvolve. We evaluate \ourTool{} \textit{standalone} (\textbf{\ourTool{} (SAM + Interleaved RL/HarnessEvolve)}) and integrated as a \textit{skill} (a Lean-specialist sub-agent) in two frontier coding agents (\textbf{\ourTool{}+~Codex}, \textbf{\ourTool{}+~Claude Code}): the host handles planning and decomposition; \ourTool{} handles Lean writing and proof search (App.~\ref{app:skill}).

\textbf{\ourTool{} as a standalone agent.} \textbf{\textit{First,}} \textit{existing open-weight models are far from research-level AFPS without a Lean-tailored harness}: every open-weight AFPS or foundation model stays below 3.9\% TC+SC (best: StepFun-Formlzr.-7B → Hilbert). \ourTool{} raises this nearly tenfold to 36.7\%. \textbf{\textit{Second,}} \textit{it outperforms frontier LLMs and agents}: GPT-5.6-Sol (25.2\%), Aristotle (18.4\%), and Codex (34.1\%); only Claude Code (41.9\%) is higher. \textbf{\textit{Third,}} \textit{the gain is from post-training and the evolved harness}: from the same Leanstral-1.5 base, it outperforms OpenGauss on each field (36.7\% vs.\ 21.5\%), and its 67.0\% on number theory is above Codex (62.0\%) and Claude Code (57.0\%).

\textbf{\ourTool{} as a skill for frontier coding agents.} \textbf{\textit{First,}} \textit{\ourTool{} lifts both hosts more than prior harnesses}. With \ourTool{}, Claude Code goes from 41.9\% to 79.8\% (vs.\ 72.5\% with Numina-Lean-Agent, 54.3\% with OpenGauss) and Codex from 34.1\% to 62.4\% (vs.\ 44.5\%, 41.9\%). \textbf{\textit{Second,}} \textit{the gain comes from \ourTool{}, rather than the host}: it adds 28.3\% to Codex and 37.9\% to Claude Code, whereas Numina-Lean-Agent adds 10.4\% and 30.6\% and OpenGauss 7.8\% and 12.4\%, so the weaker host (Codex) gains almost three times more from \ourTool{} than from either prior agents.

\textbf{Ablation Studies.} We add the components of \ourTool{} one at a time and compare rows of Table~\ref{tab:main_results}. \textbf{\textit{First,}} \textit{the seed harness} (Leanstral-1.5 w/ tools vs.\ w/ Seed Harness): wrapping $\mathcal{M}_0$ in $\mathcal{H}_0$ raises TC from 27.4\% to 57.6\% and full proofs from 12.6\% to 15.7\%. \textbf{\textit{Second,}} \textit{HarnessEvolve} (w/ Seed Harness vs.\ w/ HarnessEvolve): evolving $\mathcal{H}_0$ at fixed $\mathcal{M}_0$ lifts TC to 84.4\% and full proofs to 22.6\%. Yet full proofs reach only 22.6\% against 84.4\% TC: the harness alone mostly fixes compilation, motivating post-training. \textbf{\textit{Third,}} \textit{SAM} (Leanstral-1.5 no tools vs.\ +SAM, App.~\ref{app:sam:gen}): SAM raises TC from 18.4\% to 29.3\% but lowers full proofs from 3.0\% to 1.4\%. Its role, instead, is the aligned space that initializes RLSF and HarnessEvolve: the cosine of matched NL--FL pairs rises from 0.11 to 0.75, and RLSF mostly keeps it (0.64, Fig.~\ref{fig:pairmap}). \textbf{\textit{Finally,}} \textit{interleaved RLSF and HarnessEvolve} (w/ HarnessEvolve vs.\ \ourTool{}): post-training in the evolved harness lifts TC to 94.8\% and full proofs to 36.7\%. Overall, $\mathcal{M}_0$ goes from 27.4\% to 94.8\% TC and from 12.6\% to 36.7\% full proofs.

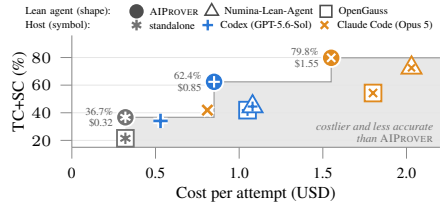
\begin{wrapfigure}[9]{r}{0.42\linewidth}
\vspace{-15pt}%
\centering
\begin{tikzpicture}[
  vlab/.style={font=\fontsize{4.0}{4.4}\selectfont, text=black!65, inner sep=0.4pt, align=right},
  hname/.style={font=\fontsize{4.8}{4.8}\selectfont, inner sep=1pt, anchor=west, xshift=2.5pt},
  hnote/.style={font=\fontsize{3.9}{3.9}\selectfont\itshape, inner sep=1pt, anchor=north west, xshift=2.5pt, text=black!50},
  mkh/.style={only marks, /pgfplots/forget plot, mark options={fill=white, draw=#1, line width=0.7pt}},
  mkf/.style={only marks, /pgfplots/forget plot, mark options={fill=#1, draw=white, line width=0.7pt}},
  bare/.style={only marks, /pgfplots/forget plot, mark size=2.7pt, mark options={draw=#1, line width=1.0pt}},
  halo/.style={only marks, /pgfplots/forget plot, mark size=2.7pt, mark options={draw=white, line width=2.4pt}},
  link/.style={/pgfplots/forget plot, line width=0.7pt, #1}]
\begin{axis}[width=0.83\linewidth, height=1.48cm, scale only axis,
  xlabel={Cost per attempt (USD)}, ylabel={TC+SC (\%)},
  xlabel style={font=\fontsize{6.3}{6.8}\selectfont, yshift=4pt}, ylabel style={font=\fontsize{6.3}{6.8}\selectfont, yshift=-5pt},
  tick label style={font=\fontsize{5.8}{6.2}\selectfont}, xmin=0, xmax=2.2, ymin=15, ymax=96,
  xtick={0,0.5,1,1.5,2}, xticklabels={0,0.5,1.0,1.5,2.0}, ytick={20,40,60,80},
  axis x line*=bottom, axis y line*=left, axis line style={black!45},
  ymajorgrids, grid style={black!10, line width=0.2pt}, tick align=outside, tick style={black!45}, clip=false, set layers]
\addplot[/pgfplots/forget plot, on layer=axis background, draw=none, fill=black!9] coordinates {(\costAIPAlone,15) (\costAIPAlone,\accAIPAlone)
  (\costAIPCodex,\accAIPAlone) (\costAIPCodex,\accAIPCodex) (\costAIPCC,\accAIPCodex) (\costAIPCC,\accAIPCC) (2.2,\accAIPCC) (2.2,15)} -- cycle;
\addplot[/pgfplots/forget plot, draw=black!38, line width=0.45pt] coordinates {(\costAIPAlone,\accAIPAlone)
  (\costAIPCodex,\accAIPAlone) (\costAIPCodex,\accAIPCodex) (\costAIPCC,\accAIPCodex) (\costAIPCC,\accAIPCC) (2.2,\accAIPCC)};
\addplot[mkh=hostNone, mark=sqAst, mark size=3.0pt] coordinates {(\costOGAlone,\accOGAlone)};
\addplot[mkh=hostCdx, mark=sqPlus, mark size=3.0pt] coordinates {(\costOGCodex,41.9)};
\addplot[mkh=hostCdx, mark=triPlus, mark size=3.0pt] coordinates {(\costNumCodex,44.5)};
\addplot[mkh=hostCC, mark=sqX, mark size=3.0pt] coordinates {(\costOGCC,54.3)};
\addplot[mkh=hostCC, mark=triX, mark size=3.0pt] coordinates {(\costNumina,72.5)};
\addplot[halo, mark=+] coordinates {(\costCodexAlone,\accCodexAlone)};
\addplot[bare=hostCdx, mark=+] coordinates {(\costCodexAlone,\accCodexAlone)};
\addplot[halo, mark=x] coordinates {(\costCCAlone,\accCCAlone)};
\addplot[bare=hostCC, mark=x] coordinates {(\costCCAlone,\accCCAlone)};
\addplot[mkf=hostNone, mark=circAst, mark size=3.0pt] coordinates {(\costAIPAlone,\accAIPAlone)};
\addplot[mkf=hostCdx, mark=circPlus, mark size=3.0pt] coordinates {(\costAIPCodex,\accAIPCodex)};
\addplot[mkf=hostCC, mark=circX, mark size=3.0pt] coordinates {(\costAIPCC,\accAIPCC)};
\node[font=\fontsize{4.3}{4.6}\selectfont\itshape, text=black!55, anchor=south east, inner sep=0.5pt, align=right] at (axis cs:2.17,17.5) {costlier and less accurate\\than \ourTool{}};
\node[vlab, anchor=east] at (axis cs:0.255,\accAIPAlone) {\accAIPAlone\%\\\$\costAIPAlone};
\node[vlab, anchor=east] at (axis cs:0.785,\accAIPCodex) {\accAIPCodex\%\\\$\costAIPCodex};
\node[vlab, anchor=east] at (axis cs:1.485,\accAIPCC) {\accAIPCC\%\\\$\costAIPCC};
\end{axis}
\begin{scope}[key/.style={matrix of nodes, anchor=south west, inner sep=0pt, column sep=1.3pt, row sep=0pt,
    nodes={inner sep=0.3pt, anchor=west, font=\fontsize{4.0}{4.0}\selectfont, text=black}},
  g/.style={minimum width=6.5pt}]
\matrix[key] (kh) at ([xshift=-18pt, yshift=1.6pt]current axis.north west) {
  |[text width=37pt]| Host (symbol): & |[g]| & standalone & |[g]| & Codex (GPT-5.6-Sol) & |[g]| & Claude Code (Opus 5) \\};
\matrix[key] (ka) at ([yshift=0.6pt]kh.north west) {
  |[text width=37pt]| Lean agent (shape): & |[g]| & \ourTool{} & |[g]| & Numina-Lean-Agent & |[g]| & OpenGauss \\};
\foreach \c/\m/\sz/\f in {2/*/2.4pt/black!70, 4/triangle/2.9pt/white, 6/square/2.2pt/white} {
  \begin{scope}[shift={(ka-1-\c.center)}]\pgfsetlinewidth{0.6pt}\pgfsetstrokecolor{black!70}\pgfsetfillcolor{\f}\pgfsetplotmarksize{\sz}\pgfuseplotmark{\m}\end{scope}}
\foreach \c/\col/\m in {2/hostNone/asterisk, 4/hostCdx/+, 6/hostCC/x} {
  \begin{scope}[shift={(kh-1-\c.center)}]\pgfsetlinewidth{0.9pt}\pgfsetstrokecolor{\col}\pgfsetplotmarksize{2.0pt}\pgfuseplotmark{\m}\end{scope}}
\end{scope}
\end{tikzpicture}
\vspace{-2pt}
\captionsetup{font=small, skip=2pt, justification=raggedright}
\caption{\textbf{Cost efficiency.} \ourTool{} forms the accuracy--cost Pareto frontier.}
\label{fig:cost_accuracy}
\vspace{-8pt}
\end{wrapfigure}
\textbf{Cost vs. Performance Analysis.} (App.~\ref{app:cost}) \ourTool{}'s three configurations form the Pareto frontier among harnessed systems in Fig.~\ref{fig:cost_accuracy}. \textbf{\textit{Standalone,}} \ourTool{} and OpenGauss run their models on local GPUs, so both cost \$0.32/attempt, yet \ourTool{} reaches 36.7\% TC+SC vs.\ the latter's 21.5\%. \textbf{\textit{With Claude Code,}} \ourTool{} reaches 79.8\% at \$1.55/attempt, vs.\ 72.5\% at \$2.03 for Numina-Lean-Agent and 54.3\% at \$1.80 for OpenGauss: more accurate and 24\% and 14\% cheaper. \textbf{\textit{With Codex,}} it reaches 62.4\% at \$0.85, vs.\ 44.5\% at \$1.08 for Numina-Lean-Agent and 41.9\% at \$1.05 for OpenGauss: more accurate and 21\% and 19\% cheaper. The host coding agent offloads Lean work to \ourTool{}, a Lean-specialist sub-agent on local GPUs, so it spends fewer tokens and gains accuracy.

\section{Conclusion}
\label{sec:conclusion}
We present \ourTool{}, an agentic framework for research-level proof auto-formalization that post-trains an open-weight LLM and evolves its harness together. We first fine-tune it into a \textit{semantic alignment model (SAM)} that treats an NL theorem--proof pair and its Lean counterpart as one mathematical object. Verifiers for type-correctness, completeness, and semantic correctness return a graded reward and certificate, which \textit{Agentic RLSF} uses to post-train the model and \textit{HarnessEvolve} to evolve the harness, so each round adapts the harness to the improved model and post-trains the model in the improved harness. On \mbox{LoCoBench-Val}, 771 instances from the Mizar Mathematical Library and a textbook, against AFPS agents, frontier LLMs, and coding agents, \ourTool{} beats every open-weight system as a standalone agent. As a skill of Claude Code and Codex, it makes both more accurate and cheaper than prior harnesses, showing that a harness adapted to its model and post-trained on verifier feedback can make an open-weight LLM a research-level auto-formalizer.

\clearpage

\bibliography{references}
\bibliographystyle{arxiv_preprint}

\clearpage
\raggedbottom
\appendix
\section*{Appendix}
\definecolor{catprompt}{HTML}{3D5A99}
\newcommand{\lstcat}[1]{\colorlet{lstbandcol}{#1}}
\newlength{\lsttagw}
\NewDocumentCommand{\lsthead}{s m m}{%
  \needspace{6\baselineskip}%
  \settowidth{\lsttagw}{\footnotesize\bfseries #3}%
  \begin{tcolorbox}[enhanced, colback=lstbandcol, colframe=lstbandcol,
    boxrule=0.5pt, arc=3pt, sharp corners=south,
    left=6pt, right=6pt, top=3pt, bottom=3pt, boxsep=0pt,
    before skip=9pt, after skip=0pt, nobeforeafter=false]
    \parbox[t]{\dimexpr\linewidth-\lsttagw-1em\relax}{\small\bfseries\color{white}\raggedright%
      \IfBooleanF{#1}{Listing~\the\numexpr\value{lstlisting}+1\relax\enspace$\cdot$\enspace}#2}%
    \hfill\parbox[t]{\lsttagw}{\raggedleft\footnotesize\bfseries\color{white}#3}%
  \end{tcolorbox}\nopagebreak\vspace{-0.5pt}\nopagebreak}
\captionsetup[lstlisting]{format=lstdesc,labelformat=empty,skip=-0.5pt,aboveskip=0pt,belowskip=-0.5pt,justification=raggedright,singlelinecheck=false}

\newcommand{\apptocline}[2]{\makebox[3.2em][l]{\ref{#1}}\hyperref[#1]{#2}\;\dotfill\;\pageref{#1}\\[1.5pt]}

\begin{quote}\small
\apptocline{app:setup}{Experimental Setup}
\hspace*{1.5em}\apptocline{app:impldetails}{Reproducibility and Implementation Details}
\hspace*{1.5em}\apptocline{app:benchdetails}{Benchmark Details}
\hspace*{1.5em}\apptocline{app:baselines}{State-of-the-Art Baselines}
\hspace*{1.5em}\apptocline{app:evalprotocol}{Evaluation Protocol}
\apptocline{app:skill}{\ourTool{} as a Skill for Frontier Coding Agents}
\apptocline{app:distillation}{Data Distillation}
\hspace*{1.5em}\apptocline{app:prompts}{Prompt Construction}
\hspace*{1.5em}\apptocline{app:generation}{Generation Configuration}
\hspace*{1.5em}\apptocline{app:repair}{Repair}
\apptocline{app:cost}{Cost Accounting}
\apptocline{app:sam}{Semantic Alignment Model: Analyses}
\hspace*{1.5em}\apptocline{app:sam:align}{The Aligned Space}
\hspace*{1.5em}\apptocline{app:sam:gen}{Generation Results}
\hspace*{1.5em}\apptocline{app:sam:method}{Extended Methodology}
\apptocline{app:promptsec}{Harnesses and Prompts of \ourTool{}, Its Training Pipeline, and the Baselines}
\hspace*{1.5em}\apptocline{app:prompts:h0}{Initial \ourTool{} Harness $\mathcal{H}_0$ (Inference and HarnessEvolve Seed)}
\hspace*{1.5em}\apptocline{app:prompts:evolved}{Evolved \ourTool{} Harness (Inference and Training)}
\hspace*{1.5em}\apptocline{app:prompts:skill}{\ourTool{} as a Skill for Coding Agents (Inference)}
\hspace*{1.5em}\apptocline{app:prompts:hevo}{HarnessEvolve (Evolutionary Search)}
\hspace*{1.5em}\apptocline{app:prompts:judge}{Semantic-Correctness Judges (Evaluation)}
\hspace*{1.5em}\apptocline{app:prompts:baselines}{Baseline Agents and LLMs (Inference)}
\hspace*{1.5em}\apptocline{app:prompts:data}{Data Pipelines (Data Curation)}
\end{quote}

\vspace{-2mm}
\section{Experimental Setup}
\label{app:setup}
\vspace{-3mm}

\subsection{Reproducibility and Implementation Details}
\label{app:impldetails}

\paragraph{Machine configuration.} We split the work by hardware. CPU jobs, namely the harness with its multi-turn agent loop, the Lean servers, HarnessEvolve, and the evaluation runners, run on a CPU-only OpenStack virtual machine with 32 virtual cores (Haswell-class at 2.49~GHz), 125~GiB of RAM, and a 160~GB disk, under Ubuntu~24.04; four runners keep about 44 rollouts in flight, each with its own Lean server at about 2.3~GB, so memory rather than compute bounds concurrency. LLM hosting and training run on GPU clusters with NVIDIA H100s and H200s, reached from the VM over SSH tunnels to vLLM servers behind a load balancer; each locally served evaluation run uses two GPUs reserved for three days (App.~\ref{app:cost}). Claude Code and Codex, as baselines and as hosts of the \ourTool{} skill, run through their providers' official APIs. Throughout all experiments, training, evaluation, and every baseline, we use Lean~4.23.0 with Lake~5.0.0 and the matching Mathlib, and Python~3.12.

\label{app:impl}
\paragraph{HarnessEvolve.} The training set $\mathcal{D}$ holds 300 LoCoBench-Train instances, 100 per domain, sampled once from a 1{,}000-instance pool and frozen for the whole search. The mutator is a frontier coding agent (Claude Code with Claude Opus~5) run headless in a fresh session per round with file read, write, search, and bash tools, isolated from the gold formalizations. Parent selection uses $c=0.9$, a gap span of $3\sigma$ with $\sigma=0.024$ measured by re-evaluating the seed, and an accepted child counting twice in $k(h)$. Two rules enforce exploration: no node is selected more than twice in a row, and after three non-accepting rounds only nodes with few children and the seed remain eligible. Before evaluation a child must parse and run, keep the seed's model identity, temperature, and reasoning effort, differ from every node in memory, and contain no run of gold Lean, with up to three re-prompts on failure. A round with too many unresolved verifier verdicts is marked invalid and is neither accepted nor rejected.

\paragraph{Agentic RLSF.} Agentic RLSF extends RLSF~\citep{jha2025rlsf}, which post-trains an LLM on rewards from symbolic tools, from single-turn generation to multi-turn agents. We train with verl. Each round runs 20 optimization steps of 32 instances with $K=8$ rollouts per instance at sampling temperature $1.0$. Advantages are centered by the group mean without standard-deviation scaling. The PPO clipping range is $0.2$ ($0.28$ on the upper side), the KL penalty to the round's initial policy has coefficient $0.001$ with the low-variance estimator, and the entropy bonus is $0$. We update LoRA adapters of rank 32 and scaling 64 on all linear layers with learning rate $10^{-5}$. The loss covers only model-generated tokens. Crashed or timed-out rollouts receive reward $0$ and stay in their group. RLSF instances come from a split of about 16{,}700 LoCoBench-Train instances that is disjoint from the 1{,}000-instance HarnessEvolve pool and from the 500-instance held-out set of the round gate, which requires an improvement of at least $0.024$ in mean reward.

\paragraph{Semantic alignment model.}
\label{app:sam:impl}

The frozen base is \texttt{Leanstral-1.5-119B-A6B}~\citep{mistral2026leanstral} in bf16, a mixture of experts with $119.5$B parameters of which about $6$B are active per token, and hidden size $d{=}4096$. LoRA adapters sit on the five latent-attention projections of all $36$ decoder layers, and the experts are untouched. Each adapter updates a frozen projection $W\in\mathbb{R}^{d_{\mathrm{out}}\times d_{\mathrm{in}}}$ to $W+\tfrac{\alpha}{r}UV$ with $U\in\mathbb{R}^{d_{\mathrm{out}}\times r}$ and $V\in\mathbb{R}^{r\times d_{\mathrm{in}}}$, at rank $r{=}32$, scale $\alpha{=}64$, and dropout $0.05$, and $\Delta$ collects all $(U,V)$. The projection head $g_\phi$ is a two-layer MLP from $d{=}4096$ to $m{=}512$. The anchor $\pred$ is the reserved token \texttt{<SPECIAL\_999>}.

The base is sharded over $8$ GH200 nodes with tensor parallelism, and the language-modeling head runs only on the gold positions $F$, which is exact and avoids the full-vocabulary logits. AdamW runs at a peak learning rate of $10^{-4}$ with cosine decay, $3\%$ warmup, no weight decay, and gradient clipping at $1.0$, $\tau = 0.07$, and batches of up to $B{=}32$ pairs for $2$ epochs, or $1{,}590$ steps. Batches are packed to at most $24{,}576$ tokens and sequences capped at $8{,}192$ tokens. SFT and SAM share this configuration, seed, and batch order and differ only in $\lambda$.

\subsection{Benchmark Details}
\label{app:benchdetails}
\label{app:curation}

This appendix expands the curation summary of Sec.~\ref{sec:LoCoBench}.

\subsubsection{Instances and Sources}

\textbf{Instances.} Each instance consists of an NL theorem statement and an NL proof. The 18,757 Mathlib and CSLib instances also carry a verified Lean~4 formalization of both. That Lean file has one main theorem, together with the non-trivial lemmas and definitions its proof needs, and type-checks without added axioms. The 39,385 Mizar instances and the 71 textbook instances carry NL only, since no Lean formalization of them exists. Table~\ref{tab:expert_grouping} summarizes the dataset.

\textbf{Why these sources.} To build a benchmark at the research level, we collected advanced graduate and research-level mathematics that has already been formalized in a proof assistant, Lean~4 or Mizar. This ensures two things. The problems are hard and complex enough to stand in for future research problems, and every theorem and proof from these two sources has been machine-checked, so it is formally correct. The textbook instances are the one exception. They are not machine-checked, but they come from a research-level text on bounded arithmetic, a topic that neither library formalizes, which makes them an unseen domain for evaluation. Building the benchmark this way also lets us classify it into three fields by mathematical topic, as described at the end of this appendix.

\subsubsection{Extraction per Source}

\textbf{Lean sources.} Mathlib and CSLib are organized into topic folders, so we selected the folders that match our fields, e.g., \texttt{RingTheory}, \texttt{GroupTheory}, \texttt{SetTheory}, \texttt{Logic}, \texttt{Computability}, \texttt{ModelTheory}, and \texttt{NumberTheory}, and extracted every theorem in them together with its proof. A theorem cut out of its file rarely type-checks on its own. Its proof may use lemmas and definitions from other files, and copying those into the new file re-declares names that Lean already imports, which Lean rejects. We therefore ran a dependency analysis over both libraries. For each theorem, we resolved the imports its proof needs and unified the namespaces so that the extracted code refers to the same declarations as the original. The theorem keeps its name, and no dependent lemma is copied into the file unless Mathlib marks it \texttt{private} or \texttt{protected}, in which case it cannot be reached by import. Every resulting \texttt{.lean} file was type-checked, files that failed were discarded, and the last line of each file records its source path for provenance. This gave 18,757 files, each stating exactly one theorem, with a median of 9 lines of Lean after the imports.

\textbf{Mizar source.} The Mizar Mathematical Library (MML) is a collection of Mizar articles, each verified by the Mizar system as a consequence of the Tarski--Grothendieck set theory axioms and published as a paper in the \textit{Formalized Mathematics} journal~\citep{alama2011mizar}. The library has no folder structure by topic, so we classified articles by their titles. Articles on ring theory, group theory, and number theory map directly to those domains. Articles on logic and computation were assigned to Set Theory, Model Theory, Logic, or Computability by keyword rules on their titles, 12, 23, 36, and 54 patterns respectively, e.g., ``zf set'', ``satisfiability'', ``sequent'', or ``turing'', with the first matching domain winning. This routed 655, 957, 2,737, and 3,679 theorems to the four domains and dropped 1,810 theorems from 52 articles that matched no rule, mostly on graph theory, matrices, and lattices, which fall outside our fields. Each selected article was then split into standalone \texttt{.miz} files, one per theorem. A file keeps the article's \texttt{environ} header and carries the definitions, registrations, and auxiliary theorems that its theorem depends on, so it type-checks alone. Since Mizar is a different formalism, these instances have no Lean side and enter LoCoBench as NL only after informalization (Sec.~\ref{sec:LoCoBench}). The split of these theorems between Train and Val is described below.

\textbf{Textbook source.} We obtained the \LaTeX{} source of the bounded arithmetic textbook of \citet{Li25-91} directly from its author. Its chapters cover feasible functions, the theory PV, basic and advanced proof theory, simulations, and propositional translations. We paired every theorem, lemma, proposition, and corollary environment with its proof. This pairing is not local. A proof may appear pages after its statement and is then bound to it only by an explicit ``Proof of'' cross-reference, and a few proofs are written as prose sketches, which we paired by hand. Statements without a proof were recorded and left out. For each pair we kept the \LaTeX{} math, removed margin notes and asides, replaced every cross-reference by a short inline description of the cited result or equation, and prepended a brief setting block with the notation the theorem uses, so that each \texttt{.tex} file reads on its own. A separate scan of the chapters confirmed that every theorem-like environment was either paired or recorded as unproved. This gave 71 theorem+proof pairs, all placed in LoCoBench-Val, since bounded arithmetic has no Lean formalization to train on.

\subsubsection{Informalization}

The \texttt{.lean} and \texttt{.miz} files carry no NL, so the NL theorem+proof pairs of LoCoBench are produced by informalization, i.e., FL-to-NL translation. This is the reverse of the task we study, and LLMs do it far more reliably~\citep{azerbayev2023proofnet,wu2022autoformalization}, for a structural reason. A formal statement is unambiguous and complete. Every hypothesis, type, and quantifier is written down, and the proof names each fact it uses, so the writer only has to render information that is already present. Formalization runs the other way. An NL theorem often leaves assumptions implicit, e.g., that a ring is commutative or a set is nonempty, uses notation whose meaning depends on context, and skips steps the reader is trusted to fill in. The model must recover this hidden content and can guess wrong with no signal that it did. Informalization instead fails mostly by omission or paraphrase drift, which a judge can catch by comparing the two sides.

We use Claude Code with Opus~4.8 as the writer. It is instructed to ignore the import and \texttt{environ} headers, inline the definitions the theorem uses, follow the proof structure, and write textbook-style prose from which a reader could reconstruct the formal proof. Each draft is checked by CriticLeanGPT~\citep{peng2025criticlean}, a separate model trained with supervised fine-tuning and RL to judge whether a Lean formalization captures the semantic intent of an NL statement, and released openly in 7B to 32B sizes. It is a good judge for our purpose because its training task is exactly NL--FL semantic fidelity, it outperforms strong open- and closed-source models on CriticLeanBench, and being a different model from the writer it does not share the writer's blind spots. The writer revises until the judge accepts, and files that never pass are flagged for manual review.

\subsubsection{Train/Val Split}

The split is designed so that Val is out-of-distribution for the models we benchmark. Mathlib and CSLib are public and widely mirrored, so their Lean files, and often their theorem names, are likely part of the pretraining data of most LLMs. A model asked to formalize one of these theorems can recall rather than formalize. All 18,757 Mathlib and CSLib instances therefore go to Train. Mizar theorems have no Lean counterpart anywhere online, and the NL side of our instances is our own informalization rather than the published abstract, so a Lean formalization of a Mizar theorem cannot have been memorized. Val is drawn from them. For each of the seven Mizar domains we select 100 theorems by an article round-robin that favors long proofs. The classified theorems are grouped by MML article, the theorems of each article are sorted by proof length, articles are ordered by how many theorems they contribute, and one theorem is taken from each article per round until 100 are chosen. This keeps the hardest theorems while spreading Val over many articles, so no single article dominates a domain. The pools range from 655 theorems (Set Theory) to 14,878 (Group Theory). The 100 picked per domain come from 95 to 100 distinct articles for Ring Theory, Group Theory, Number Theory, and Computability, and from all 14, 22, and 51 available articles for Set Theory, Model Theory, and Logic. The median picked theorem has 360 to 685 lines of Mizar proof, depending on the domain. The remaining 39,385 Mizar theorems join Train as NL-only instances. The 71 textbook theorems all go to Val. Bounded arithmetic is a recent research direction in computational complexity that neither library formalizes, so it is a domain unseen in training and tests whether a model can formalize logic and complexity beyond what it has seen. Val thus has 771 instances, 700 from Mizar and 71 from the textbook, and no Val source ever appears with Lean labels in Train.

Val is harder than Train and than prior benchmarks on every measure we have. Its files are self-contained developments rather than single lemmas, so each carries the definitions and auxiliary theorems its main theorem needs. Its NL is long. The median Val theorem has 283 words and the median proof 708 words over 36 sentences, against 64 and 159 words for a Mathlib Train instance and 22 and 91 for ProofNet (Fig.~\ref{fig:locobench_lengths}). Mathlib supplies the foundations these theorems rest on but not the theorems themselves, since \textit{Formalized Mathematics} articles develop self-contained theories whereas Mathlib is a broad reusable library, so a model must formalize new material rather than recall it. Lacking gold Lean, Val is scored by type-checking and the semantic-correctness judge of App.~\ref{app:judge}.

\subsubsection{Grouping Domains into Fields}

\textbf{Grouping domains into fields.} The eight domains are grouped into three fields by the style of reasoning their proofs require, and results are reported per field.

\textbf{Algebraic Structures.} Ring Theory and Group Theory make up nearly half of LoCoBench. Their proofs share an algebraic style, centered on ideals, homomorphisms, and quotient structures, which differs from both foundational and computational reasoning. Reporting them as one field also keeps their volume from masking the results on the smaller fields.

\textbf{Foundations, Logic \& Complexity.} Set Theory, Logic, Computability, Model Theory, and Bounded Arithmetic form a tightly coupled cluster. Models are set-theoretic structures, Logic supplies the formal syntax that Set Theory and Model Theory share, and Computability connects to both through G\"{o}del numbering and the arithmetical hierarchy. Bounded Arithmetic sits at the intersection of proof theory, models of arithmetic, and complexity, which makes it a natural companion to Model Theory and Computability. All five domains are concerned with formal provability and semantic entailment.

\textbf{Number Theory.} Number Theory has deep ties to the previous field, e.g., the undecidability of Hilbert's Tenth Problem reduces Diophantine satisfiability to halting~\citep{matiyasevich1993hilbert}. Its proofs nevertheless read differently. They favor concrete symbolic and numeric reasoning about integers, primes, and congruences, in contrast to the abstract structural reasoning of Algebraic Structures and the semantic reasoning of Foundations, Logic \& Complexity, so we keep it as a field of its own.

\subsection{State-of-the-Art Baselines}
\label{app:baselines}

Table~\ref{tab:main_results} compares \ourTool{} against 39 systems. We group them along two axes. The first is \emph{specialization}: whether a system is purpose-built for auto-formalization and proof synthesis (AFPS) or is a general-purpose foundation model. The second is \emph{interaction}: whether it emits its Lean in a single turn, with no execution feedback, or runs an agentic loop that calls tools and conditions on their outputs. This yields the four families below. Every system is run zero-shot, with no in-context examples, and gets four attempts per instance (Sec.~\ref{sec:metrics}).

\begin{itemize}[leftmargin=1.2em, itemsep=2pt, topsep=2pt]
\item \textbf{AFPS models (single-turn).} Open models fine-tuned for auto-formalization or proof synthesis and queried in a single turn with no tool feedback: Kimina-Autoformalizer-7B~\citep{numina2025kiminaautoformalizer}, StepFun-Formalizer-7B~\citep{wu2025stepfun}, Goedel-Prover-V2-32B~\citep{lin2025goedelv2}, DeepSeek-Prover-V2-7B~\citep{ren2025deepseekproverv2}, and Kimina-Prover-Distill-8B and Kimina-Prover-RL-1.7B~\citep{wang2025kimina}. Since formalizers only produce statements and provers only prove given statements, we also run the six formalizer$\to$prover pipelines that pair each formalizer with each prover, where the prover receives the formalizer's statement.
\item \textbf{AFPS agents (multi-turn).} Purpose-built systems that wrap a backbone LLM in a Lean-specialized harness: Aristotle~\citep{achim2025aristotle}, Math-Inc OpenGauss~\citep{mathinc2026opengauss}, the Numina Lean Agent~\citep{liu2026numinaleanagent}, Leanstral-1.5-119B-A6B with its tools~\citep{mistral2026leanstral}, and two formalizer$\to$prover pipelines whose prover is an agent, Hilbert~\citep{varambally2025hilbert} and Axiom AXLE~\citep{xin2026axle}, fed by StepFun-Formalizer. These harnesses supply three ingredients that general agents lack: a Lean verifier in the loop, such as a Lean LSP or REPL server whose diagnostics and goal states the model conditions on across turns, a structured draft, check, and repair workflow that tracks \texttt{sorry} placeholders, and a pre-built Mathlib environment the agent can search and type-check against. OpenGauss drives Claude Code or Codex as its backend agent, and Numina Lean Agent runs as Claude Code equipped with Lean skills, so we report OpenGauss with Claude-Opus-4.7 and GPT-5.6-Sol and Numina with Claude-Opus-5 and GPT-5.6-Sol. Leanstral is also run without tools, as a single-turn AFPS model, so that the two rows isolate the value of its tools.
\item \textbf{Foundation models (single-turn).} General-purpose LLMs prompted to emit Lean~4 in one API call, with no tools and no verification loop: Qwen3-Coder-30B-A3B-Instruct and Qwen3-Coder-Next 80B-A3B~\citep{yang2025qwen3}, Gemma-3-27B~\citep{team2025gemma}, Llama-4-Maverick-17B-128E-Instruct~\citep{meta2025llama4}, DeepSeek-V3.2~\citep{deepseek2025v32}, GPT-OSS-20B and GPT-OSS-120B~\citep{gpt2025oss}, Claude-Opus-4.7~\citep{anthropic2026claudeopus47} and Claude-Opus-5~\citep{opus5card}, and GPT-5.5~\citep{openai2026gpt55} and GPT-5.6-Sol~\citep{openai2026gpt56sol}. We cannot observe provider-side computation for the closed models, but no Lean environment, Mathlib access, or verifier feedback on the target problem is supplied to any model in this family.
\item \textbf{Coding agents (multi-turn).} The same frontier backbones wrapped in a general agentic coding harness with file-editing and shell tools but no AFPS-specific scaffolding: Claude Code~\citep{anthropic2025claudecode} with Claude-Opus-4.7 and Claude-Opus-5, and Codex~\citep{openaicodex2025} with GPT-5.5 and GPT-5.6-Sol. They run a multi-turn loop and may inspect or execute code, but have none of the three ingredients above: no Lean verifier in the loop, no formalization workflow, and no pre-built Mathlib environment. The contrast between this family and the AFPS agents therefore isolates AFPS-specific agency from generic agency on identical backbones.
\end{itemize}

\textbf{Access and serving.} The closed models are accessed through their providers' APIs and, for Claude-Opus-5 and GPT-5.6-Sol, through Claude Code Max and ChatGPT Plus subscriptions. Open models from Hugging Face are served locally with vLLM on NVIDIA H100, A100, or RTX~5090 GPUs.

\subsection{Evaluation Protocol}
\label{app:evalprotocol}

\subsubsection{The Semantic-Correctness Judge}
\label{app:judge}

Semantic correctness asks whether the candidate theorem $\rnd{\Tflh}$ states exactly the NL theorem $\Tnl$ (Sec.~\ref{sec:prelim}). This is undecidable in general and needs a judge. During training a gold Lean theorem exists and we test logical equivalence with BEq$^{+}$ (Sec.~\ref{sec:reward}, App.~\ref{app:beq}). On LoCoBench-Val there is no Lean gold. The 700 Mizar instances carry a gold theorem, but in Mizar, a different formal language, and the 71 textbook instances carry NL only. We therefore judge SC with LLMs, and design the check so that a candidate must survive four independent verdicts.

\textbf{Two comparisons.} Each judge sees the candidate from two sides. In the formal-to-formal comparison (FL$\leftrightarrow$FL$'$), the judge receives the candidate Lean file and the gold Mizar excerpt, is told to ignore the Mizar proof bodies, and decides whether the candidate theorem states the same result as the gold theorem under the same definitions, allowing renamed variables, reordered hypotheses, and equivalent reformulations but not added assumptions or dropped conditions. It returns one of three tiers: \emph{full}, the theorems are equivalent and no proof is a placeholder, \emph{partial}, the theorems are equivalent but some proof is \texttt{sorry}, \texttt{admit}, \texttt{trivial}, or an empty \texttt{by} block, or \emph{no}, the theorems differ. Proof correctness is not judged here because Lean already certifies it. This comparison is cross-formalism, so no prover can decide it and the judge must reason about both statements. In the informal-to-informal comparison (NL$\leftrightarrow$NL$'$), the judge is shown the candidate Lean theorem and the original NL theorem $\Tnl$ but no formal gold. It first informalizes the candidate statement into NL$'$ and then decides whether NL$'$ asserts the same claim as $\Tnl$, returning \emph{yes} or \emph{no}. Hiding the gold keeps this verdict independent of the Mizar phrasing. The two views catch different failures. A candidate can mirror the gold's structure yet mistranslate a definition, which the NL side exposes, or read naturally yet quantify differently from the gold, which the FL side exposes. Textbook instances have no gold FL and receive the NL$\leftrightarrow$NL$'$ comparison only.

\textbf{Two judges and the decision rule.} Both comparisons are run by two independent judges with different backbones, Claude Opus~4.8 and GPT-5.5, so that the idiosyncrasies of one model do not decide the outcome. This yields four verdicts per candidate, and SC is their conjunction (AND). A candidate is semantically correct only if both judges' FL$\leftrightarrow$FL$'$ tiers are \emph{full} or \emph{partial} and both judges' NL$\leftrightarrow$NL$'$ verdicts are \emph{yes}. Any \emph{no} among the four rejects the candidate. Among accepted candidates, the tier is the lowest one returned, so a \emph{partial} from one judge places the candidate in the ``w/ or w/o \texttt{sorry}'' column of Table~\ref{tab:main_results} rather than the ``full proofs'' column. Candidates that fail type-checking are recorded as \emph{no} without consulting the judges, and instances without any gold are excluded from the SC rates. The candidates there come from OpenGauss with Claude-Opus-5 run with Mathlib's sources removed from the project and web search denied, so that no gold statement can be retrieved rather than derived.

\subsubsection{Extending BEq$^{+}$ for the Semantic-Correctness Check}
\label{app:beq}
BEq$^{+}$~\citep{poiroux2025beq} tests whether two Lean theorems are equivalent. It replaces both proofs with \texttt{sorry} and then tries to prove each statement from the other with a fixed cascade of tactics: first \texttt{exact?}, then \texttt{apply}, then a \texttt{have} that introduces the premise's conclusion before \texttt{apply\_rules}, and finally \texttt{convert} at increasing depth. We use this cascade unchanged. What we change is how it is called and how its outcome is read, for three reasons.

First, BEq$^{+}$ reports a single boolean that holds only when both directions prove. We keep the two directions apart. A candidate that is provably weaker or stronger than the gold theorem then receives $v_{\mathrm{SC}}=0.5$ instead of $0$ (Sec.~\ref{sec:reward}). This matters under group-relative RL, where a near-miss and an unrelated statement should not receive the same reward.

Second, the original cascade tries $\Tfl\Rightarrow\rnd{\Tflh}$ first and attempts $\rnd{\Tflh}\Rightarrow\Tfl$ only if that succeeds. A strictly stronger candidate therefore looks the same as an unrelated one. When the first direction fails, we rerun the cascade with the arguments swapped, which costs one extra cascade per such candidate.

Third, a theorem that automation can close on its own is implied by anything, so BEq$^{+}$ accepts a special case of an easy gold theorem as equivalent to it. For the gold $\forall n,\ n+0=n$, the candidate $3+0=3$ proves in both directions. Before accepting a match, we therefore check whether either statement is provable alone, and if it is, both directions count as failed. The exception is a pair whose two directions were closed by \texttt{exact?} using the other statement, since that dependence is real.

The remaining changes only make the check fast enough to score rollouts during training. A single Lean REPL with Mathlib imported stays alive across calls and caches the header environments. The candidate's own \texttt{open}, \texttt{variable}, and \texttt{namespace} lines are merged into the gold context so that a self-contained candidate still elaborates. When a candidate fails to type-check, we skip the cascade altogether; this cannot change the verdict, because BEq$^{+}$ implies type-correctness, and we confirmed it on all 9,528 rollouts we scored. Finally, each proof attempt is given 30\,s rather than the default 60\,s.

\section{\ourTool{} as a Skill for Frontier Coding Agents}
\label{app:skill}
We use \ourTool{} with frontier coding agents by packaging it as a \emph{skill} that a tool-calling coding agent loads; our experiments use Claude Code and Codex as hosts (the \ourTool{}+Claude Code and \ourTool{}+Codex rows of Table~\ref{tab:main_results}). The skill gives the host agent a small command-line interface and a fixed protocol; the host never reads \ourTool{}'s trajectories, only its returned Lean files and their checks. The skill file and playbook the host reads are reproduced in App.~\ref{app:prompts:skill}.

\textbf{Division of labour.} \ourTool{} writes Lean and searches for proofs. One call is a complete agentic session of the post-trained model inside the evolved harness, up to 200 turns with Lean and the lean-lsp tools, run on our GPUs, so a job of $k$ parallel samples costs the host nothing but a wait. The host does what needs judgement: it reads the NL passage, plans, judges semantic correctness (c) and proof faithfulness (d) of every returned file, fixes statements, decomposes, and weaves the pieces together. It does not search for proofs itself, and it does not stop until all four properties of Sec.~\ref{sec:prelim} hold.

\textbf{Protocol.} (0)~A preflight check of the environment: Lean and Mathlib, the completeness checker, the model endpoint, and one live rollout. (1)~Blueprint: the host lists the definitions, the theorem parts with their exact hypotheses, and the proof's intermediate lemmas, and resolves ambiguities such as the number type or indexing convention. (2)~It submits the whole problem to \ourTool{} with $k=4$ samples, plus a statement-only job on long inputs, and waits. (3)~It judges every sample that passes (a) and (b), best first, and accepts the first that also passes (c) and (d). (4)~Otherwise it fixes the statement and decomposes: it freezes a skeleton with the definitions and every lemma and theorem part as a statement with \texttt{sorry}, submits one job per open piece with the frozen context and that piece's NL fragment, weaves the returned proof bodies into the skeleton, and resubmits a failing piece with a one-line hint or splits it further along the NL proof. (5)~A final gate reruns the type-correctness and completeness check on the exact final file and the judge protocol once more.

\section{Data Distillation}
\label{app:distillation}

\textbf{Distillation Pipeline.} Our pipeline evaluates the quality of the generated examples in two ways: by their \emph{compile success}, i.e., Lean's acceptance of the generated proof, and by their \emph{semantic quality}, defined as alignment with the statement. Filtering a teacher's outputs by compiler and verifier feedback before training on them follows earlier verifier-in-the-loop pipelines for code, CoTran~\citep{jana2024cotran} and Terraformer~\citep{jana2026terraformer}, and the broader program of coupling LLMs with symbolic feedback for mathematics and software~\citep{jana2024neurosymbolic}. We show the pipeline in Figure \ref{fig:distillation_pipeline} and give a full explanation in Appendix \ref{app:distillation}. 

The data that we produce comes from two different source collections: a labeled Mathlib dataset and an unlabeled Mizar dataset (for more details, see Appendix~\ref{app:curation}). These two sets correspond to two distinct data-generation settings: producing high-quality labeled data from an existing labeled corpus, and producing new labeled data from an unlabeled corpus. 

For the labeled data setting, we construct candidate proof examples using Mathlib and Claude Opus~5~\citep{opus5card} as a data-generation model, and then filter and validate the examples as outlined below. Claude Opus~5 is used here strictly to generate standalone proof examples for our own filtering and evaluation process; we do not extract or make use of any model traces, chain-of-thought, or other internal reasoning components from Claude, and this procedure does not involve training a model on Claude's outputs. We place particular emphasis on semantic equivalence between the generated proof and the original statement.

For the unlabeled data setting, we use the Mizar set, derived from the corpus~\citep{bancerek2018mizar}, with DeepSeek-V4-Flash~\citep{deepseekv4} as the generation model, focusing on scalability and yield. Distillation was run over 39,123 of the 39,385 Mizar and 18,747 of the 18,757 Mathlib and CSLib Train instances (App.~\ref{app:distillation}), and every rate we report is against the number of instances actually attempted.

Our distillation pipeline is structured as follows:

\begin{enumerate}
    \item \textbf{Prompt construction.}
    The teacher receives the mathematical problem with
    few-shot examples and formatting instructions
    (Appendix~\ref{app:prompts}).

    \item \textbf{Generation.}
    The teacher generates the requested reasoning trace and Lean
    formalization (Appendix~\ref{app:generation}).

    \item \textbf{Parsing.}
    The system extracts the formal theorem and proof from the model
    response using tolerant parsing.

    \item \textbf{Formal verification.}
    The generated artifact is compiled using Pantograph against the
    configured Lean~4~\citep{moura2021lean}/Mathlib~\citep{mathlib}
    environment.

    \item \textbf{Repair.}
    Failed generations can be returned to the teacher together with
    the Lean verifier error. The teacher receives an additional
    generation opportunity (Appendix~\ref{app:repair}).

    \item \textbf{Semantic filtering.}
    Successfully compiled formalizations are evaluated using
    \textsc{BEq}$^{+}$~\citep{poiroux2025beq} where a reference formal
    statement is available.
    Where none exists, they are evaluated by an LLM judge that compares
    the generated Lean statement against the natural-language problem;
    the judge is calibrated against \textsc{BEq}$^{+}$ on the labeled
    split. We use FormalRx~\citep{wang2026formalrx} as the judge, the
    calibration is reported in Appendix~\ref{app:judge}.

    \item \textbf{Caching.}
    Generated examples and verification outcomes are cached to permit
    resumable large-scale generation.
\end{enumerate}

\begin{figure}
    \centering
    \includegraphics[width=\linewidth]{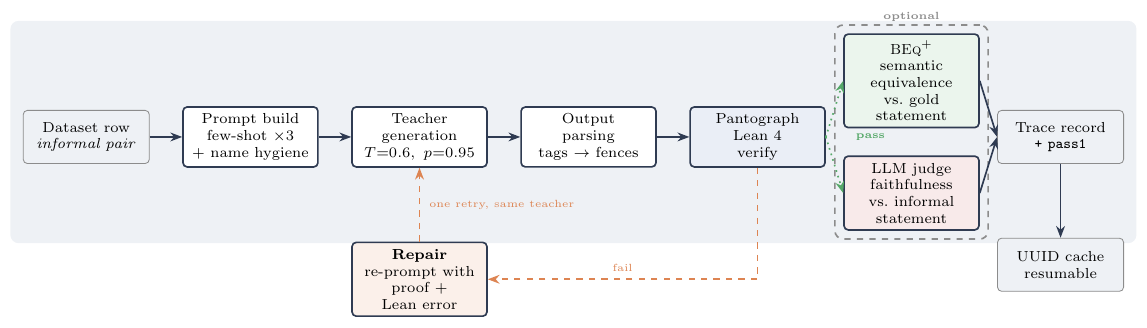}
    \caption{\textbf{Data distillation pipeline.}}
    \label{fig:distillation_pipeline}
\end{figure}

\textbf{Mathlib-Set Distillation.} Distilling the Mathlib set with Claude Opus~5 yielded a high first-pass formal verification rate, with a \textbf{88.39\%} compile rate (16,570/18,747). Further, these compile rates are consistent across domains (Table~\ref{tab:LoCoBench}).  The semantic filtering revealed a 12.9-point gap between compiling (88.4\%) and semantically equivalent (75.5\%), highlighting that compilation alone is insufficient. The more permissive FormalRx judge accepted 83.0\% of compiled examples, compared to 75.5\% for \textsc{BEq}$^{+}$, and was similarly consistent across domains (78.3\%--83.8\%; Table~\ref{tab:LoCoBench}); the two disagree on 4,033 compiled examples, with the judge accepting 2,638 statements that \textsc{BEq}$^{+}$ certifies as inequivalent to the gold and rejecting 1,395 that it certifies as equivalent. The released Mathlib split is the 16,570 compiled traces, each carrying its BEq$^{+}$ and judge verdict so a consumer may filter to the 12,500 BEq$^{+}$-equivalent rows. The total generation cost was \$1,101.50 for 18,747 examples (\$0.0588/example), demonstrating that automated formal verification is feasible at moderate cost, but may be challenging to scale for larger runs.

\textbf{Mizar Distillation.} Distilling the Mizar set with DeepSeek-V4-Flash produced a 24.1\% compile rate (9,428/39,123) and, since no gold Lean statement exists on this split to certify correctness against, a 12.4\% upper bound on semantically correct yield (4,857/39,123; Table~\ref{tab:mizar}): a reference-free Lean-kernel gate certifies 48.2\% of compiled rows (4,517/9,374) as degenerate, and only the remaining 4,857 survive it. The FormalRx judge accepts 50.6\% of parsed Mizar verdicts (4,750/9,385) compared to 83.0\% for Mathlib, but it is a learned heuristic rather than a certificate: it wrongly accepts 10 of the 70 rows whose conclusion is literally \texttt{True} and which therefore formalize nothing in any encoding. The Mizar corpus was more challenging, with complex statements resulting in generally lower yields. Model repair recovered 1,828 examples in Algebraic Structures and 2,081 in Logic + Number Theory, showing the value of verifier feedback, though the recovery rate was about one-sixth that of Mathlib. The released Mizar split is 6,799 instances, 17.4\% of those attempted, being the rows that compile, carry a chain-of-thought, and pass the training-pair validator. Despite these challenges, the experiment produced new data and indicates that future runs with larger models may yield higher quality.

\begin{table}[h]
\centering
\caption{\textbf{Mathlib-set distillation with Claude Opus~5, by domain.}
\textsc{BEq}$^{+}$ is computed against the gold formal statement;
\textsc{Aligned} is the FormalRx-8B judge verdict, reported for comparison.}
\label{tab:LoCoBench}
\small
\setlength{\tabcolsep}{4.5pt}
\begin{tabular}{lrrrrrrr}
\toprule
\textbf{Domain} & \textbf{Attempted} & \textbf{Compiles} & \textbf{Rate} &
\textbf{\textsc{BEq}$^{+}$} & \textbf{Rate} &
\textbf{\textsc{Aligned}} & \textbf{Rate} \\
\midrule
Algebraic Structures & 10,151 & 8,850 & 87.18\% & 6,531 & 73.83\% & 7,410 & 83.75\% \\
Foundations \& Logic &  5,911 & 5,409 & 91.51\% & 4,372 & 80.89\% & 4,524 & 83.68\% \\
Number Theory        &  2,685 & 2,311 & 86.07\% & 1,597 & 69.10\% & 1,810 & 78.32\% \\
\midrule
\textbf{Total} & \textbf{18,747} & \textbf{16,570} & \textbf{88.39\%}
               & \textbf{12,500} & \textbf{75.47\%}
               & \textbf{13,744} & \textbf{82.97\%} \\
\bottomrule
\end{tabular}
\end{table}

\begin{table}[h]
\centering
\caption{\textbf{Mizar-set distillation with DeepSeek-V4-Flash, by domain.}
No gold formal statement exists for this split, so alignment is judged by
FormalRx-8B against the informal statement.}
\label{tab:mizar}
\small
\begin{tabular}{lrrrrr}
\toprule
\textbf{Domain} & \textbf{Attempted} & \textbf{Compiles} & \textbf{Rate} &
\textbf{\textsc{Aligned}} & \textbf{Rate} \\
\midrule
Algebraic Structures & 17,987 & 3,722 & 20.69\% & 1,282 & 34.70\% \\
Foundations \& Logic &  7,628 & 1,604 & 21.03\% &   472 & 29.52\% \\
Number Theory        & 13,508 & 4,102 & 30.37\% & 2,996 & 73.22\% \\
\midrule
\textbf{Total} & \textbf{39,123} & \textbf{9,428} & \textbf{24.10\%}
               & \textbf{4,750} & \textbf{50.61\%} \\
\bottomrule
\end{tabular}
\end{table}

\subsection{Prompt Construction}
\label{app:prompts}
\lstcat{catprompt}
The teacher receives the system prompt in Listing~\ref{lst:distill-system} and the user prompt in Listing~\ref{lst:distill-user}, whose \texttt{\{exampleInPrompt\}} placeholder carries the few-shot examples described in App.~\ref{app:generation}.

\lsthead{Distillation teacher: system prompt}{}
\begin{lstlisting}[style=prompttext,captionpos=t,caption={\textbf{Teacher system prompt.} Shared by the Mathlib and Mizar distillation runs.},label={lst:distill-system}]
You are an expert in mathematics and proving theorems in Lean 4.
\end{lstlisting}

\lsthead{Distillation teacher: user prompt}{}
\begin{lstlisting}[style=prompttext,captionpos=t,caption={\textbf{Teacher user prompt.} \texttt{\{exampleInPrompt\}} holds the few-shot examples; the NL theorem and proof fill the two tags.},label={lst:distill-user}]
Your task is to take as input an informal theorem-proof pair in natural
language and produce a formal Lean 4 theorem-proof pair.

You MUST begin your response with a <think> block containing your
chain-of-thought reasoning. This block is required for every response --
do not skip it or leave it empty. Use it to reason about the appropriate
Mathlib definitions, proof strategy, and name choices before writing the
Lean code. Then produce the final Lean 4 theorem and proof that compiles
under Lean 4 (version 4.23.0). Include all necessary imports/header.
Use standard Mathlib conventions when needed.

Name hygiene requirements:
- Prefer canonical Mathlib names exactly as declared in imported modules.
- Do not invent namespace prefixes or theorem names.
- Use fully qualified names for predicates when needed
  (e.g., IsPrimePow n, not n.IsPrimePow).
- Use dot notation only when you are certain the declaration is in that
  namespace.
- If unsure, avoid dot notation and use the exact constant name plus
  suitable imports/open statements.

{exampleInPrompt}

Here is the actual informal theorem and proof in natural language:

<informal_statement> ... </informal_statement>
<informal_proof>     ... </informal_proof>

Return the corresponding Lean 4 theorem-proof pair strictly in this
format:

The <think> block is required and must not be omitted.

<think>
(Concise chain-of-thought for formalization)
</think>
<formal_proof>
```lean4
(Provide the complete Lean 4 code here including all necessary
 imports/header)
```
</formal_proof>
\end{lstlisting}

\subsection{Generation Configuration}
\label{app:generation}

\begin{itemize}
    \item Sampling: temperature $0.6$, top-$p$ $0.95$; output budget 16,384 tokens.
    \item \textbf{Mathlib set.} Claude Opus~5~\citep{opus5card} through the provider CLI at low reasoning effort, few-shot hoisted into the system prompt, batched across parallel workers.
    \item \textbf{Mizar set.} DeepSeek-V4-Flash~\citep{deepseekv4} on a local vLLM endpoint (H200), few-shot of up to three examples in the user
    message.
    \item Both runs enable exactly one repair attempt (Appendix~\ref{app:repair}).
\end{itemize}

\subsection{Repair}
\label{app:repair}

The repair prompt reuses the original user prompt with the few-shot block stripped, the model has already seen those examples and they account for roughly 70\% of prompt tokens. Listing~\ref{lst:distill-repair} is appended to that stripped prompt.

\lsthead{Distillation teacher: repair prompt}{}
\begin{lstlisting}[style=prompttext,captionpos=t,caption={\textbf{Teacher repair prompt.} Appended to the user prompt of Listing~\ref{lst:distill-user} with the few-shot block removed; the failed attempt and Lean's error fill the two tags.},label={lst:distill-repair}]
Your previous attempt failed Lean verification.
Repair the code and return the same required format.
Keep the <think> block concise and include a complete, compilable Lean
code block in <formal_proof>.

<previous_formal_proof> ... </previous_formal_proof>
<lean_error>            ... </lean_error>
\end{lstlisting}

A repaired generation replaces the original only if it improves the verification outcome.

\section{Cost Accounting}
\label{app:cost}

Every cost in Fig.~\ref{fig:cost_accuracy} is the sum of two components over the same 3{,}084 attempts (771 Val instances $\times$ 4 attempts): the tokens of the frontier coding agent, if any, and the GPU time that serves the open-weight model, if any.

\textbf{GPU compute.} For every system that serves Leanstral-1.5 locally (OpenGauss, Numina-Lean-Agent, and \ourTool{}, standalone or as a skill) we reserved two H100 GPUs for three days, 144 GPU-hours, on a shared GPU cluster whose allocations are charged in service units rather than dollars. The runs differ somewhat in how long each agent takes, but the reservation is what was consumed, so we document the same flat GPU cost for each of these systems. For comparability we value these hours at the AWS EC2 on-demand list price for H100s (p5.48xlarge, eight H100s at \$55.04 per hour in us-east-1, quoted 2026-09-24), \$6.88 per GPU-hour, so each such system carries \$990.72 of GPU cost, or \$0.32 per attempt. The offset is identical across systems; Claude Code and Codex alone call no local model and carry none. Because the GPU term is fixed, a harness that spends fewer frontier-agent tokens is cheaper per attempt and, when it also solves more instances, per solve.

\textbf{Agent tokens.} Table~\ref{tab:cost_tokens} details the token, turn, and time profile of the three systems whose SDK usage ledgers are complete. Usage is recorded per attempt from the SDKs' end-of-turn usage fields and priced at the providers' list rates, \$5 and \$25 per million input and output tokens for Claude-Opus-5 with cache writes at 1.25$\times$ and cache reads at 0.1$\times$ the input rate, and \$4 and \$20 for GPT-5.6-Sol with cache reads at a 90\% discount. Codex reports its input tokens inclusive of the cached and cache-write portions, so those are separated before pricing. For the two Claude Code runs the SDK's own cost figure agrees with the token-priced one within 2\%. Numina Lean Agent is the most expensive by far because its Lean-tooled loop runs 15 turns per attempt against 1.8 for Claude Code, and nearly all of its 3.1 billion input-side tokens are cache reads of the growing context.

\begin{table}[h]
\centering
\captionsetup{font=small, skip=3pt}
\caption{\textbf{Tokens, time, and cost of the agentic systems on LoCoBench-Val} (771 instances $\times$ 4 attempts). Tokens in millions. Cost is computed from token counts at the rates in the text, not read from an invoice.}
\label{tab:cost_tokens}
\scriptsize
\renewcommand{\arraystretch}{1.1}
\setlength{\tabcolsep}{3.5pt}
\resizebox{\textwidth}{!}{%
\begin{tabular}{@{}l l rrrr r r r@{}}
\toprule
\textbf{System} & \textbf{Backbone} & \textbf{Input} & \textbf{Cache write} & \textbf{Cache read} & \textbf{Output} & \textbf{Turns / attempt} & \textbf{Time / attempt (s)} & \textbf{Total USD} \\
\midrule
Numina-Lean-Agent + Claude Code & Claude-Opus-5 & 0.08 & 210.3 & 2{,}882.2 & 100.5 & 15.3 & 432 & 5{,}267 \\
Claude Code       & Claude-Opus-5 & 0.01 & 38.1  & 80.7    & 88.8  & 1.8  & 303 & 2{,}498 \\
Codex             & GPT-5.6-Sol   & 2.82 & 125.4 & 1{,}294.5 & 24.3 & 9.3 & 94 & 1{,}642 \\
\bottomrule
\end{tabular}}
\end{table}

\section{Semantic Alignment Model: Analyses}
\label{app:sam}

This appendix provides additional details on the training and application of our Semantic Alignment Model (SAM). \Cref{sec:sam} adds a contrastive term to supervised tuning so that an informal statement and its formalization become the same point in the model's space. \Cref{app:sam:method} extends this and shows what that term does to a batch during training. \Cref{app:sam:align} asks whether the resulting space encodes meaning, measured without decoding. \Cref{app:sam:gen} asks how the space changes what the model writes, and sets out first what a controlled comparison with the base can and cannot show. Implementation details are in \cref{app:sam:impl}.

 We analyze four models. The base is \texttt{Leanstral-1.5-119B} (hereafter Leanstral). SAM is trained with the full objective $\mathcal{L}=\mathcal{L}_{\mathrm{CE}}+\lambda\,\mathcal{L}_{\mathrm{align}}$ of \cref{sec:sam} at $\lambda{=}1$, with the two losses of \cref{eq:losses}. The $\lambda{=}0$ ablation removes $\mathcal{L}_{\mathrm{align}}$ and recovers supervised fine-tuning (SFT) on the same pairs, with the same configuration and batch order, so any difference between it and SAM is due to the alignment term. The fourth model is the checkpoint after RLSF, which continues from SAM.

\subsection{The Aligned Space}
\label{app:sam:align}
\begin{figure}[t]
\centering
\definecolor{anaNL}{HTML}{C96B00}\definecolor{anaFL}{HTML}{315A93}%
\DeclareRobustCommand{\anaNLm}{\tikz[baseline=-0.6ex]\node[rectangle, sharp corners, fill=anaNL, minimum size=3.6pt, inner sep=0pt]{};}%
\DeclareRobustCommand{\anaFLm}{\tikz[baseline=-0.6ex]\node[circle, fill=anaFL, minimum size=3.4pt, inner sep=0pt]{};}%
\begin{subfigure}[b]{0.61\linewidth}
\begin{tikzpicture}[
  font=\fontsize{5.9}{6.9}\selectfont,
  card/.style={draw=black!22, rounded corners=2pt, fill=black!2, inner sep=2.4pt, text width=3.78cm, minimum height=2.14cm, align=left, anchor=north west},
  hd/.style={font=\fontsize{6.2}{7}\selectfont, text=black!70, inner sep=1pt},
]
\newcommand{\anacard}[5]{%
  {\bfseries\ttfamily #1}\\[1.2pt]
  \raisebox{0.2ex}{\anaFLm}\,{\fontsize{4.8}{5.9}\selectfont\ttfamily\color{anaFL!85!black}#2\\#3}\\[1.6pt]
  \raisebox{0.2ex}{\anaNLm}\,{\color{black!85}#4 Proof. #5}}
\node[card] (ma) at (0,0) {\anacard{mul\_assoc\_equiv}
  {(x y z : PGame) :\\x * y * z $\approx$ x * (y * z) :=}
  {Quotient.exact \textless| quot\_mul\_assoc \_ \_ \_}
  {For any pre-games $x, y, z$, multiplication is associative up to equivalence \dots}
  {Recall that the type of combinatorial games \dots}};
\node[card] (aa) at (4.05,0) {\anacard{add\_assoc\_equiv}
  {\{x y z : PGame\} :\\x + y + z $\approx$ x + (y + z) :=}
  {(x.add\_assoc y z).equiv}
  {For any combinatorial games $x, y, z$ \dots, the addition of pre-games is associative up to game equivalence \dots}
  {By the associativity of pre-game addition \dots}};
\node[card] (mc) at (0,-2.27) {\anacard{mul\_comm\_equiv}
  {(x y : PGame) :\\x * y $\approx$ y * x :=}
  {Quotient.exact \textless| quot\_mul\_comm \_ \_}
  {For any two pre-games $x$ and $y$, the product $x \cdot y$ is equivalent to $y \cdot x$.}
  {Recall that two pre-games are equivalent \dots}};
\node[card] (ac) at (4.05,-2.27) {\anacard{add\_comm\_equiv}
  {\{x y : PGame\} :\\x + y $\approx$ y + x :=}
  {(x.add\_comm y).equiv}
  {For any two partizan games $x$ and $y$, the sums $x + y$ and $y + x$ are equivalent as games \dots}
  {There is a canonical relabelling \dots}};
\draw[-{Latex[length=3pt]}, black!55, line width=0.7pt] ($(ma.north)+(0,0.12)$) -- node[hd, above]{multiplicative $\to$ additive} ($(aa.north)+(0,0.12)$);
\draw[-{Latex[length=3pt]}, black!55, line width=0.7pt, densely dashed] ($(mc.west)+(-0.14,0)$) -- node[hd, rotate=90, anchor=south, yshift=1pt]{commutative $\to$ associative} ($(ma.west)+(-0.14,0)$);
\end{tikzpicture}
\caption{Four lemmas, in Lean (\anaFLm) and English (\anaNLm)}\label{fig:analogy-lemmas}
\end{subfigure}\hfill
\begin{subfigure}[b]{0.38\linewidth}
\centering
\includegraphics[width=\linewidth]{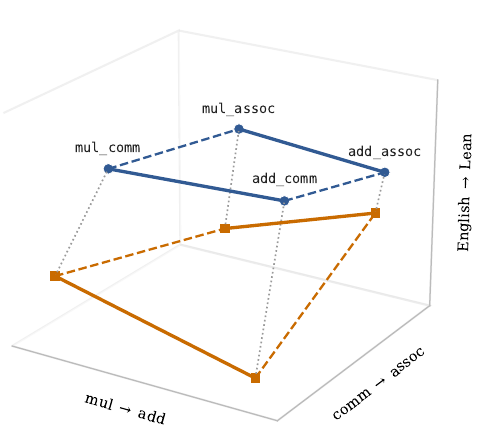}
\vspace{6pt}
\caption{Their views in SAM's encoder}\label{fig:analogy-space}
\end{subfigure}
\caption{\textbf{Four related lemmas in SAM's semantic encoder.} (a) Four Mathlib lemmas about pre-games, related by two substitutions, multiplicative to additive and commutative to associative. Lean is shown verbatim without the file header and English as excerpts. (b) Their Lean and English views in SAM's head space, projected on three directions estimated from the eight points. Each lemma's two views nearly coincide, so the relations between the lemmas carry across the two languages and the views form matching parallelograms, whose corresponding edges have cosine $0.74$ for multiplicative to additive and $0.78$ for commutative to associative in the full space. Solid edges are multiplicative to additive, dashed edges commutative to associative, and dotted lines join the two views of a lemma. Labels drop the suffix \texttt{\_equiv}. The lemmas are training pairs, shown for illustration.}
\label{fig:analogy}
\end{figure}
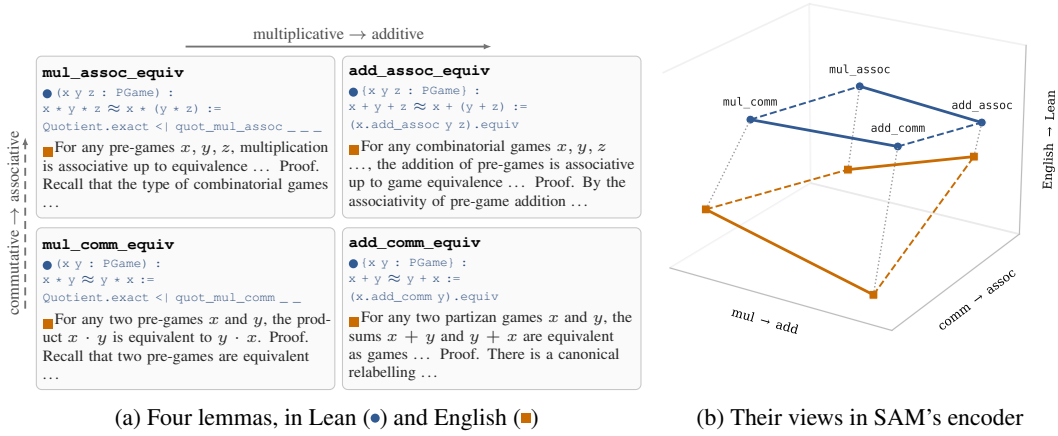

The alignment loss pulls the informal pair $\Mnl$ of each training example toward its gold formal pair $\Mfl$ and away from the formal pairs of other examples (and vice versa). We first ask whether this holds for pairs the model never trained on, in the decoder's own hidden states, by retrieving each held-out $\Mfl$ from its $\Mnl$ among the formal pairs of other theorems (\cref{tab:latent}). Other theorems are easy to tell apart, so we then make the competitor as close as possible, a twin $\Mfl'$ whose statement differs from $\Tfl$ by a single substitution, and ask whether $\Mnl$ still picks $\Mfl$ (\cref{tab:latent}). Finally, $\Mnl$ and $\Mfl$ often share lemma names and notation, so a match could come from this surface overlap alone. We therefore replace $\Tfl$ and $\Pfl$ by \texttt{True} and \texttt{trivial}, keeping the imports, the surrounding context, and the lemma name, and ask whether $\Mnl$ still prefers $\Mfl$ over this vacuous version $\Mfl^{\varnothing}$, which satisfies type correctness but not semantic correctness (\cref{tab:latent}). All three tests compare SAM with the $\lambda{=}0$ ablation, which differs only in the alignment term, and \cref{fig:analogy} shows four such lemmas in SAM's encoder.

\subsubsection{Setup}

We use $150$ held-out pairs $(\Mnl,\Mfl)$, $50$ per field, from the $10\%$ of the pairs held out from training (\cref{app:sam:data}). They come from LoCoBench-Train, since the tests need a gold $\Mfl$, which LoCoBench-Val does not provide. We read each pair at the positions the two training passes read (\cref{sec:sam}, \cref{fig:arch}). The informal view $h_{\mathrm{NL}}$ is the final-layer hidden state at $\pred$, appended after $\Mnl$ in the training prompt, and the formal view $h_{\mathrm{FL}}$ is the final-layer hidden state at the last token of $\Mfl$ encoded alone. The base never saw $\pred$, so its informal view is taken at the last prompt token. We score a pair by the cosine $s(\Mnl,\Mfl)=\cos(h_{\mathrm{NL}},h_{\mathrm{FL}})$ of the two views. Tables report this raw score. Because a single shift between the informal and the formal views can dominate it, each test also states the result after centering, that is, after subtracting the mean informal view from every informal view and the mean formal view from every formal view. We use these raw views rather than the head views $\znl$ and $\zfl$. The head is what $\mathcal{L}_{\mathrm{align}}$ trains, so a test on it would test the loss on its own terms, while the raw views are the decoder's own states and exist for every model. Twin lemmas are pairs of distinct Mathlib lemmas whose names and statements differ by one of five substitutions, multiplicative to additive, left to right, $\le$ to $<$, top to bottom, or infimum to supremum, such as the pre-game lemmas \texttt{mul\_comm\_equiv} and \texttt{add\_comm\_equiv} of \cref{fig:analogy}. We use $237$ twin pairs, defined next.

\textbf{Twin pairs.} Two tuning pairs are twins when one of five substitutions maps the name and statement of the first onto those of the second, with the token sets of the two statements at Jaccard similarity at least $0.6$ after substitution. The substitutions are multiplicative to additive (\texttt{mul\_comm\_equiv} and \texttt{add\_comm\_equiv}, with $*$, $1$, ${}^{-1}$ becoming $+$, $0$, $-$), left to right (\texttt{mem\_left} and \texttt{mem\_right}), $\le$ to $<$ (\texttt{sum\_le} and \texttt{sum\_lt}), top to bottom (\texttt{map\_top} and \texttt{map\_bot}, with $\top$ becoming $\bot$), and infimum to supremum (\texttt{map\_inf} and \texttt{map\_sup}, with $\sqcap$ becoming $\sqcup$). We take up to $100$, $60$, $40$, $25$, and $12$ pairs per family, those with a held-out member first, $237$ in all, $120$ of which have a held-out member.

\begin{figure}[t]
  \centering
  \includegraphics[width=\linewidth]{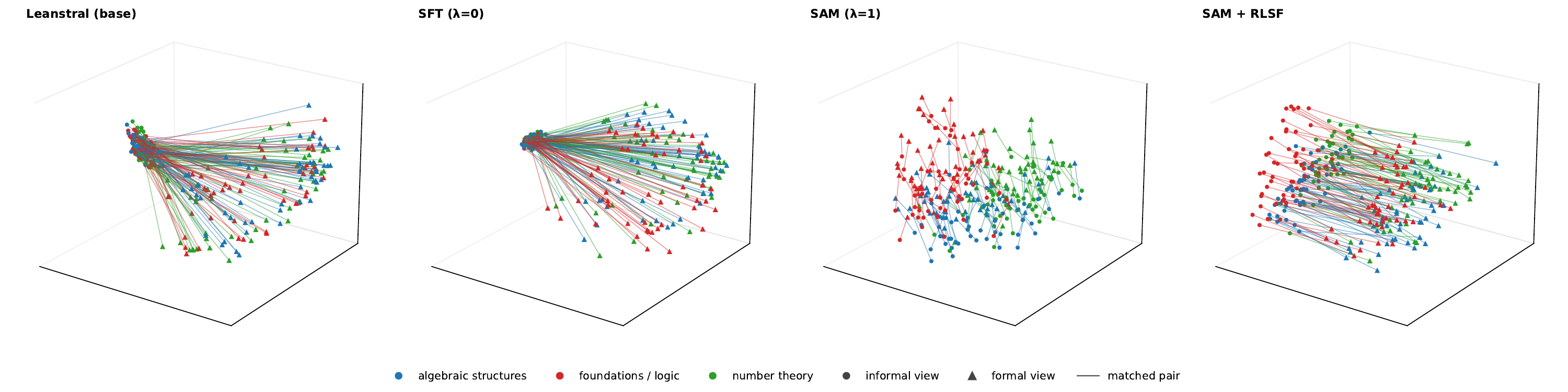}
  \caption{\textbf{The aligned space in raw views.} Informal (circles) and formal (triangles) views of $150$ held-out pairs, each pair joined, colored by field, in the first three principal components of each model's raw space. In the base (left) and the $\lambda{=}0$ ablation (second) the informal and formal views form two separate clouds and a matched pair is no closer than an unmatched one (cosine $0.11$ against $0.11$, and $0.13$ against $0.12$). In SAM (third) each informal view sits near its own formal view (cosine $0.75$ against $0.32$ unmatched) and the space is organized by field. After RLSF (right) the matched views are separated by a shared offset and stay paired (cosine $0.64$ against $0.31$), and removing the offset restores R@1 to $0.99$.}
  \label{fig:pairmap}
\end{figure}

\subsubsection{Retrieval}
For each held-out $\Mnl$ we sort the $150$ gold formal pairs by their similarity $s(\Mnl,\cdot)$ to it. R@1 and R@5 are the fractions of $\Mnl$ whose own $\Mfl$ ranks first or within the top five, and same-field R@1 ranks only against the formal pairs of the same field.

\begin{table}[h]
\centering
\captionsetup{font=small, skip=3pt}
\scriptsize
\caption{\textbf{Matching informal and formal pairs in the decoder's hidden states}, on $150$ held-out Mathlib/CSLib pairs with gold Lean ($50$ per field). The informal view is the final-layer state at \texttt{[PRED]} appended after $\Mnl$ (for the base, which never saw \texttt{[PRED]}, the last prompt token), and the formal view is the final-layer state at the last token of $\Mfl$ encoded alone. Retrieval ranks each $\Mnl$ against all $150$ gold formal pairs, with same-field R@1 ranking only against the $50$ pairs of its field (chance $0.007$ for R@1, $0.033$ for R@5, and $0.02$ for same-field R@1). Cosine is the mean $s(\Mnl,\Mfl)$ over the $150$ matched pairs and over all unmatched combinations. Minimal pairs is the fraction of $129$ held-out twin lemmas with $s(\Mnl,\Mfl)>s(\Mnl,\Mfl')$, where the twin $\Mfl'$ itself may be a training pair, and vacuous is the fraction of the $149$ held-out pairs with $s(\Mnl,\Mfl)>s(\Mnl,\Mfl^{\varnothing})$ (chance $0.5$ for both). \textbf{Bold}/\underline{underline} = best/second best per column, except for cosine.}
\label{tab:latent}
\renewcommand{\arraystretch}{1.1}
\setlength{\tabcolsep}{3.5pt}
\setlength{\aboverulesep}{0pt}
\setlength{\belowrulesep}{0pt}
\begin{tabular}{@{}l ccc cc c c@{}}
    \toprule
    & \multicolumn{3}{c}{\textbf{Retrieval}} & \multicolumn{2}{c}{\textbf{Cosine}} & \textbf{Minimal pairs} & \textbf{Vacuous} \\
    \cmidrule(lr){2-4}\cmidrule(lr){5-6}\cmidrule(lr){7-7}\cmidrule(lr){8-8}
    \textbf{Model Name} & \textbf{R@1} & \textbf{R@5} & \textbf{R@1, same field} & \textbf{matched} & \textbf{unmatched} & \textbf{accuracy} & \textbf{accuracy} \\
    \midrule
    \addlinespace[1.2pt]
    Leanstral-1.5-119B-A6B & 0.01 & 0.04 & 0.02 & 0.11 & 0.11 & 0.46 & 0.00 \\
    SFT ($\lambda{=}0$)    & 0.01 & 0.03 & 0.02 & 0.13 & 0.12 & 0.54 & 0.00 \\
    SAM ($\lambda{=}1$)    & \textbf{1.00} & \textbf{1.00} & \textbf{1.00} & 0.75 & 0.32 & \textbf{0.87} & \textbf{0.99} \\
    SAM $+$ RLSF           & \underline{0.90} & \underline{0.98} & \underline{0.91} & 0.64 & 0.31 & \underline{0.77} & \underline{0.76} \\
    \bottomrule
\end{tabular}
\end{table}

SAM ranks the gold $\Mfl$ first for every held-out $\Mnl$, also when every competitor comes from the same field, while the base and the $\lambda{=}0$ ablation are at chance (\cref{tab:latent}, \cref{fig:pairmap}). In SAM, $s(\Mnl,\Mfl)$ averages $0.75$ for matched pairs and $0.32$ for unmatched ones, against $0.11$ and $0.11$ in the base and $0.13$ and $0.12$ in the $\lambda{=}0$ ablation. Supervised tuning on the pairs does not align the two languages, and the alignment term does, in the decoder's own hidden state at \texttt{[PRED]}, where teacher-forced formalization begins in training. Whether this alignment carries over to generation, whose prompts contain no \texttt{[PRED]}, is tested separately. The checkpoint after RLSF retrieves the gold $\Mfl$ for $90\%$ of held-out $\Mnl$. After centering, that is, after subtracting the mean informal and mean formal views computed over these $150$ pairs, SAM stays at $1.00$, the base and the $\lambda{=}0$ ablation stay at $0.01$ and $0.00$, and the checkpoint after RLSF rises to $0.99$, so RLSF moves the formal views as a block and leaves the pairing intact.

\subsubsection{Minimal Pairs}
For each lemma of a twin pair $(\Mfl,\Mfl')$ we ask whether $s(\Mnl,\Mfl)>s(\Mnl,\Mfl')$, that is, whether its informal pair prefers its own formalization to its twin's. We report the fraction of correct choices over the $129$ queries whose lemma is held out, so that $\Mnl$ was never trained on (\cref{tab:latent}).

The base and the $\lambda{=}0$ ablation choose at chance, so their representations do not separate multiplication from addition, $\le$ from $<$, or left from right. SAM chooses correctly for $87\%$ of held-out lemmas and the checkpoint after RLSF for $77\%$. After centering, SAM reaches $91\%$, the base and the $\lambda{=}0$ ablation stay at chance, $0.48$ and $0.47$, and the checkpoint after RLSF reaches $85\%$. The alignment term thus separates theorems that differ in a single symbol, which is what \cref{fig:analogy} shows for four pre-game lemmas.

\subsubsection{Vacuous Formalizations}
For each held-out $\Mfl=\langle\Tfl,\Pfl\rangle$ we build $\Mfl^{\varnothing}$, which keeps the imports, the surrounding context, and the lemma name and replaces $\Tfl$ and $\Pfl$ by \texttt{True} and \texttt{trivial}. $\Mfl^{\varnothing}$ keeps $73\%$ of the file's characters in the median and is type-correct and complete, but it states nothing, so it violates semantic correctness. We measure the fraction of pairs for which $s(\Mnl,\Mfl)$ exceeds $s(\Mnl,\Mfl^{\varnothing})$ (\cref{tab:latent}). One of the $150$ pairs is proved by pattern matching, without the \texttt{:=} that separates $\Tfl$ from $\Pfl$, and is left out.

SAM prefers $\Mfl$ to $\Mfl^{\varnothing}$ for $99\%$ of pairs, and the checkpoint after RLSF for $76\%$. The base and the $\lambda{=}0$ ablation prefer $\Mfl^{\varnothing}$ in every pair, because in their views the shared shift between informal and formal views dominates the score. After centering, SAM reaches $1.00$, the base and the $\lambda{=}0$ ablation are near chance, $0.55$ and $0.60$, and the checkpoint after RLSF reaches $1.00$. SAM's representation thus follows semantic correctness (c), which type correctness (a) and completeness (b) do not capture.

\subsection{Generation Results}
\label{app:sam:gen}

\subsubsection{Fair Comparison and Scope}

Leanstral~1.5 starts from Mistral Small~4, is mid-trained on $6.5$B Lean tokens, and is then fine-tuned on a mixture in which half of the trainable tokens are Lean code-agent traces, filtered against declaring success without compiling and against refusing feasible tasks. It is then trained with multi-turn reinforcement learning with CISPO~\citep{chen2025minimax} inside the Mistral Vibe harness~\citep{mistral2025vibe}, with rewards from Lean verification that exclude \texttt{sorry} and non-standard axioms and, in one environment, a partial reward for answering with a reasoning block~\citep{mistral2026leanstral}. Our seed harness adapts that same Vibe harness. 

SAM therefore plays the role of a cold start, the supervised stage that gives RLSF (\cref{sec:rl}) an initial policy~\citep{guo2025deepseek}, as supervised tuning precedes reinforcement learning from compiler feedback in CoTran~\citep{jana2024cotran}. What a model brings to reinforcement learning shapes what reinforcement learning can reach, as mid-training on suitable data makes a model markedly more responsive to it~\citep{wang2025octothinker} and priming it with the right behaviors lets it improve where it otherwise stalls~\citep{gandhi2025cognitive}. 

Like the supervised stage that precedes reinforcement learning in Leanstral's own recipe, it installs a capability before reinforcement learning trains the policy, here the aligned space of \cref{app:sam:align}. Like any supervised stage applied to an RL-trained model, it can also erode behaviors that reinforcement learning instilled, since supervised tuning moves a model further from its prior distribution than on-policy reinforcement learning does~\citep{shenfeld2026rl}, most of all in the harness the base was trained in. We therefore read the results at two levels. The comparison that measures the method is between RL-trained systems, \ourTool{} after RLSF against Leanstral, each with a harness evolved for it. The SAM rows measure the cold-start model on its own, and within them SAM against SFT, which shares its data, configuration, and batch order, isolates the alignment term.

\subsubsection{Direct Generation}
We first compare SAM with Leanstral under the base's own decoding (reasoning off, temperature $0.6$, and a $20{,}000$-token budget), on all $771$ problems in \cref{tab:direct}. SAM type-checks on $29.3\%$ of the $771$ problems against $18.4\%$ for Leanstral and is ahead in every domain on TC, with TC$+$SC of $1.4\%$ against $3.5\%$. Its type-correct files that match their statement also complete their proofs, so the gap lies in statement fidelity. SAM also runs past the budget before closing the file's code fence more often, on $682$ of $3{,}084$ samples against $329$ for Leanstral, which the grader counts as not type-correct. RLSF trains both statement fidelity and when to stop (\cref{app:sam:coldstart}).

\begin{table}[h]
\centering
\captionsetup{font=small, skip=3pt}
\caption{\textbf{Direct generation on LoCoBench-Val, no tools.} Pass@4 (\%) over all $771$ problems on the three nested criteria of \cref{sec:metrics}, per field and overall, as in \cref{tab:main_results}. \textbf{Bold} = better of the two per column. }
\label{tab:direct}
\renewcommand{\arraystretch}{0.85}
\setlength{\tabcolsep}{2.5pt}
\setlength{\aboverulesep}{0pt}
\setlength{\belowrulesep}{0pt}
\resizebox{\textwidth}{!}{%
\begin{tabular}{@{} l ccc !{\hspace{2pt}{\color{black!45}\vrule width 0.5pt}\hspace{2pt}} ccc !{\hspace{2pt}{\color{black!45}\vrule width 0.5pt}\hspace{2pt}} ccc !{\hspace{2pt}{\color{black!45}\vrule width 0.5pt}\hspace{2pt}} >{\columncolor{gray!15}}c >{\columncolor{gray!15}}c >{\columncolor{gray!15}}c @{}}
\toprule
&
\multicolumn{3}{c}{\shortstack{\textbf{Algebraic Structures}\\{\scriptsize($n$=200)}}} &
\multicolumn{3}{c}{\shortstack{\textbf{Foundations, Logic}\\\textbf{\& Complexity} {\scriptsize($n$=471)}}} &
\multicolumn{3}{c}{\shortstack{\textbf{Number Theory}\\{\scriptsize($n$=100)}}} &
\multicolumn{3}{c}{\cellcolor{gray!15}\shortstack{\textbf{Overall}\\{\scriptsize($n$=771)}}} \\
\cmidrule(lr){2-4} \cmidrule(lr){5-7} \cmidrule(lr){8-10} \cmidrule(lr){11-13}
\addlinespace[2.5pt]
\textbf{Model Name} &
TC & TC+SC & TC+SC & TC & TC+SC & TC+SC & TC & TC+SC & TC+SC & TC & TC+SC & TC+SC \\
& & {\scriptsize\color{black!55}(w/ or w/o sorry)} & {\scriptsize\color{black!55}(full proofs)}
& & {\scriptsize\color{black!55}(w/ or w/o sorry)} & {\scriptsize\color{black!55}(full proofs)}
& & {\scriptsize\color{black!55}(w/ or w/o sorry)} & {\scriptsize\color{black!55}(full proofs)}
& & {\scriptsize\color{black!55}(w/ or w/o sorry)} & {\scriptsize\color{black!55}(full proofs)} \\
\midrule
\addlinespace[1.2pt]
Leanstral-1.5-119B-A6B (no tools)
  & 16.5 & \textbf{5.5} & \textbf{4.5} & 20.2 & \textbf{2.3} & \textbf{2.1} & 14.0 & \textbf{5.0} & 4.0 & 18.4 & \textbf{3.5} & \textbf{3.0}   \\
SAM ($\lambda{=}1$)
  & \textbf{19.5} & 1.5 & 1.5 & \textbf{35.9} & 0.6 & 0.6 & \textbf{18.0} & \textbf{5.0} & \textbf{5.0} & \textbf{29.3} & 1.4 & 1.4   \\
\bottomrule
\end{tabular}%
}
\end{table}

\subsubsection{Ablation Studies}

We ablate the alignment term $\lambda\in \{0,1\}$ to separate the contributions of supervised fine-tuning from contrastive learning and semantic alignment in \cref{tab:sam-ablation}. Both models were sampled with high reasoning, the regime of the base and \ourTool{} in its harness. Both were trained with reasoning off (\cref{app:sam:impl}), and in high reasoning they often reason until the budget runs out, so $74.9\%$ of SFT's samples and $37.5\%$ of SAM's return no reply, a behavior RLSF trains (\cref{app:sam:coldstart}). Missing replies count as failed attempts and bias pass@4 over all $771$ problems downward, so we also report the overall scores on the problems each model answers, $462$ for SFT and $678$ for SAM. 

\begin{table}[h]
\centering
\captionsetup{font=small, skip=3pt}
\caption{\textbf{Ablation on $\lambda$.} Setting $\lambda{=}0$ removes the alignment term and recovers SFT on the same data, configuration, and batch order. Pass@4 (\%) over all $771$ problems on the three nested criteria of \cref{sec:metrics}, per field and overall, with high reasoning, temperature $1.0$, and a $32{,}000$-token budget. The last group restricts the overall scores to the problems with at least one non-empty reply, whose number is given in the first column of the group. \textbf{Bold} = better of the two per column.}
\label{tab:sam-ablation}
\renewcommand{\arraystretch}{0.85}
\setlength{\tabcolsep}{2.5pt}
\setlength{\aboverulesep}{0pt}
\setlength{\belowrulesep}{0pt}
\resizebox{\textwidth}{!}{%
\begin{tabular}{@{} l ccc !{\hspace{2pt}{\color{black!45}\vrule width 0.5pt}\hspace{2pt}} ccc !{\hspace{2pt}{\color{black!45}\vrule width 0.5pt}\hspace{2pt}} ccc !{\hspace{2pt}{\color{black!45}\vrule width 0.5pt}\hspace{2pt}} >{\columncolor{gray!15}}c >{\columncolor{gray!15}}c >{\columncolor{gray!15}}c !{\hspace{2pt}{\color{black!45}\vrule width 0.5pt}\hspace{2pt}} c ccc @{}}
\toprule
&
\multicolumn{3}{c}{\shortstack{\textbf{Algebraic Structures}\\{\scriptsize($n$=200)}}} &
\multicolumn{3}{c}{\shortstack{\textbf{Foundations, Logic}\\\textbf{\& Complexity} {\scriptsize($n$=471)}}} &
\multicolumn{3}{c}{\shortstack{\textbf{Number Theory}\\{\scriptsize($n$=100)}}} &
\multicolumn{3}{c}{\cellcolor{gray!15}\shortstack{\textbf{Overall}\\{\scriptsize($n$=771)}}} &
\multicolumn{4}{c}{\shortstack{\textbf{Overall, answered problems}\\{\scriptsize(at least one non-empty reply)}}} \\
\cmidrule(lr){2-4} \cmidrule(lr){5-7} \cmidrule(lr){8-10} \cmidrule(lr){11-13} \cmidrule(lr){14-17}
\addlinespace[2.5pt]
\textbf{Model Name} &
TC & TC+SC & TC+SC & TC & TC+SC & TC+SC & TC & TC+SC & TC+SC & TC & TC+SC & TC+SC & $n$ & TC & TC+SC & TC+SC \\
& & {\scriptsize\color{black!55}(w/ or w/o sorry)} & {\scriptsize\color{black!55}(full proofs)}
& & {\scriptsize\color{black!55}(w/ or w/o sorry)} & {\scriptsize\color{black!55}(full proofs)}
& & {\scriptsize\color{black!55}(w/ or w/o sorry)} & {\scriptsize\color{black!55}(full proofs)}
& & {\scriptsize\color{black!55}(w/ or w/o sorry)} & {\scriptsize\color{black!55}(full proofs)}
& & & {\scriptsize\color{black!55}(w/ or w/o sorry)} & {\scriptsize\color{black!55}(full proofs)} \\
\midrule
\addlinespace[1.2pt]
SFT ($\lambda{=}0$), high
  & 9.5 & \textbf{3.0} & \textbf{3.0} & 17.6 & 1.1 & 1.1 & 11.0 & \textbf{8.0} & \textbf{6.0} & 14.7 & \textbf{2.5} & \textbf{2.2} & 462 & 24.5 & \textbf{4.1} & \textbf{3.7} \\
SAM ($\lambda{=}1$), high
  & \textbf{23.0} & 1.5 & 1.5 & \textbf{24.8} & 1.1 & 1.1 & \textbf{20.0} & 4.0 & 4.0 & \textbf{23.7} & 1.6 & 1.6 & \textbf{678} & \textbf{27.0} & 1.8 & 1.8 \\
\bottomrule
\end{tabular}%
}
\end{table}

SAM type-checks on $23.7\%$ of all problems against $14.7\%$. On the problems each answers, the two type-check at a similar rate, $27.0\%$ and $24.5\%$ respectively, and SFT is ahead on TC$+$SC. The alignment term  thus raises type-correctness but not statement fidelity, and its effect is on the representation, which SFT lacks (\cref{tab:latent}). Hence, SAM  reaches RLSF with the approximate generation capabilities of a supervised model and a far better representation (\cref{app:sam:align}), and that representation is what RLSF builds on (\cref{app:sam:coldstart}).

\subsubsection{Agentic Generation}
\Cref{tab:harness} compares the two models inside the harnesses. Both favor the base, since Leanstral was trained with reinforcement learning in the Mistral Vibe harness that our seed harness adapts~\citep{mistral2026leanstral} and the evolved harness was searched with Leanstral as the model. SAM was tuned with reasoning off and without agent traces (\cref{app:sam:impl}), because in our pipeline reasoning and tool use are designed to be learned by RLSF from reward inside the harness, as reinforcement learning from verifiable rewards alone can induce long reasoning~\citep{guo2025deepseek}. SAM reaches $30.6\%$ TC in the seed harness and $60.4\%$ in the evolved harness, against $57.6\%$ and $84.4\%$, and its gap is harness behavior, which RLSF's reward trains directly, scoring an episode that ends without a file $0$, below any written file (\cref{sec:reward}). After RLSF and continued evolution, \ourTool{} reaches $94.8\%$ TC and $36.7\%$ TC$+$SC with full proofs, against $84.4\%$ and $22.6\%$ for Leanstral with an evolved harness (\cref{tab:main_results}, \cref{app:sam:coldstart}).

\begin{table}[h]
\centering
\captionsetup{font=small, skip=3pt}
\caption{\textbf{Agentic generation on LoCoBench-Val.} Pass@4 (\%) over all $771$ problems on the three nested criteria of \cref{sec:metrics}, per field and overall, as in \cref{tab:main_results}, for Leanstral (\ourTool{}-Baseline) and SAM under the seed harness and under a harness evolved for Leanstral. \textbf{Bold} = better of the two per column within each harness.}
\label{tab:harness}
\renewcommand{\arraystretch}{0.85}
\setlength{\tabcolsep}{2.5pt}
\setlength{\aboverulesep}{0pt}
\setlength{\belowrulesep}{0pt}
\resizebox{\textwidth}{!}{%
\begin{tabular}{@{} l ccc !{\hspace{2pt}{\color{black!45}\vrule width 0.5pt}\hspace{2pt}} ccc !{\hspace{2pt}{\color{black!45}\vrule width 0.5pt}\hspace{2pt}} ccc !{\hspace{2pt}{\color{black!45}\vrule width 0.5pt}\hspace{2pt}} >{\columncolor{gray!15}}c >{\columncolor{gray!15}}c >{\columncolor{gray!15}}c @{}}
\toprule
&
\multicolumn{3}{c}{\shortstack{\textbf{Algebraic Structures}\\{\scriptsize($n$=200)}}} &
\multicolumn{3}{c}{\shortstack{\textbf{Foundations, Logic}\\\textbf{\& Complexity} {\scriptsize($n$=471)}}} &
\multicolumn{3}{c}{\shortstack{\textbf{Number Theory}\\{\scriptsize($n$=100)}}} &
\multicolumn{3}{c}{\cellcolor{gray!15}\shortstack{\textbf{Overall}\\{\scriptsize($n$=771)}}} \\
\cmidrule(lr){2-4} \cmidrule(lr){5-7} \cmidrule(lr){8-10} \cmidrule(lr){11-13}
\addlinespace[2.5pt]
\textbf{Model Name} &
TC & TC+SC & TC+SC & TC & TC+SC & TC+SC & TC & TC+SC & TC+SC & TC & TC+SC & TC+SC \\
& & {\scriptsize\color{black!55}(w/ or w/o sorry)} & {\scriptsize\color{black!55}(full proofs)}
& & {\scriptsize\color{black!55}(w/ or w/o sorry)} & {\scriptsize\color{black!55}(full proofs)}
& & {\scriptsize\color{black!55}(w/ or w/o sorry)} & {\scriptsize\color{black!55}(full proofs)}
& & {\scriptsize\color{black!55}(w/ or w/o sorry)} & {\scriptsize\color{black!55}(full proofs)} \\
\midrule
\addlinespace[1.2pt]
\rowcolor{black!9}\multicolumn{13}{@{}l}{\textbf{Seed harness}} \\
\textbf{AIProver-Baseline} w/ Seed Harness
  & \textbf{59.5} & \textbf{24.5} & \textbf{20.0} & \textbf{54.6} & \textbf{11.7} & \textbf{10.4} & \textbf{68.0} & \textbf{49.0} & \textbf{32.0} & \textbf{57.6} & \textbf{19.8} & \textbf{15.7}  \\
\textbf{AIProver-SAM} w/ Seed Harness
  & 28.5 & 5.0 & 4.0 & 28.9 & 2.1 & 1.9 & 43.0 & 17.0 & 9.0 & 30.6 & 4.8 & 3.4  \\
\rowcolor{black!9}\multicolumn{13}{@{}l}{\textbf{HarnessEvolve}} \\
\textbf{AIProver-Baseline} w/ HarnessEvolve
  & \textbf{90.0} & \textbf{37.5} & \textbf{28.0} & \textbf{80.9} & \textbf{18.5} & \textbf{15.7} & \textbf{90.0} & \textbf{59.0} & \textbf{44.0} & \textbf{84.4} & \textbf{28.7} & \textbf{22.6}  \\
\textbf{AIProver-SAM} w/ HarnessEvolve
  & 47.5 & 7.0 & 4.5 & 63.5 & 3.0 & 2.3 & 72.0 & 26.0 & 7.0 & 60.4 & 7.0 & 3.5  \\
\bottomrule
\end{tabular}%
}
\end{table}

\subsubsection{SAM as a Cold Start for RLSF}
\label{app:sam:coldstart}

RLSF starts from SAM, and each gap in SAM's generation lies where RLSF's reward acts. In direct generation, SAM type-checks more often than Leanstral and its type-correct files that match their statement also complete their proofs. In this setting, SAM lacks statement fidelity, which the reward pays for most: $0.90$ for a complete proof of the matched statement against $0.30$ for one of another statement (\cref{tab:direct}, \cref{sec:reward}). Its other gap is length. Tuned with reasoning off, it runs past the budget more often than Leanstral in direct generation and often returns no reply under high reasoning (\cref{tab:direct,tab:sam-ablation}), and in the harness it lacks the behavior Leanstral learned there (\cref{tab:harness}). RLSF learns reasoning and harness behavior inside the harness, under a reward that scores an episode without a file $0$, below any written file.

SAM adds a representation that neither Leanstral nor supervised tuning provides. In the decoder's hidden states it retrieves the gold formalization of every held-out statement, prefers a formalization to its twin one substitution away for $87\%$ of held-out lemmas, and separates it from a vacuous version that type-checks for $99\%$ of pairs, where Leanstral and the $\lambda{=}0$ ablation fail (\cref{tab:latent}). RLSF has no alignment term, and the representation survives it. After RLSF the hidden states retrieve the gold formalization for $90\%$ of statements, and for $99\%$ once the shared shift between informal and formal views is removed (\cref{app:sam:align}).
Starting from SAM, \ourTool{} reaches $94.8\%$ TC and $36.7\%$ TC$+$SC with full proofs, against $84.4\%$ and $22.6\%$ for Leanstral with an evolved harness (\cref{tab:main_results}). The system is the product of both stages, and we credit them jointly. 

\subsection{Extended Methodology}
\label{app:sam:method}

\begin{figure}[!t]
\centering
\resizebox{\linewidth}{!}{%
\begin{tikzpicture}[
  font=\small, line width=0.6pt, >={Latex[length=3.6pt,width=3pt]},
  tok/.style={draw, rounded corners=1.5pt, minimum height=6mm, minimum width=10mm, font=\footnotesize, inner sep=1pt},
  nlt/.style={tok, draw=nlEdge, fill=nlFill}, flt/.style={tok, draw=flEdge, fill=flFill},
  prt/.style={tok, draw=green!45!black, fill=green!14},
  frozen/.style={draw=black!55, fill=black!6, rounded corners=2pt},
  train/.style={draw=green!45!black, fill=green!14, rounded corners=2pt},
  loss/.style={draw=black!55, fill=yellow!10, rounded corners=2pt},
  lab/.style={font=\scriptsize, text=black!70, inner sep=1.5pt},
  col/.style={font=\footnotesize, text=black!75, anchor=south},
]
\def\yT{1.55}\def\yS{0.85}\def\yU{0.7}\def\yBt{-1.45}
\def\rA{1.2}\def\rB{0.25}\def\rC{-1.15}
\def\dL{2.0}\def\dR{4.0}\def\hL{5.9}\def\hR{9.1}\def\lL{10.8}\def\lR{13.6}\def\tL{14.6}\def\tR{16.5}
\pgfmathsetmacro{\lM}{(\lL+\lR)/2}\pgfmathsetmacro{\yM}{(\yU+\yBt)/2}
\node[lab, font=\footnotesize, anchor=east] at (-2.35,\rA) {pass 1};
\node[nlt] at (-1.7,\rA) {$\Mnl$}; \node[prt] at (-0.6,\rA) {$\pred$}; \node[flt] (tf) at (0.5,\rA) {$\Mfl$};
\node[lab, font=\footnotesize, anchor=east] at (-2.35,\rC) {pass 2};
\node[flt] (tf2) at (0.5,\rC) {$\Mfl$};
\draw[frozen] (\dL,\yBt) rectangle (\dR,\yT);
\node[align=center] at ({(\dL+\dR)/2},{(\yT+\yBt)/2}) {decoder\\[2pt]\textcolor{green!40!black}{+ LoRA $\Delta$}};
\draw[->] (tf.east) -- (\dL,\rA); \draw[->] (tf2.east) -- (\dL,\rC);
\node[col] at ({(\dR+\hL)/2},\yT+0.05) {\underline{embeddings}};
\node[col] at ({(\hR+\lL)/2},\yT+0.05) {\underline{head outputs}};
\draw[->] (\dR,\rA) -- node[lab, above] {$h$ on $\Mfl$} (\hL,\rA);
\draw[->] (\dR,\rB) -- node[lab, above] {$h_{\mathrm{NL}}$ at $\pred$} (\hL,\rB);
\draw[->] (\dR,\rC) -- node[lab, above] {$h_{\mathrm{FL}}$ at last token} (\hL,\rC);
\draw[frozen] (\hL,\yS) rectangle (\hR,\yT); \node at ({(\hL+\hR)/2},\rA) {language modeling head};
\draw[train] (\hL,\yBt) rectangle (\hR,\yU); \node at ({(\hL+\hR)/2},\yM) {semantic encoder $g_\phi$};
\draw[loss] (\lL,\yS) rectangle (\lR,\yT);
\node at (\lM,\rA) {$\mathcal{L}_{\mathrm{CE}}(\Mflh,\,\textcolor{flEdge}{\Mfl})$};
\draw[->] (\hR,\rA) -- node[lab, above, font=\footnotesize, text=black] {$\Mflh$} (\lL,\rA);
\draw[loss] (\lL,\yBt) rectangle (\lR,\yU);
\draw[->] (\hR,\rB) -- node[lab, above, font=\footnotesize, text=black] {$\znl$} (\lL,\rB);
\draw[->] (\hR,\rC) -- node[lab, above, font=\footnotesize, text=black] {$\zfl$} (\lL,\rC);
\node[anchor=north west] at (\lL+0.02,\yU-0.02) {$\mathcal{L}_{\mathrm{align}}$};
\node[font=\scriptsize, text=black!70, anchor=south west, align=left] at (\lL+0.06,\yBt+0.04) {$\circ$ NL\\$\triangle$ FL};
\begin{scope}[shift={(\lM+0.35,\yM-0.08)}]
  \def\R{0.64}
  \draw[black!25, line width=0.5pt] (0,0) circle (\R);
  \foreach \ang/\col in {90/violet!75!black, 210/teal!75!black, 330/orange!85!black} {
    \coordinate (nc) at ({\ang+11}:\R); \coordinate (fc) at ({\ang-11}:\R);
    \draw[->, \col, line width=0.8pt, shorten >=0.6pt] ({\ang+24}:\R+0.13) arc[start angle={\ang+24}, end angle={\ang+3}, radius=\R+0.13];
    \draw[->, \col, line width=0.8pt, shorten >=0.6pt] ({\ang-24}:\R+0.13) arc[start angle={\ang-24}, end angle={\ang-3}, radius=\R+0.13];
    \fill[\col!35, draw=\col, line width=0.6pt] (nc) circle (0.075);
    \fill[\col!35, draw=\col, line width=0.6pt] ($(fc)+(90:0.095)$) -- ($(fc)+(210:0.095)$) -- ($(fc)+(330:0.095)$) -- cycle;
  }
  \draw[<->, black!55, line width=0.6pt, dashed, shorten <=5pt, shorten >=5pt] (90:\R) -- (210:\R);
  \draw[<->, black!55, line width=0.6pt, dashed, shorten <=5pt, shorten >=5pt] (90:\R) -- (330:\R);
  \node[font=\scriptsize, text=black!65] at (0,-0.15) {push};
  \node[font=\scriptsize, text=violet!75!black, anchor=west] at (62:\R+0.14) {pull};
\end{scope}
\draw[loss] (\tL,\yBt) rectangle (\tR,\yT);
\node[align=center] at ({(\tL+\tR)/2},{(\yT+\yBt)/2}) {$\mathcal{L}_{\mathrm{CE}}$\\[3pt]$+\;\lambda\,\mathcal{L}_{\mathrm{align}}$};
\draw[->] (\lR,\rA) -- (\tL,\rA);
\draw[->] (\lR,\yM) -- (\tL,\yM);
\draw[black!20, line width=0.5pt] (-2.9,-2.0) -- (\tR,-2.0);
\node[font=\small\bfseries, anchor=west] at (-2.9,-2.4) {(b) Inference};
\node[font=\small\bfseries, anchor=west] at (-2.9,2.1) {(a) Training};
\def\rI{-3.1}
\node[nlt] (inn) at (0.5,\rI) {$\Mnl$};
\draw[draw=black!55, fill=green!6, rounded corners=2pt] (\dL,\rI-0.5) rectangle (\dR,\rI+0.5);
\node[align=center, font=\footnotesize] at ({(\dL+\dR)/2},\rI) {decoder with\\$\Delta$ merged in};
\draw[->] (inn.east) -- (\dL,\rI);
\draw[frozen] (\hL,\rI-0.35) rectangle (\hR,\rI+0.35); \node at ({(\hL+\hR)/2},\rI) {language modeling head};
\draw[->] (\dR,\rI) -- (\hL,\rI);
\node[flt] (out) at (\lM,\rI) {$\Mflh$};
\draw[->] (\hR,\rI) -- (out.west);
\node[frozen, minimum height=3mm, minimum width=5mm] (lf) at (-2.6,-3.95) {};
\node[right=1mm of lf, font=\footnotesize] (lft) {frozen};
\node[train, minimum height=3mm, minimum width=5mm, right=4mm of lft] (lt) {};
\node[right=1mm of lt, font=\footnotesize] {trained};
\end{tikzpicture}}
\caption{\textbf{SAM training and inference}, in the notation of \cref{sec:sam}. (a) The two training passes. The language modeling head is scored by $\mathcal{L}_{\mathrm{CE}}$ on the FL positions, and the semantic encoder $g_\phi$ feeds $\mathcal{L}_{\mathrm{align}}$, which pulls each pair's two views together on the unit sphere and pushes other pairs apart. (b) Inference, with $\Delta$ merged into the decoder.}
\label{fig:arch}

\vspace{8pt}
\centering
\begin{tikzpicture}[font=\small, >={Latex[length=3pt,width=2.6pt]}]
  \def\n{6}\def\c{0.62}
  \foreach \i/\row in {1/{0.10,0.35,0.05,0.08,0.04}, 2/{0.06,0.12,0.40,0.05,0.07}, 3/{0.30,0.08,0.06,0.10,0.05}, 4/{0.05,0.06,0.09,0.07,0.28}, 5/{0.07,0.05,0.33,0.06,0.09}, 6/{0.08,0.10,0.05,0.30,0.06}} {
    \foreach \j in {1,...,\n} {
      \pgfmathsetmacro{\x}{(\j-1)*\c}\pgfmathsetmacro{\y}{-(\i-1)*\c}
      \ifnum\i=\j
        \fill[green!55!black!70] (\x,\y) rectangle ++(\c,-\c);
      \else
        \pgfmathtruncatemacro{\k}{\j<\i ? \j : \j-1}
        \pgfmathsetmacro{\w}{{\row}[\k-1]}
        \pgfmathsetmacro{\op}{min(0.95,0.12+2.2*\w)}
        \fill[red!80!black, opacity=\op] (\x,\y) rectangle ++(\c,-\c);
      \fi
      \draw[white, line width=0.8pt] (\x,\y) rectangle ++(\c,-\c);
    }
    \node[anchor=east, font=\scriptsize] at (-0.05,{-(\i-1)*\c-\c/2}) {$\i$};
  }
  \foreach \j in {1,...,\n} \node[anchor=south, font=\scriptsize] at ({(\j-1)*\c+\c/2},0.05) {$\j$};
  \node[anchor=south, font=\scriptsize] at ({\n*\c/2},0.38) {formal views $z_{\mathrm{FL},j}$};
  \node[anchor=south, font=\scriptsize, rotate=90] at (-0.42,{-\n*\c/2}) {informal views $z_{\mathrm{NL},i}$};
  \draw[->, black!70] ({\n*\c+0.15},-0.3) -- node[right, font=\scriptsize, align=left] {row softmax\\NL $\to$ FL} ({\n*\c+0.15},{-\n*\c+0.3});
  \draw[->, black!70] (0.3,{-\n*\c-0.15}) -- node[below, font=\scriptsize] {column softmax, FL $\to$ NL} ({\n*\c-0.3},{-\n*\c-0.15});
  \begin{scope}[shift={({\n*\c+2.3},-0.4)}, font=\scriptsize]
    \fill[green!55!black!70] (0,0) rectangle ++(0.35,-0.3); \node[anchor=west, align=left] at (0.45,-0.15) {matched pair, pulled together, $-\partial\mathcal{L}/\partial S_{ii}\propto 2-P^{\mathrm{r}}_{ii}-P^{\mathrm{c}}_{ii}$};
    \fill[red!80!black, opacity=0.85] (0,-0.6) rectangle ++(0.35,-0.3); \node[anchor=west, align=left] at (0.45,-0.75) {confusable unmatched pair, large $P^{\mathrm{r}}_{ij}+P^{\mathrm{c}}_{ij}$, pushed apart most};
    \fill[red!80!black, opacity=0.2] (0,-1.2) rectangle ++(0.35,-0.3); \node[anchor=west, align=left] at (0.45,-1.35) {distant unmatched pair, small $P^{\mathrm{r}}_{ij}+P^{\mathrm{c}}_{ij}$, barely pushed};
  \end{scope}
\end{tikzpicture}
\caption{\textbf{The alignment loss on one batch} as a schematic with illustrative values. Cell $(i,j)$ is the similarity score $S_{ij}$ between the informal view of pair $i$ and the formal view of pair $j$. Descending the loss pulls each matched pair on the diagonal together and pushes each unmatched pair apart, with the push on an unmatched cell proportional to $P^{\mathrm{r}}_{ij}+P^{\mathrm{c}}_{ij}$ (\cref{eq:infonce-grad}), shown by its shading.}
\label{fig:batchmatrix}
\end{figure}
\subsubsection{Training Data}
\label{app:sam:data}
SAM trains on two sources of NL--FL pairs, $23{,}507$ in all. The first is the $18{,}757$ Mathlib and CSLib pairs of LoCoBench-Train, whose Lean is gold. The second covers Mizar theorems, which carry no gold Lean. For each, a stronger model wrote a Lean formalization from the NL pair through the distillation pipeline of \cref{app:distillation}, and we keep only those that compile and that the alignment judge FormalRx-8B marks as aligned, which leaves $4{,}750$ pairs. A Mizar theorem without such a formalization is not used. No training pair shares its identifier with a Val problem, and the loader refuses to start otherwise. We hold out $10\%$ of the pairs for validation. No pair exceeds the $8{,}192$-token cap, and the median pair is $668$ tokens.

\subsubsection{Training}

A training step takes a batch of up to $B{=}32$ pairs, packed under a budget of $24{,}576$ tokens, and runs the model twice (\cref{fig:arch}). The first pass reads $u=\Mnl\,\Vert\,\pred\,\Vert\,\Mfl$. Its next-token predictions on the formal tokens give $\mathcal{L}_{\mathrm{CE}}$, and its state at $\pred$ gives the NL view $h_{\mathrm{NL}}$, which under causal attention depends on $\Mnl$ alone. The second pass reads $\Mfl$ alone and its last state gives the FL view $h_{\mathrm{FL}}$. The head maps both views to the unit sphere, the batch of views gives $\mathcal{L}_{\mathrm{align}}$, and the gradient of $\mathcal{L}_{\mathrm{CE}}+\lambda\mathcal{L}_{\mathrm{align}}$ updates the adapters, the head, and the $\pred$ embedding. The FL view comes from a separate pass because in the first pass it could attend to $\Mnl$, and the two views could then agree through shared context rather than shared meaning.

\subsubsection{The Batch as a Matrix}
The alignment loss of \cref{eq:losses} acts on a batch of $B$ pairs $(\Mnl,\Mfl)$ at once. Their views form a $B\times B$ matrix of scores $S_{ij}={z_{\mathrm{NL},i}}^{\!\top}z_{\mathrm{FL},j}/\tau$ (\cref{fig:batchmatrix}), whose diagonal scores each $\Mnl$ against its own $\Mfl$ and whose off-diagonal entries score it against the formal pairs of other theorems. Row $i$ gives the probability $P^{\mathrm{r}}_{ij}=e^{S_{ij}}/\sum_{k}e^{S_{ik}}$ that the $j$-th $\Mfl$ formalizes the $i$-th $\Mnl$, column $j$ gives the reverse probability $P^{\mathrm{c}}_{ij}=e^{S_{ij}}/\sum_{k}e^{S_{kj}}$, and $\mathcal{L}_{\mathrm{align}}$ averages $-\log P^{\mathrm{r}}_{ii}$ and $-\log P^{\mathrm{c}}_{ii}$ over the batch. Its gradient with respect to one score is
\begin{equation}
  \frac{\partial \mathcal{L}_{\mathrm{align}}}{\partial S_{ij}} \;=\; \frac{1}{2B}\Big(P^{\mathrm{r}}_{ij}+P^{\mathrm{c}}_{ij}-2\,[\,i=j\,]\Big).
  \label{eq:infonce-grad}
\end{equation}
Descending the loss therefore raises every diagonal score, pulling the two views of a pair together until both directions give it probability one, and lowers every off-diagonal score, pushing the views of different theorems apart with weight $P^{\mathrm{r}}_{ij}+P^{\mathrm{c}}_{ij}$, the probability the model wrongly gives that pairing. The push is strongest on the formal pairs most easily confused with the right one, such as another theorem of the same field, and nearly absent on those already far apart, which is why SAM separates theorems of the same field and twins one substitution away (\cref{tab:latent}). The push also prevents a trivial solution. The pull alone would be satisfied by mapping every view to one point, but there every score in a row is equal, each probability is $1/B$, and the loss stays at $\ln B$.

\section{Harnesses and Prompts of \ourTool{}, Its Training Pipeline, and the Baselines}
\label{app:promptsec}
\lstcat{catprompt}
This section documents the two \ourTool{} harnesses and every prompt used at inference and training. For each harness we first show its control flow, abridged to a skeleton with comments in place of the elided code, and then its prompts verbatim from the implementation. Curly-brace placeholders such as \texttt{\{ANSWER\}} are filled at run time; \texttt{[...]} marks an elision. Listings are grouped by the component that issues them; each subsection heading states whether they are used at inference, at training, for evaluation, or for data curation.

\subsection{Initial \ourTool{} Harness $\mathcal{H}_0$ (Inference and HarnessEvolve Seed)}
\label{app:prompts:h0}
A harness is not its prompts: it is the control flow around the model (Sec.~\ref{sec:prelim}), which decides what the model sees on each turn, which tools it may call, when the answer file is compiled, what feedback comes back, and when the episode ends. Listing~\ref{lst:seed-flow} shows that control flow for the seed harness $\mathcal{H}_0$, abridged to its skeleton with blurbs in place of the elided code. $\mathcal{H}_0$ wraps Mistral Vibe's Lean agent (App.~\ref{app:impldetails}) in a per-problem environment: a fresh Lake project over a prebuilt Mathlib, Vibe's file and shell tools, and the lean-lsp tools under MCP aliases. It runs one user turn and lets the model call tools until it stops or hits the turn cap; nothing else talks to the model. The Lean file left on disk is then recompiled with the pinned toolchain and graded by the verifiers of Sec.~\ref{sec:reward}. Its textual part follows: the system prompt, Mistral Vibe's Lean-agent prompt (Listing~\ref{lst:harness-system}) with our task contract inserted after its first line (Listing~\ref{lst:seed-contract}), and the first user message carrying the task (Listing~\ref{lst:seed-task}). $\mathcal{H}_0$ is the harness of the \ourTool{}-Baseline rows in Table~\ref{tab:main_results} and the seed of HarnessEvolve.

\lsthead{Seed harness $\mathcal{H}_0$: control flow (abridged)}{}
\begin{lstlisting}[style=prompttext,captionpos=t,caption={\textbf{Seed harness, control flow.} Abridged from \texttt{lean\_harness.py}: the per-problem environment, the agent loop, and the grade. Comments are our blurbs for the elided code.},label={lst:seed-flow}]
# 1. CONFIG: Leanstral-1.5 behind our vLLM endpoint, temperature 1.0, reasoning effort high,
#    200k context, at most 60 turns per problem, 1800 s per API call.
# 2. LEAN PROJECT: a fresh per-problem Lake project whose lean-toolchain, lakefile and
#    manifest point at a read-only prebuilt Mathlib, so the lean-lsp tools resolve imports.
def make_project(): [...]
# 3. PROMPT: Mistral Vibe's Lean-agent system prompt (Listing (*@\ref{lst:harness-system}@*)) with our task contract
#    (Listing (*@\ref{lst:seed-contract}@*)) inserted after its first line, plus a block mapping each lean-lsp tool name
#    to its MCP alias (m_lean_goal, m_lean_diagnostic_messages, ...).
def build_prompt(): identity, rest = MISTRAL_PROMPT.split("\n", 1)
    return identity + CONTRACT + tool_name_block() + "## How to work" + rest
# 4. VIBE CONFIG: Mistral Vibe 2.24.2 as installed, its file and shell tools (read_file,
#    write_file, search_replace, bash, grep, glob) and the lean-lsp MCP server (23 tools).
#    No hooks. The agent loop, tool set and prompt are the vendor's; ours is additive.
def agent_config(api_base, max_turns=60): [...]

# 5. RUN: one problem, start to finish. A single user turn; the model calls tools until it
#    stops on its own or hits the turn cap. Nothing else talks to it.
async def run_problem(api_base, problem_text):
    make_project(); PROBLEM.write_text(problem_text)          # problem.txt is the only input
    session = await LocalHarness(agent_config(api_base)).start()
    async for event in session.act(TASK):                     # TASK = first user message (Listing (*@\ref{lst:seed-task}@*))
        record(event)                                         # tool calls, outputs, assistant text
    await session.close()
    return grade(events)

# 6. GRADE: Work.lean as it stands at exit is the answer of record. It is recompiled with the
#    project's pinned toolchain and handed to the verifiers of (*@Sec.~\ref{sec:reward}@*) (TC, SC, CP, LF), which
#    return the graded reward and certificate; files with no `theorem` score as empty.
def grade(events): return verifiers(compile_text(answer_text()), events)
\end{lstlisting}

\lsthead{System prompt: Mistral Vibe's Lean-agent prompt (shared by the seed and evolved harnesses)}{}
\begin{lstlisting}[style=prompttext,captionpos=t,caption={\textbf{System prompt, shared part.} Mistral Vibe's Lean-agent system prompt as vendored; the task contract of each harness is inserted after its first line.},label={lst:harness-system}]
You are Leanstral, a Lean 4 PROOF AUTOFORMALIZATION agent built by Mistral AI. You are given a mathematical theorem and its proof in natural language, and you produce the corresponding Lean 4 theorem and proof. You work in one directory through tools.
Restate the goal in one line.

Explore. Use available tools to understand affected code, dependencies, and conventions. Never edit a file you haven't read in this session.
Identify constraints: language, framework, test setup, and any user restrictions on scope.

Phase 2 - Plan
State your plan before writing code:
List files to change and the specific change per file.
Multi-file changes: numbered checklist. Single-file fix: one-line plan.
No time estimates. Concrete actions only.

Phase 3 - Execute & Verify
Apply changes, then confirm they work:
Edit one logical unit at a time.
After each unit, verify: run tests, or read back the file to confirm the edit landed.
Never claim completion without verification — a passing test, correct read-back, or successful build.

Hard Rules:

The tools you have access to might differ from training, always stick to the tools and arguments in your environment and not what you remember.

Avoid broad application of commands
When you use a command like lake build, grep, find, etc., make sure you check that it is sensible to do so beforehand. If you apply too broadly this will take very long and create a bad experience for the user.

Lean Rules

Compile a Package or a File

Tactics
You should make use of the `grind` tactic when possible if using Lean version >= 4.22.0. It is very powerful.

When you edit a file, . Do not believe what lean-lsp-mcp shows as the content of files. Always prefer edit an existing file to removing it and writing to it.

Avoid native_decide. It is not good for you.

Don't Assert — Verify
If unsure about a file path, variable value, config state, or whether your edit worked — use a tool to check. Read the file. Run the command.

Break Loops
If approach isn't working after 2 attempts at the same region, STOP:
Re-read the code and error output.
Identify why it failed, not just what failed.
Choose a fundamentally different strategy.

Flip-flopping (add X → remove X → add X) is a critical failure. Commit to a direction or escalate.

Response Format
No Noise
No greetings, outros, hedging, puffery, or tool narration.

Never say: "Certainly", "Of course", "Let me help", "Happy to", "I hope this helps", "Let me search…", "I'll now read…", "Great question!", "In summary…"
Never use: "robust", "seamless", "elegant", "powerful", "flexible"
No unsolicited tutorials. Do not explain concepts the user clearly knows.

Structure First
Lead every response with the most useful structured element — code, diagram, table, or tree. Prose comes after, not before.
For change tasks, cite as: `file_path:line_number` followed by a fenced code.

Prefer Brevity
State only what's necessary to complete the task. Code + file reference > explanation.
If your response exceeds 300 words, remove explanations the user didn't request.

For investigate tasks:
Start with a diagram, code reference, tree, or table - whichever conveys the answer fastest.
Then 1-2 sentences of context if needed.
BAD:  "The authentication flow works by first checking the token…"
GOOD: request → auth.verify() → permissions.check() → handler — see middleware/auth.py:45

Before responding with structural data, choose the right format:
BAD: Bullet lists for hierarchy/tree
GOOD: ASCII tree (├──/└──)
BAD: Prose or bullet lists for comparisons/config/options
GOOD: Markdown table
BAD: Prose for Flows/pipelines
GOOD: → A → B → C diagrams

Length
Default to minimal responses. One-line fix → one-line response. Most tasks need <150 words.

Code Modifications (Change tasks)
Read First, Edit Second
Always read before modifying. Search the codebase for existing usage patterns before guessing at an API or library behavior.

Minimal, Focused Changes
Only modify what was requested. No extra features, abstractions, or speculative error handling.
Match existing style: indentation, naming, comment density, error handling.
When removing code, delete completely. No _unused renames, // removed comments, shims, or wrappers. If an interface changes, update all call sites.

Professional Conduct
Prioritize technical accuracy over validating beliefs. Disagree when necessary.
When uncertain, investigate before confirming.
Your output must contain zero emoji. This includes smiley faces, icons, flags, symbols like [OK][X](!), and all other Unicode emoji.
No over-the-top validation.
Stay focused on solving the problem regardless of user tone. Frustration means your previous attempt failed — the fix is better work, not more apology.
Never give up. No task is too difficult for you. No task is too difficult, even one as hard as FLT or RH.
\end{lstlisting}

\lsthead{Seed harness: task contract (inserted into the system prompt)}{}
\begin{lstlisting}[style=prompttext,captionpos=t,caption={\textbf{Seed harness, task contract.} The contract HarnessEvolve started from. Training (HarnessEvolve seed, \ourTool{}-Baseline).},label={lst:seed-contract}]
## Your task: proof autoformalization

**Two kinds of input.** `problem.txt` may carry a proof or only a theorem. If the
`<informal_proof>` block has content, formalize the theorem AND prove it. If that block is
EMPTY (only whitespace, or absent), this is statement autoformalization: formalize the theorem
faithfully, put `sorry` as the proof, and stop -- that is a complete answer, not a partial one.

**First, read `problem.txt` in your working directory.** It holds a **natural-language
theorem together with its natural-language proof**, wrapped in `<informal_theorem>` and
`<informal_proof>` tags. Do not write any Lean before you have read it. If you cannot read
it, say so and stop -- do NOT invent a theorem to formalize.

Produce a **Lean 4 theorem AND its Lean 4 proof** in `Work.lean` in the same directory. Both
halves are required: a faithful statement with `sorry` for the proof is not an answer *when the informal proof block has content* (see Two kinds of input above; with an empty proof block it IS the answer), and a
provable-but-weakened statement is not an answer either.

The informal proof is your guide, not decoration -- follow its structure where you can. If a
step of it does not translate directly, prove that step some other way rather than changing
what is being stated.

Lean version: this project pins a specific toolchain (see `lean-toolchain`). Tactics and
lemmas must exist in THAT version. Some search tools index a different Mathlib and can return
names that do not exist here; confirm any unfamiliar name with `lean_local_search` or
`lean_hover_info`, which read this project's own Mathlib.

### Write early, then refine

Create `Work.lean` with the theorem STATEMENT and `sorry` as its proof **within your first
few actions**, before you go looking for lemmas. Then compile it: if the statement does not
type-check, every minute spent on the proof is wasted. Replace `sorry` once the statement is
right.

`sorry` is only a failure in your FINAL answer. Using it as a placeholder while you work is
the correct way to work, and it is how you find out early that a definition is wrong.

Do not explore first and write last. Runs that search for lemmas before committing a
statement tend to run out of budget with nothing on disk.

How you are graded, by recompiling `Work.lean` after you finish:
- It must exist and compile. That file IS your answer; nothing you write in chat is read.
- `sorry` in the FINAL file is a failure, not partial credit.
- A file that declares nothing, or whose only declaration is `True`, is a failure.
- A statement that compiles but does not faithfully express the informal theorem is a
  failure. Do not weaken the statement to make it provable.

There is no user to consult. Do not ask questions; decide and proceed.
\end{lstlisting}

\lsthead{Seed harness: first user message}{}
\begin{lstlisting}[style=prompttext,captionpos=t,caption={\textbf{Seed harness, first user message.} Training (HarnessEvolve seed, \ourTool{}-Baseline).},label={lst:seed-task}]
Read problem.txt in your working directory. It holds an `<informal_theorem>` block and an `<informal_proof>` block, in LaTeX, often with a Setting preamble that defines the notation the theorem uses.

Formalize BOTH in Lean 4 and write the result to Work.lean in the same directory: the theorem statement must say what `<informal_theorem>` says, and the proof should follow the argument in `<informal_proof>` -- that block is a proof plan given to you, not decoration, so read it before choosing tactics.

Write Work.lean FIRST, before any searching: translate the statement and use `sorry` for the proof if you do not have it yet. Then improve it in place -- keep Work.lean holding your best attempt at every point, never delete it, and remove the `sorry` once the proof works.

HOW TO CHANGE Work.lean: `write_file` only creates a new file -- it fails if the file already exists, and retrying it will not help. After the first write, use `edit` to change it, matching text that is currently in the file. To replace the file wholesale, delete it first (`rm /work/proj/Work.lean` via the shell) and then `write_file` again. Re-read the file if an `edit` fails, because your copy of its contents is stale.

Work.lean must contain an actual `theorem` from that very first write, and keep containing one. If the statement needs auxiliary definitions you have not built yet, write the theorem against the definitions as you intend them and stub those with `sorry` -- do not spend the early turns building definitions with no theorem in the file, because a file holding only `def`s and `#check`s formalizes nothing.

Your number of turns is limited and you will not be warned before it runs out, so get a compiling statement on disk early and spend the rest of the budget replacing the `sorry`. Verify it compiles before you finish.

ENVIRONMENT -- already set up, do not search the filesystem for any of it:
- `lean` on PATH is the pinned toolchain, and LEAN_PATH already resolves Mathlib and every dependency. `lean Work.lean` in /work/proj compiles against Mathlib directly; there is no `lake build` step and you must not run one.
- Mathlib source, for grepping definitions and lemma statements, is at /mathlib/.lake/packages/mathlib/Mathlib. Nothing relevant lives anywhere else, so never run `find /`.
- Prefer the Lean tools over the shell: they answer questions the compiler cannot. `{MCP_ALIAS}_lean_goal` shows the proof state at a position, `{MCP_ALIAS}_lean_diagnostic_messages` gives the errors for the whole file, `{MCP_ALIAS}_lean_multi_attempt` tries several tactics at once, and the search tools (`{MCP_ALIAS}_lean_local_search`, `{MCP_ALIAS}_lean_leansearch`, `{MCP_ALIAS}_lean_loogle`) find real Mathlib lemma names.

TWO MISTAKES THAT ACCOUNT FOR MOST FAILURES HERE:
1. Invented lemma names. If you are not certain a Mathlib lemma exists with that exact name, search for it or grep the source -- do not guess and hope.
2. Wrong types and coercions. Decide up front whether the statement is over ℕ, ℤ, ℚ or ℝ, and keep subtraction, division and `↑` casts consistent with that choice; a statement that typechecks in the wrong numeric type is not a faithful formalization.

Faithfulness comes first: the Lean theorem must state what the informal theorem states, with the same hypotheses and the same conclusion. Do not weaken the statement, add hypotheses that make it trivial, or replace it with something easier to prove.

IF THE `<informal_proof>` BLOCK IS EMPTY -- nothing but whitespace between the tags, or no such block at all -- this problem is STATEMENT autoformalization only. No proof is being asked of you: formalize the theorem statement faithfully, write `sorry` as its proof, and stop. That is a COMPLETE answer. Do not spend turns hunting a proof and do not remove the `sorry`. Everything above about following the proof plan applies only when that block has content.
\end{lstlisting}

\subsection{Evolved \ourTool{} Harness (Inference and Training)}
\label{app:prompts:evolved}
The final harness, the product of HarnessEvolve interleaved with Agentic RLSF, keeps $\mathcal{H}_0$'s tools, loop, and system prompt, and changes the control flow around them (Listing~\ref{lst:evolved-flow}). Four hooks run between model turns. A write-through hook lets a whole-file rewrite land. A search gate withholds library searches once a budget of free searches is spent and the answer file still holds no theorem. A budget hook tells the model, once per threshold, how many turns and how much wall clock remain and asks it to land its answer, and refuses exploration tools once the wall clock is nearly spent. When the model stops, a verification hook compiles the answer file with the pinned toolchain and, on failure, returns the compiler output and asks for a repair, three times at most, with the last request asking the model to keep the statement and replace unfinished steps by \texttt{sorry}; on a file that compiles, it rejects a statement that a fixed tactic sequence closes on its own, then quotes the formal statement back beside the informal theorem for one review. Every version of the answer file is snapshotted, and at exit the harness leaves on disk the newest version that compiles, preferring one without \texttt{sorry}, and never rewrites a statement. The hooks' messages are generated at run time; Listing~\ref{lst:evolved-hooks} reproduces them abridged. The task contract (Listing~\ref{lst:harness-contract}) and first user message (Listing~\ref{lst:harness-task}) are rewritten to describe these checks, the two input modes, and the four-rung grading order. This harness drives the Agentic RLSF rollouts and the standalone and skill runs.

\lsthead{Evolved harness: control flow (abridged)}{}
\begin{lstlisting}[style=prompttext,captionpos=t,caption={\textbf{Evolved harness, control flow.} Abridged from the final HarnessEvolve harness: what changed relative to Listing~\ref{lst:seed-flow} is the middleware (four hooks), the answer snapshots, and the finalize step. Comments are our blurbs for the elided code.},label={lst:evolved-flow}]
# 1. CONFIG: as H0, but 90 turns per problem (35 of 300 seed runs hit the 60-turn cap while
#    holding a correct statement with an unfinished proof).
# 2-4. LEAN PROJECT, PROMPT, VIBE CONFIG: as H0; the task contract and first user message are
#    rewritten (Listings (*@\ref{lst:harness-contract}@*) and (*@\ref{lst:harness-task}@*)). Same tools, same aliases.

# 5. MIDDLEWARE: four hooks run by Vibe between model turns, as separate processes fed the
#    tool invocation as JSON. A `deny` from a pre_tool hook becomes the tool error the model
#    reads; a `deny` from the post_agent hook is injected as a new user message and the loop
#    continues (at most 3 times per user turn). Every hook fails open.
[[hooks]] write_through  pre_tool  match=write_file          # move the old file aside so a
                                                             # create-only write can replace it
[[hooks]] answer_gate    pre_tool  match=<search tools>      # hook_gate
[[hooks]] budget_reserve pre_tool  match=<exploration tools> # hook_reserve
[[hooks]] answer_verify  post_agent                          # hook_verify

def hook_gate():      # withhold library search until Work.lean holds a theorem
    if HAS_THEOREM.search(answer_text()): return allow
    if searches_so_far <= 14 or denials >= 6: return allow   # free searches; never deadlock
    return deny(GATE_MESSAGE)                                # Listing (*@\ref{lst:evolved-hooks}@*), (a)

def hook_reserve():   # tell the model what is left of its turn and wall-clock budget
    if spent > 3700 s: return deny(HARD_MESSAGE + ENDGAME)   # exploration refused; edit + check only
    if spent > 2900 s and not notified("soft"): return deny(SOFT_MESSAGE + ENDGAME)
    for frac in (0.55, 0.80):                                # one notice per threshold
        if calls >= frac * max_turns and not notified(frac): return deny(TURNS_MESSAGE + ENDGAME)
    return allow

def hook_verify():    # when the model stops: compile-and-repair, then two content checks
    if denials >= 3: return allow                            # bounded: 3 denials, 3 compiles
    statement_only = not proof_block_present()               # empty <informal_proof>: sorry is the answer
    text = answer_text()
    if not text: return deny(HEAD + "That file is missing or empty..." + TAIL)
    if not HAS_THEOREM.search(text): return deny(HEAD + "...declares no `theorem`..." + TAIL)
    why = vacuity(text)                                      # syntactic: sorry / axiom / empty proof
    ok, out = compile_text(text)                             # pinned toolchain, 240 s cap, memoized
    if ok and why is None:
        if statement_is_trivial(text):                       # probe: `tauto; simp_all_arith!;
            return deny(PROBE_MESSAGE)                       #  noncomm_ring` closes the statement
        if not audited: return deny(audit_reason(text))      # statement review, once (Listing (*@\ref{lst:evolved-hooks}@*), (c))
        return allow
    msg = HEAD + (f"Problem: {why}." if why else "") + ("Compiler output:\n" + out[:4000] if not ok else "")
    return deny(msg + (LAND if last_call and not ok else TAIL))   # last call asks to land, not repair

# 5c. ANSWER OF RECORD: every distinct version of Work.lean seen during the run is kept.
class Snapshots: [...]
def finalize_answer(snaps):          # at exit, leave the best version the agent itself wrote
    for band in (compiles_without_sorry, compiles_with_sorry):   # newest first within a band
        if any compiles: sync_solution(it); return
    for text in newest_first(snaps):                             # nothing elaborates: keep the
        if compile_text(sorry_text(text)): sync_solution(...)    # statement, drop the proof
    # never a rewrite of the statement, never a synthesis; at most 3 + 2 compiles, before 3700 s

# 6. RUN: as H0, plus snapshots on every event and finalize before grading.
async def run_problem(api_base, problem_text):
    [...] ; state_write(t0=now, max_turns=90)
    async for event in session.act(TASK): snaps.observe(); record(event)
    finalize_answer(snaps); return grade(events)
\end{lstlisting}

\lsthead{Evolved harness: messages the hooks send to the model (abridged)}{}
\begin{lstlisting}[style=prompttext,captionpos=t,caption={\textbf{Evolved harness, hook messages.} The environment's side of the conversation, generated at run time from the state of \texttt{Work.lean}; \texttt{\{...\}} marks values filled in per run and \texttt{[...]} an elision. (a) search gate, (b) budget notices, (c) post-agent verification.},label={lst:evolved-hooks}]
(a) answer_gate, on a search tool call past the free budget with no theorem on disk:
Withheld: {n} searches have been made and {ANSWER} still contains no `theorem`. Write your current
best statement there now -- `sorry` for the proof is fine, a rough statement is fine, and you may
rewrite it completely afterwards. The search tools come back as soon as that file holds a theorem.
Searching with nothing on disk is scored the same as writing nothing at all.

(b) budget_reserve, once per threshold (55% and 80% of the turn budget; 2900 s; 3700 s):
Budget notice (automatic, not an error): you have made {calls} exploration calls against a budget
of {turns} turns, so roughly {turns - calls} turns remain and you get no further warning after the
last one. Nothing is wrong with the call you just made -- repeat it if you still need it. But from
here, finishing beats exploring. Land it now: {ANSWER} is compiled exactly as it stands when you
stop. Make sure it holds the FULL statement of the informal theorem and as much of the proof as you
have. If one step will not close, leave `sorry` in that step only and keep the statement complete
-- an unfinished proof of the right theorem is worth far more than a finished proof of a weaker
one. Check it with m_lean_diagnostic_messages, then stop.
[the wall-clock variants replace the first sentence: "... {spent}s of wall clock spent, and it is the
clock rather than your turn count that will end this run." / "Wall clock exhausted: ... Searching,
reading and shell commands are refused from here on; writing, editing and the Lean diagnostics
still work."]

(c) answer_verify, when the model stops. Head, then the finding, then the request:
Not accepted yet. {ANSWER} is your answer and it is checked with this project's pinned toolchain
before you are finished.

Compiler output:
{lean output, paths renamed to ANSWER, first 4000 chars}

Fix it in that file. Repair the PROOF and the syntax -- do not make the theorem say less in order
to make it compile, and do not delete hypotheses or specialise the statement. A file that
elaborates with a `sorry` in the step you cannot close is worth more than a file that does not
elaborate at all, and much less than a finished proof, so prefer that over weakening what is
stated. Lean often names the right identifier in its hint. Check with m_lean_diagnostic_messages
before you stop again.
[on the last of the three checks, when the file still does not elaborate, the request changes:]
This is the LAST automatic check of this run: whatever is in that file when you stop next is the
answer, and nothing will ask you again. So stop repairing and LAND it. Keep the statement exactly
as it stands -- every binder, every hypothesis, the whole conclusion, the same names -- and replace
each proof step that will not close with `sorry`, as many as it takes, until the file elaborates
with no errors. [...] Do NOT change the statement to make the errors go away -- that is scored as a
wrong statement, which is worse.
[statement-only inputs (empty <informal_proof>) get the same head and compiler output, but:]
Fix it in that file. This problem gives you no informal proof, so `sorry` is the proof to leave
behind -- do not go hunting for one. Repair whatever stops the file from elaborating, keep the
statement complete and faithful [...]

    triviality probe, when the file compiles but `tauto ; simp_all_arith! ; noncomm_ring` closes
    the statement without its proof (asked once per problem):
It compiles, and that part is done. The problem is the STATEMENT: with the same context and its own
hypotheses available, `tauto ; simp_all_arith! ; noncomm_ring` closes it on its own, without using
anything you assumed. A statement the automation proves by itself carries no mathematical content,
so it is scored as if no theorem had been formalized -- the same as an empty file -- however cleanly
it compiled. This is what it usually means: the statement has been narrowed to a special case, a
hypothesis it needs has been dropped, the conclusion has been stated for one object instead of all
of them, or the whole thing has been restated in the exact shape of a library lemma that the simp
set already knows and then proved by citing it. Here is the theorem you were asked to formalize,
again, verbatim: <informal_theorem>{...}</informal_theorem> Go clause by clause [...] State the
theorem that was actually given to you.

    statement review, when the file compiles and passes the probe (asked once per problem):
Automatic statement review. Every answer gets this one; it is not a verdict on yours and it is not
a compile error. {ANSWER} elaborates, declares a theorem, and is not closed by the automation. The
proof is done and is not in question here -- only what it proves is.
One question is left, and it decides more answers here than the compiler does: does that theorem
SAY what the informal theorem says? Both halves are below, so you do not have to remember either
of them.
THE THEOREM YOU WERE ASKED TO FORMALIZE, verbatim: <informal_theorem>{...}</informal_theorem>
THE STATEMENT YOU WROTE, as it stands in that file: ```lean {statement} ```
Walk the English one clause at a time and say, for each clause, which part of that signature
carries it: 1. every object it introduces, AND the structure it gives that object [...]; 2. every
hypothesis it assumes [...]; 3. every quantifier [...]; 4. the conclusion, whole [...]. Then the
reverse direction, which is the half that gets skipped: is there anything in your signature the
English does not ask for -- an added hypothesis, a stronger structure, a fixed value where the text
quantifies? These mismatches all survive the compiler, so they are worth checking with a tool
rather than from memory: [five named mismatch patterns, each with the lean-lsp tool that exposes
it: structure on a carrier, a near-namesake constant, a notion replaced by its parenthetical gloss,
a pointwise vs. functional equality, a bound variable vs. a subtype] [...] If every clause is
carried, stop; the file is your answer. If one is not, fix the STATEMENT [...] and stop.
\end{lstlisting}

\lsthead{Evolved harness: task contract (inserted into the system prompt)}{}
\begin{lstlisting}[style=prompttext,captionpos=t,caption={\textbf{Evolved harness, task contract.} Adds the two input modes, the four-rung grading order, and the compile-and-repair loop.},label={lst:harness-contract}]
## Your task: proof autoformalization

**Two kinds of input.** `problem.txt` may carry a proof or only a theorem. If the
`<informal_proof>` block has content, formalize the theorem AND prove it. If that block is
EMPTY (only whitespace, or absent), this is statement autoformalization: formalize the theorem
faithfully, put `sorry` as the proof, and stop -- that is a complete answer, not a partial one.

**First, read `problem.txt` in your working directory.** It holds a **natural-language
theorem together with its natural-language proof**, wrapped in `<informal_theorem>` and
`<informal_proof>` tags. Do not write any Lean before you have read it. If you cannot read
it, say so and stop -- do NOT invent a theorem to formalize.

Produce a **Lean 4 theorem AND its Lean 4 proof** at this exact path:

    {?}

That path is the answer. A Lean file anywhere else -- including one directory above it, in
`{ not}` -- is not read, not compiled and not graded, however good it is.

Both halves are wanted and both halves are the aim. But they are not worth the same, and if
you cannot get both, what you leave behind is graded in this order, best first:

1. a faithful statement with a finished proof -- the answer;
2. a faithful statement in a file that ELABORATES with no errors, whose unfinished step is a
   `sorry` -- most of the credit, because the statement is the part that is checked against
   the informal theorem;
3. a statement you narrowed or weakened until you could prove it -- worth much less, even
   with a flawless proof;
4. a file that does not elaborate at all -- worth almost nothing, whatever is in it.

So: never trade the statement for the proof. 2 beats 3, and 3 beats 4.

The informal proof is your guide, not decoration -- follow its structure where you can. If a
step of it does not translate directly, prove that step some other way rather than changing
what is being stated.

Lean version: this project pins a specific toolchain (see `lean-toolchain`). Tactics and
lemmas must exist in THAT version. Some search tools index a different Mathlib and can return
names that do not exist here; confirm any unfamiliar name with `lean_local_search` or
`lean_hover_info`, which read this project's own Mathlib.

### Write early, then refine

Create `Work.lean` with the theorem STATEMENT and `sorry` as its proof **within your first
few actions**, before you go looking for lemmas. Then compile it: if the statement does not
type-check, every minute spent on the proof is wasted. Replace `sorry` once the statement is
right.

`sorry` is how you work, not a confession. Keep the file ELABORATING at every point: if a
step will not close, put `sorry` in that step so the file is error-free, and then attack the
step from there. That gives you a checkpoint you can always fall back to. A file full of
half-written tactic blocks has no such floor -- if the budget ends there, it is worth nothing.

Do not explore first and write last. Runs that search for lemmas before committing a
statement tend to run out of budget with nothing on disk.

How you are graded, by recompiling that file after you finish:
- It must exist and elaborate. That file IS your answer; nothing you write in chat is read.
- `sorry` in the FINAL file means the proof is unfinished, and that is graded well below a
  finished proof -- so remove it if you can. Do not remove it by changing what is stated.
- A file that declares nothing, or whose only declaration is `True`, is a failure.
- A statement that compiles but does not faithfully express the informal theorem is a
  failure. Do not weaken the statement to make it provable.
- The statement is also checked for content: the automation tactics are run against it with
  nothing else in scope, and if they close it by themselves it is treated as formalizing
  nothing, whatever the compiler said. A faithful statement of a real theorem essentially
  never falls to automation alone, so if yours does, something in it has gone missing.

When you stop, the file is compiled with the pinned toolchain BEFORE your answer is accepted.
If it does not compile, or still contains `sorry`, you are handed the compiler's own output
and asked to fix it -- so stopping early buys nothing, and you will see the real errors rather
than your guess at them. That loop is finite: after a few attempts the file is taken as it
stands, so the last thing you do with a proof you cannot finish is make the file elaborate
around it, not leave it broken.

There is no user to consult. Do not ask questions; decide and proceed.
\end{lstlisting}

\lsthead{Evolved harness: first user message}{}
\begin{lstlisting}[style=prompttext,captionpos=t,caption={\textbf{Evolved harness, first user message.} Sent once per instance; \texttt{<probe tactics>} stands for the automation used by the triviality check.},label={lst:harness-task}]
Read problem.txt in your working directory. It holds an `<informal_theorem>` block and an `<informal_proof>` block, in LaTeX, often with a Setting preamble that defines the notation the theorem uses.

Formalize BOTH in Lean 4 and write the result to {ANSWER} -- that exact path, which is `{ANSWER_NAME}` in your working directory: the theorem statement must say what `<informal_theorem>` says, and the proof should follow the argument in `<informal_proof>` -- that block is a proof plan given to you, not decoration, so read it before choosing tactics.

That path IS your answer. Nothing else on the filesystem is graded, so a finished file written one directory up (in {WORK}) scores zero. If you are unsure where you are, `pwd` -- but the absolute path above always works.

Write {ANSWER_NAME} FIRST, before any searching: translate the statement and use `sorry` for the proof if you do not have it yet. Then improve it in place -- keep it holding your best attempt at every point, never delete it, and remove the `sorry` once the proof works.

KEEP {ANSWER_NAME} ELABORATING AT ALL TIMES. This is the one rule that decides more runs than any other. Treat an error-free file as a checkpoint: before you attempt a risky proof edit, the version on disk should elaborate. When a step will not close, put `sorry` in THAT STEP so the whole file is error-free again, then go on attacking the step -- with the rest of the proof, and the whole statement, safely on disk. Never leave a half-written tactic block, a dangling `:= by`, or an unfinished `have` in the file while you go and search for something: if your budget ends there, a file that does not elaborate is worth almost nothing, while the same file with one `sorry` in place of the step you had not finished is worth a large part of the task.

HOW TO CHANGE {ANSWER_NAME}: just call `write_file` on it again with the complete new contents. In this environment `write_file` REPLACES an existing `.lean` file in your workspace rather than refusing it, so a whole-file rewrite is one call and needs no `rm` first. `edit` is still there and is cheaper for a small change; if an `edit` fails on a stale match, re-read the file or rewrite it whole. Either way {ANSWER_NAME} must never be left empty: `write_file` with the full file is the supported way to replace it, not deleting it and writing later.

{ANSWER_NAME} must contain an actual `theorem` from that very first write, and keep containing one. If the statement needs auxiliary definitions you have not built yet, write the theorem against the definitions as you intend them and stub those with `sorry` -- do not spend the early turns building definitions with no theorem in the file, because a file holding only `def`s and `#check`s formalizes nothing.

You have about {MODEL['max_turns']} turns. You will get an automatic notice partway through and again near the end, but nothing after that, so get a compiling statement on disk early and spend the rest of the budget replacing the `sorry`. Verify it compiles before you finish.

THINGS THIS ENVIRONMENT DOES ON ITS OWN:
- The Mathlib search tools are withheld once you have searched many times and {ANSWER_NAME} still holds no `theorem`. Writing a first draft statement -- even a rough one, even with `sorry` -- brings them straight back, and you may rewrite that draft completely afterwards. Searching without a file on disk is what produces a score of zero.
- When you stop, {ANSWER_NAME} is compiled with the pinned toolchain. If it does not compile, or still contains `sorry`, you get the compiler's own output back and another chance -- a few times, and then the file is taken as it stands. Use those errors; do not answer them by making the theorem say less, and do not spend the last of them on a proof step that will not close -- `sorry` that one step and hand back a file that elaborates.
- Your statement is also checked for CONTENT once it compiles: the automation tactics (`{' ; '.join(PROBE_TACTICS)}`) are run against it with nothing else in scope. If they close it on their own, the statement is treated as formalizing nothing -- scored the same as an empty file -- and you are told so and asked to state the theorem in full. The way you walk into that is by NARROWING the theorem: one object instead of all of them, a dropped hypothesis, a special case, a weaker conclusion.
  The way NOT to respond to it is to change what you state. Do not unfold, rename, inline or substitute the definitions the informal theorem names in order to get past this check -- naming the objects the informal theorem names is the formalization. If the faithful statement happens to be one automation can close, state it faithfully anyway: an honest statement is always the better answer, and a statement built to defeat a check is scored as a wrong statement. State what the informal theorem states, with its own quantifiers, its own hypotheses and its own vocabulary, and prove THAT.

- Once {ANSWER_NAME} compiles and passes that content check, you get ONE more automatic message, and it is not an error either: your own statement is quoted back to you beside the `<informal_theorem>` you were given, and you are asked to walk the two against each other clause by clause before the answer is taken. Expect it. Most answers that are thrown away are thrown away for a single clause -- a carrier given a structure the text did not grant it, a definition swapped for a similarly-named neighbour, an equality of maps stated at one point instead -- and that message is the last moment at which such a clause is still cheap to fix. If the two do agree, saying so and stopping is the correct answer to it.

ENVIRONMENT -- already set up, do not search the filesystem for any of it:
- `lean` on PATH is the pinned toolchain, and LEAN_PATH already resolves Mathlib and every dependency. `lean {ANSWER_NAME}` in {PROJECT} compiles against Mathlib directly; there is no `lake build` step and you must not run one.
- Mathlib source, for grepping definitions and lemma statements, is at {mathlib_source()}. Nothing relevant lives anywhere else, so never run `find /`.
- Prefer the Lean tools over the shell: they answer questions the compiler cannot. `{MCP_ALIAS}_lean_goal` shows the proof state at a position, `{MCP_ALIAS}_lean_diagnostic_messages` gives the errors for the whole file, `{MCP_ALIAS}_lean_multi_attempt` tries several tactics at once, and the search tools (`{MCP_ALIAS}_lean_local_search`, `{MCP_ALIAS}_lean_leansearch`, `{MCP_ALIAS}_lean_loogle`) find real Mathlib lemma names.

THREE MISTAKES THAT ACCOUNT FOR MOST FAILURES HERE:
1. Invented lemma names. If you are not certain a Mathlib lemma exists with that exact name, search for it or grep the source -- do not guess and hope.
2. Wrong types and coercions. Decide up front whether the statement is over ℕ, ℤ, ℚ or ℝ, and keep subtraction, division and `↑` casts consistent with that choice; a statement that typechecks in the wrong numeric type is not a faithful formalization.
3. Formalizing the GLOSS instead of the NOTION. These informal theorems keep naming a standard notion and then glossing it in passing -- "e is idempotent (i.e., e^2 = e)", "S is countable", "f is injective (distinct inputs have distinct images)", "v lies in the subgroup generated by c". The gloss is there to tell you WHICH notion is meant. It is not the thing to transcribe. If the library has a name for that notion, the faithful statement assumes the NAME; a statement that inlines the spelled-out property instead is a different statement from the one you were asked for, even when the two are mathematically equivalent, and it is scored as one.
   So for each named property in the informal theorem: look for the library's predicate for it (`{MCP_ALIAS}_lean_local_search` by name, `{MCP_ALIAS}_lean_hover_info` to read what it is actually defined as) and use that predicate if -- and only if -- its definition IS the property the text describes. If you cannot confirm that, write the property out explicitly rather than guess at a name: an explicit hypothesis is much better than a similarly-named neighbour. The same discipline applies to HOW a hypothesis is bound -- "let v be an element of S" is a variable together with a membership hypothesis unless the text really talks about the subtype, and a property the library carries as a class or a structure should be assumed the way the library assumes it.

Faithfulness comes first: the Lean theorem must state what the informal theorem states, with the same hypotheses and the same conclusion. Do not weaken the statement, add hypotheses that make it trivial, or replace it with something easier to prove.

IF THE `<informal_proof>` BLOCK IS EMPTY -- nothing but whitespace between the tags, or no such block at all -- this problem is STATEMENT autoformalization only. No proof is being asked of you: formalize the theorem statement faithfully, write `sorry` as its proof, and stop. That is a COMPLETE answer. Do not spend turns hunting a proof and do not remove the `sorry`. Everything above about following the proof plan applies only when that block has content.
\end{lstlisting}

\subsection{\ourTool{} as a Skill for Coding Agents (Inference)}
\label{app:prompts:skill}
When \ourTool{} runs inside Claude Code or Codex (Sec.~\ref{sec:mainresults}, App.~\ref{app:impldetails}), the host agent loads the skill file below (Listing~\ref{lst:skill}); the file tells the host what to delegate, how to judge the returned Lean, and when to decompose, and points it to a playbook (Listing~\ref{lst:playbook}) for harder instances. The Codex copy differs only in how the host waits for jobs and in the sandbox notes.

\lsthead{Skill file loaded by the host coding agent (\texttt{SKILL.md})}{}
\begin{lstlisting}[style=prompttext,captionpos=t,caption={\textbf{The \ourTool{} skill.} The host's instructions: division of labour, the judge protocol for (c) and (d), and the decomposition procedure.},label={lst:skill}]
---
name: aiprover-autoformalize
description: Proof auto-formalization into Lean 4 (v4.23.0 + Mathlib) with the AIProver agent -- our fine-tuned Leanstral prover driven by an evolved agentic harness. Use whenever you are given a natural-language theorem with its proof (e.g. <informal_theorem>/<informal_proof> blocks) and must produce a Lean 4 formalization that compiles, has no sorry, states exactly the theorem, and follows the proof. Delegate Lean writing and proof search to AIProver; you plan, judge, decompose and weave.
---

# AIProver: proof auto-formalization

You turn a natural-language pair  M = ⟨T, P⟩  (a theorem T, possibly in several parts, with the
definitions and assumptions it relies on, and its proof P, possibly through intermediate lemmas)
into ONE self-contained Lean 4 file ⟨T^, P^⟩ that is **equivalent** to it. The NL pair is your
only input. Four properties, ALL required:

| | property | who certifies |
|---|---|---|
| (a) | **Type-correct**: Lean's kernel accepts the file (Lean v4.23.0, Mathlib v4.23.0, `import Mathlib`) | `aiprover check` |
| (b) | **Complete**: no `sorry`, `admit`, empty `by`, or any other placeholder -- anywhere, including helper lemmas and `def`s | `aiprover check` (kernel axiom probe) |
| (c) | **Semantically correct**: T^ states exactly T under the same definitions -- no added assumptions, no dropped conditions, every part | **you**, with the judge protocol below |
| (d) | **Proof-faithful**: P^ follows P's strategy, including its intermediate lemmas, with no new assumptions | **you**, with the judge protocol below |

**You do not stop until all four hold.** (a) and (b) are mechanical; (c) and (d) are undecidable
and you are the judge. A file that compiles is not an answer.

## Who does what

**AIProver** = our fine-tuned Leanstral model inside the evolved hevo harness (champion
`d01_r04`). One call is a full agentic Lean session of up to 100 turns: it writes the file,
compiles it, searches Mathlib, reads goals and repairs, with the lean-lsp tools. It is strong at
writing Lean and closing proofs. On 509 hard training problems it produced a solved, faithful
formalization in 205 cases -- and in **173 more its file compiled but stated the WRONG theorem**.
It cannot reliably judge its own statement. Five evolution rounds tried to teach it to; all failed.

**Division of labour -- this is how you save frontier tokens without losing accuracy:**
- AIProver writes Lean and searches for proofs. Send it the whole problem first, then pieces.
  Sampling is cheap for it (our GPUs) and expensive for you, so ask for several samples rather
  than doing the search yourself.
- You read the NL carefully, plan, **judge (c) and (d)**, fix statements, decompose, weave the
  pieces, and make small repairs. Do not grind through tactic search yourself while AIProver can
  do it. Do not read AIProver's trajectories. `aiprover result` gives you everything you need.
- While jobs run, WAIT (`aiprover wait`). Waiting costs nothing, and polling or idle exploration
  costs tokens.

## The commands

`AIP=<this skill's directory>/scripts/aiprover` -- the skill's base directory is given to you when
this skill loads. Every command prints a short summary; add `--json` where noted for the full record.

```
$AIP doctor                        # once per session, first. All PASS or read STARTUP.md
$AIP submit --problem FILE -k 4 --name main      # prints JOB id, returns at once
$AIP submit --theorem-text "..." --proof-text "..." [--context C.lean] [--lean-statement S.lean] [--hint "..."] [--statement-only] -k 2
$AIP wait JOB [JOB...] --timeout 540 [--any]     # blocks <= 9 min; rc 0 done, rc 3 still running
$AIP result JOB [--all] [--json]   # per-sample status + best Lean file + its check
$AIP check FILE.lean [--statement-only] [--fixed C.lean]   # (a)+(b); rc 0 = PASS
$AIP workspace --new NAME          # path for YOUR Lean file inside the lean-lsp project
$AIP status | list | cancel JOB
```

- A job runs `-k` independent AIProver samples in parallel, typically 5-40 min each (up to 90).
  **Run `wait` as a normal foreground command with the Bash timeout set to 600000 ms**, and repeat
  it while it returns rc 3. Nothing is lost between waits; jobs run detached.
- Sample status, best first: `verified` (passes (a)+(b); still needs YOUR (c)/(d) judgement),
  `sorry` (elaborates but has a placeholder), `rejected` (compiles, but a disqualifying
  construct or axiom), `error` (does not compile), `empty`, `infra`.
- `--context C.lean`: declarations the answer must contain VERBATIM (your fixed definitions,
  lemma statements with `sorry` as given facts). `--lean-statement S.lean`: the fixed statement
  to prove (`... := by sorry`). `result` reports `fixed-code:kept` or `fixed-code:CHANGED`. Treat
  CHANGED as a failed sample unless the change is harmless and you adopt it deliberately.
- `--statement-only` sends an empty proof block: AIProver formalizes the statement alone
  (a `sorry` proof is the complete answer there). This is useful for drafting T^ quickly.
- `--hint` passes one line of guidance, e.g. the compiler error to fix or the Mathlib lemma to use.

**lean-lsp MCP tools** (`lean_diagnostic_messages`, `lean_goal`, `lean_multi_attempt`,
`lean_run_code`, `lean_local_search`, `lean_loogle`, `lean_leansearch`, `lean_hover_info`,
`lean_verify`, ...) are available to you for quick local work: checking a goal, trying a tactic,
confirming a lemma name, reading a definition. Files must live in the project that
`$AIP workspace --new NAME` points into. Use them for cheap targeted checks, not for long proof
searches (delegate those).

## The procedure

**0. Preflight.** `$AIP doctor` (about 40 s). If anything FAILs, stop and follow STARTUP.md
(doctor prints its path). Do not work around a broken environment.

**1. Read and blueprint.** Write the NL input to a file (`problem.txt`, with the
`<informal_theorem>`/`<informal_proof>` blocks exactly as given). Then list, briefly:
definitions D1..; theorem parts T1..Tk with their exact hypotheses; the proof's intermediate
lemmas L1..Lm and how the main argument uses them. Resolve ambiguity now: which number type
(ℕ/ℤ/ℚ/ℝ), what "positive"/"nonzero" range, 0- or 1-indexing, which Mathlib notion each named
concept is.

**2. Delegate the whole problem at once.** `$AIP submit --problem problem.txt -k 4 --name whole`,
then `wait`. For a long problem (many parts or lemmas) ALSO submit
`--statement-only -k 2 --name stmt` at the same time, so you get statement drafts to judge early.

**3. Judge every `verified` sample**, best first, with the judge protocol. Take the first that
passes (c) and (d). If none pass, keep what IS right: a faithful statement, correct definitions,
lemma statements, and proofs of some parts.

**4. If no sample is complete and faithful: fix the statement, then decompose.**
  a. Write the skeleton yourself (usually by correcting the best sample): definitions, every
     lemma Li of the NL proof and every theorem part Tj as a Lean statement with `:= by sorry`.
     `$AIP check --statement-only skeleton.lean` must PASS, and the skeleton must pass the judge
     for (c) and for the lemma structure of (d). **Freeze it** -- from now on the statements do
     not change unless the judge finds a fault.
  b. Submit one job per open piece, all in parallel: `--context` = the frozen definitions plus
     the statements of the lemmas that piece may use (as `sorry`d givens); `--lean-statement` =
     the piece's own statement; `--theorem-text`/`--proof-text` = the NL for that piece only (its
     statement and its part of P). `-k 2`..`-k 4`. Give the main theorem's job the lemmas as
     givens, so its proof follows P's structure instead of re-deriving everything.
  c. Weave: paste each returned proof into the skeleton (take only the proof bodies; the
     statements are frozen). `$AIP check`. Fix small breakage yourself (a name clash, an `open`,
     a missing `import`) and check again.
  d. A piece that fails: read its best sample's diagnostics, then resubmit with a `--hint` naming
     the error or a useful lemma. If it fails again, split it further along P's own reasoning
     (sub-lemmas), recurse, and weave back. Only prove a small step yourself when it is a few
     lines and AIProver has already failed on it.

**5. Final gate** -- on the exact final file, in this order:
  1. `$AIP check final.lean` -> PASS (covers (a) and (b); warnings about a bare `trivial` proof
     must also be fixed, because the evaluation judge counts `trivial` as a placeholder).
  2. The full judge protocol, (c) and (d), on the final file.
  3. Only then answer.

Never deliver a file that fails any of the four. If you are stuck, go back to step 4 with a finer
split. Do not weaken the statement to make a proof go through; that trades an (a)/(b) failure
for a (c) failure, which is just as fatal and harder to see.

## Judge protocol for (c) and (d)

Do it in this order, because order matters: judging after re-reading the NL anchors you to what
the text meant rather than what the Lean says.

**(c) Semantic correctness.**
1. **Blind back-translation.** From the Lean alone -- every `def`, `structure`, instance argument,
   binder, hypothesis and conclusion -- write in plain English what T^ asserts. Unfold local
   definitions. Do this BEFORE comparing with T. (With several candidates, or for the final gate,
   delegate this step to a fresh subagent given ONLY the Lean file, so it cannot be anchored by T.)
2. **Compare clause by clause** with T (read P too, since it can fix the meaning of notation):
   - every part of a multi-part theorem present, each with its own hypotheses;
   - same quantifiers, in the same order and scope; ∃ vs ∃!; for-all vs for-some;
   - no extra hypothesis (including a stronger typeclass: `Field` for a ring, `Fintype`,
     `DecidableEq` that changes meaning, `Nonempty`, positivity) and no dropped one;
   - the conclusion neither weaker nor stronger (≤ vs <, ↔ vs →, pointwise vs as functions);
   - number types: ℕ subtraction truncates and ℕ/ℤ division floors -- is that what T means?
     Casts sit where T's arithmetic happens;
   - named notions: use the library's notion for what T NAMES ("idempotent", "countable",
     "subgroup generated by") and check its definition really is T's meaning. Formalizing the
     parenthetical gloss instead of the named notion, or a similarly-named neighbour, is the most
     common way AIProver's files compile yet are wrong;
   - definitions: T's own definitions formalized as defined (not replaced by an unfolded copy, not
     stubbed, conventions such as `1/0 = 0` preserved); indexing (0- vs 1-based, `Fin n` vs a
     list of length n) and edge cases (empty, zero, n = 1) agree;
   - not vacuous: hypotheses are satisfiable, and the statement is not closed by
     `simp`/`tauto`/`decide` alone because it was narrowed or specialised.
3. Verdict: equivalent, or list each mismatch and fix it (in the skeleton, then re-run 4b for the
   affected pieces).

**(d) Proof faithfulness.**
- Each intermediate lemma of P appears as a Lean lemma (or a clearly labelled `have`) whose
  statement matches it, and the main proof USES them where P does.
- P's method is followed: induction where P inducts, contradiction where P argues by
  contradiction, the same case split and the same key constructions. Tactics like `simp`,
  `omega`, `linarith` or `positivity` finishing a routine step P also treats as routine are
  fine. Replacing P's whole argument with a different one (brute-force `decide` over a case P
  handles conceptually, or a Mathlib lemma that IS the theorem when P proves it) is not.
- No new assumptions: no `axiom`, no extra hypotheses on helper lemmas that P does not have, no
  `sorry` anywhere (already checked by (b)).

## The answer

- ONE self-contained Lean 4 file: `import Mathlib` (or Mathlib submodules), your definitions,
  lemmas and theorems, with no reference to files of your own. It must PASS `$AIP check`.
- If the request specifies an output format, follow it EXACTLY. The step1 evaluation, for
  example, wants nothing but
  `<formal_proof>` + one ```` ```lean4 ```` fence holding the whole file + `</formal_proof>`.
  Otherwise print the file in a ```` ```lean4 ```` block and state that all four checks pass.

More detail -- decomposition worked through on an example, and AIProver's measured failure
modes -- is in `references/playbook.md`. Read it the first time you decompose.
\end{lstlisting}

\lsthead{Playbook the skill refers the host to}{}
\begin{lstlisting}[style=prompttext,captionpos=t,caption={\textbf{Playbook.} Read by the host the first time it decomposes a problem.},label={lst:playbook}]
# AIProver playbook

Read this the first time you decompose a problem. SKILL.md has the procedure; this has the
detail behind it.

## 1. What AIProver is, in numbers

The harness is `hevo_mixed_v1` champion `d01_r04`, selected over 10 rounds of harness evolution
on 509 hard LoCoLib training problems (algebraic structures, foundations/logic, number theory).
One sample of it, on that set:

| outcome | count | what it means for you |
|---|---|---|
| solved (faithful + complete) | 205 | still judge it, but usually right |
| compiles, complete, **unfaithful** | 173 | the trap: (a)+(b) pass, (c) fails |
| faithful statement, proof has `sorry` | 9 | keep the statement, delegate the proof piecewise |
| elaborates with `sorry`, statement unverified | 61 | salvage the parts that are right |
| does not elaborate | 61 | usually still has a useful statement draft |

So roughly 4 of 10 samples are right as they stand, and many more carry a correct statement or a
correct part. Several samples plus your judgement plus decomposition beats any single sample.
Median session is ~13 min (p90 ~40 min). The harness spends its last 18% of turns making the
file elaborate, so even a failed sample usually hands back a compiling skeleton with `sorry`s.

## 2. AIProver's measured failure modes -- check these first when judging

From the evolution run's error analysis, ordered by how often they cost a correct answer:

1. **Gloss instead of notion.** T says "e is idempotent (i.e. e^2 = e)" and the Lean inlines
   `e * e = e` instead of `IsIdempotentElem e`, or the reverse with the wrong library notion.
   Formalize the notion the text NAMES, and check with `lean_hover_info` that the library
   definition really is it.
2. **Under-specified / narrowed.** It proves one instance, a special case, or drops a hypothesis
   or a whole part of a multi-part theorem. It is especially tempted to do this when the
   automation check calls a statement trivial.
3. **Over-generalized.** A carrier given more structure than T grants (`Field` for a ring,
   `LinearOrder` added, a universe-polymorphic `Type*` where T fixes a concrete set), or a
   hypothesis silently strengthened.
4. **Wrong numeric type / coercion.** ℕ where T means ℤ or ℝ; truncated subtraction; floor
   division; casts in the wrong place.
5. **Definitions replaced.** A named definition from T replaced by an unfolded copy, or by a
   `def` that is stubbed or does not match (Mizar-style conventions such as `1/0 = 0` must be
   preserved).
6. **Representation drift.** `Fin n → α` vs `List α` of length n, 0- vs 1-based indexing,
   sets vs types, `Finset.range (n+1)` vs `Finset.Icc 1 n`.
7. **Disqualified constructs** (`native_decide`, `axiom`, `maxHeartbeats 0`): `check` catches
   these and marks the sample `rejected`.

## 3. Decomposition, worked through

Input (abridged): *"Let R be an abelian group. Define NatMul(n,a) = a+…+a (n times) and
IntMul(i,a) = NatMul(i,a) if i ≥ 0, NatMul(-i,-a) if i < 0. Then (1) if i ≤ j and k = j-i then
NatMul(k,a) = NatMul(j,a) - NatMul(i,a); (2) -NatMul(i,a) = NatMul(i,-a); (3) IntMul(i+j,a) =
IntMul(i,a) + IntMul(j,a). Proof: (1) by distributivity … (2) by induction on n … (3) by cases
on the signs of i and j, using (1) and (2)."*

**Step 2** (`submit --problem problem.txt -k 4`): suppose the best `verified` sample states
`NatMul`/`IntMul` as `def`s correctly and proves (1) and (2), but its (3) replaces `IntMul` by
`zsmul` (definition replaced -> (c) fails), and sample 2 has the right (3) statement with `sorry`.

**Step 4a**, `skeleton.lean` (you write it from the two samples):
```lean
import Mathlib

variable {R : Type*} [AddCommGroup R]

def NatMul : ℕ → R → R                             -- as T defines it: a + ... + a, n times
  | 0, _ => 0
  | n + 1, a => NatMul n a + a
def IntMul (i : ℤ) (a : R) : R := if 0 ≤ i then NatMul i.toNat a else NatMul (-i).toNat (-a)

theorem part1 (a : R) (i j k : ℕ) (hij : i ≤ j) (hk : k = j - i) :
    NatMul k a = NatMul j a - NatMul i a := by sorry
theorem part2 (a : R) (i : ℕ) : -NatMul i a = NatMul i (-a) := by sorry
theorem part3 (a : R) (i j : ℤ) : IntMul (i + j) a = IntMul i a + IntMul j a := by sorry
```
`check --statement-only skeleton.lean` passes and the skeleton passes your judge. Freeze it.

**Step 4b**, three jobs in parallel (parts 1 and 2 already have verified proofs in sample 1, so
only part 3 is actually needed here -- shown in full for the pattern):
```
# context for part3 = the defs + part1/part2 statements as givens
$AIP submit --name part3 -k 4 \
   --context ctx_part3.lean --lean-statement stmt_part3.lean \
   --theorem-text "Statement (3): for all integers i, j, IntMul(i+j,a) = IntMul(i,a) + IntMul(j,a)." \
   --proof-text  "By cases on the signs of i, j and i+j. When all are ≥ 0 this is additivity of NatMul ... use part (1) when the signs differ and part (2) to move negation inside."
```
`ctx_part3.lean` holds the `variable`, both `def`s and the `part1`/`part2` statements ending in
`:= by sorry`. `stmt_part3.lean` holds `theorem part3 ... := by sorry`. The returned file
contains all of these verbatim plus a real proof of `part3`. It may call `part1`/`part2`, which
is exactly P's structure.

**Step 4c**: paste the proof bodies into the skeleton, `check final.lean`, judge, answer.

Rules of thumb:
- One job per NL lemma is the natural grain. Split further only when a piece fails twice.
- Always hand a piece the lemmas P uses for it, as givens in `--context`, so AIProver follows
  P instead of inventing another route (that protects (d)) and has less to do.
- Keep NL for a piece self-contained: restate the notation it needs in `--theorem-text`. The
  context Lean pins the meaning, so the NL can be brief.
- A piece's proof may need a helper lemma P does not state. Accept it only if it is a routine
  step of P's argument and adds no hypothesis.
- `--statement-only -k 2` is a cheap second opinion when you are unsure how to state something.
  Judge its outputs like any other.

## 4. Reading `result`

```
job 2026...-whole  (theorem+proof)  counts: {'verified': 2, 'sorry': 1, 'error': 1}
  s2: verified  611.2s turns=34
  s0: verified  1022.4s turns=51
  s3: sorry     1840.0s turns=88 problems: placeholder (sorry/admit) reaches: main_thm
  s1: error     2011.7s turns=100 TIMEOUT problems: does not compile (see diagnostics)
===== s2 [verified]  ~/.aiprover/jobs/<job>/s2.lean
<the Lean file>
```
- Only the best sample's Lean is printed. `--all` prints every sample's file. Use it when the
  best one fails your judge, because a lower-ranked sample may have the faithful statement.
- Files stay on disk (`s<i>.lean`), so pass paths around rather than re-printing files.
- `infra` means the model endpoint failed (already retried once). Run `$AIP tunnel up` and
  `$AIP doctor`, then resubmit.

## 5. Small repairs you should do yourself

It is cheaper to fix these directly than to resubmit: a missing `open`/namespace, a renamed
Mathlib lemma (`lean_local_search` / `lean_loogle` to find the 4.23 name), a universe or
implicit-argument annotation, a duplicated helper name between woven pieces, and weaving glue
(`exact partK ...`). Anything that needs real proof search, delegate.

## 6. Lean / Mathlib version

Lean `v4.23.0`, Mathlib `v4.23.0`. Hosted search tools (`lean_leansearch`, `lean_leanfinder`,
`lean_loogle`) may return names from a newer Mathlib. Confirm any name with `lean_local_search`
or `lean_hover_info` before relying on it. `grind` exists on 4.23. `native_decide` is forbidden.
\end{lstlisting}

\subsection{HarnessEvolve (Evolutionary Search)}
\label{app:prompts:hevo}
HarnessEvolve's mutator is a frontier coding agent run headless once per round (Sec.~\ref{sec:coevo}, App.~\ref{app:impldetails}). It receives a static guide as its system prompt (Listing~\ref{lst:hevo-guide}, abridged after its first four sections) and a per-round prompt (Listing~\ref{lst:hevo-round}) whose double-brace placeholders are filled with the parent's source path, its certificates and rung table, the lineage memory, and the acceptance bar.

\lsthead{Mutator guide (system prompt of the mutator agent)}{}
\begin{lstlisting}[style=prompttext,captionpos=t,caption={\textbf{HarnessEvolve mutator guide}, abridged: Sections 1--3b in full, later sections by heading and opening lines. \texttt{\{\{LADDER\_TABLE\}\}} is filled from the reward configuration.},label={lst:hevo-guide}]
# Harness mutation — Lean 4 proof autoformalization

## 1. Your job this round

Produce **one** child harness and **one** `mutation.json`, at the two absolute paths the round
prompt gives you. You do not run rollouts; the loop spends the GPU hours.

The round is consumed either way. Do **not** write "the current harness is optimal", do not
propose skipping, do not stop to ask a question, do not hand back an explanation instead of a
file. A child that fails validation wastes a round; a child that crashes on rollout 1 wastes
~2 GPU-hours. **Correctness of the file beats ambition of the idea.**

There is no human in this session and there is no user to consult.

## 2. The goal

Given an informal theorem **and its informal proof** in LaTeX, the harness must make a
**fixed** LLM produce a Lean 4 file containing a theorem that *says what the informal theorem
says* **and** a proof of it that compiles under the pinned toolchain with no `sorry`.

The model, its temperature (1.0) and its reasoning effort (high) are the experiment's fixed
point. You are searching over the harness: prompt, control flow, tool wiring, scaffolding,
per-problem project setup, any new tool you implement in Python inside the harness file.

## 3. How you are scored, exactly

Each of 300 problems lands on one rung. Fitness is the mean over all 300; a missing result
scores 0.00.

{{LADDER_TABLE}}

Equivalence is BEq+ with the grader's triviality guard (statement only; proofs ignored).

### 3a. The asymmetry you are actually designing against

**The reward compares the agent's Lean against a GOLD LEAN FILE. The agent never sees it.**

    the agent sees      <informal_theorem> + <informal_proof>          English, LaTeX
    the reward compares the Lean it wrote   vs   the gold Lean         FL vs FL, BEq+

Nothing in the score reads the English. BEq+ asks only whether the generated statement and the
gold statement each follow from the other. So an instance fails `solved` for one of two very
different reasons, and the certificate tells you which:

* the agent misread the English and formalized the wrong claim, **or**
* it read the English correctly but expressed it in Lean in a way that does not line up with
  how the gold expresses it.

**This is what you are optimising: make a model that sees only English produce the Lean a
mathematician would have written.** Predicting the gold IS the task. Recognising that the
English describes a result already in Mathlib and citing it is a *correct* and often ideal
answer — one observed `solved` rollout proved a hard `RingTheory` statement in 84 characters by
naming the right library lemma against a 2335-character gold.

What is not available is *retrieving* the gold. The runner writes the English into the agent's
workspace and the gold to a separate file the grader alone reads, so at rollout time there is
nothing to read. Your own session is given the memory root and nothing else; the gold FILE sits
outside it.

**You DO see the reference STATEMENT, on purpose.** When an instance misses `solved`, its
certificate's SEMANTIC section shows the reference statement (never its proof) beside the
candidate's, under `STATEMENTS COMPARED`. Use it to name the CLASS of mistranslation the harness
keeps producing — a Mathlib notion re-invented as a new `def`, a hypothesis dropped, `ℕ` where
the reference quantifies over a ring, a bespoke structure where Mathlib has one — and then change
how the harness reasons from English, so the next problem of that class comes out right. Read
twenty of them and look for the pattern; one instance is an anecdote. What you learn from them
must generalise to a problem you have never seen: the agent will not see these statements at
rollout time, and a harness that only works because it quotes them is worthless. The
distinction that matters:

| | |
|---|---|
| **the task** | make the harness better at INFERRING the formal statement from the English |
| **not the task** | building the answer in — a table of Mathlib lemma names keyed by problem, or reading `gold.jsonl` |

The second is checked for. A long run of reference text in your diff is scanned, your transcript
is flagged if it mentions the gold path, and a lookup table keyed by problem contains no long
run of anything yet would carry a harness most of the way to fitness 1.0 — which is why the
structural boundary exists rather than only the scan.

**So: never write a harness change whose mechanism is "know the answer". Write one whose
mechanism is "read the English more carefully, and say it in Lean the way Mathlib says it".**
Concretely, the levers that move this are things like: making the model restate the English
claim before formalizing it; making it enumerate the quantifiers, hypotheses and edge cases the
English actually carries; making it search Mathlib for the standard spelling of the objects
involved; making it re-read its own statement back into English and compare. All of those work
from the English alone, which is the only thing the agent will have.

**Where the exposure is, so you can avoid walking into it.** Every rung below `solved` is
reachable *without* the statement being verified against the gold, so the `compiles` value is
the ceiling of any policy that produces type-correct Lean without formalizing anything. Two such
policies are cheap, and both are closed by admission tests on that rung:

- `instance : Inhabited Nat := ⟨0⟩` compiles clean — the vacuity gate accepts `instance` — but
  the equivalence grader can extract no statement from it. **A theorem must be extractable.**
- `theorem t (n : Nat) : n = n := rfl` passes the vacuity gate too. **The statement must not be
  closable by `tauto` / `simp_all_arith!` / `noncomm_ring` with nothing else in scope**, checked
  against your file's own context.

Closing those is what lets the `compiles` rung be worth enough to give the search a real
gradient. What is *not* closed: getting the model to state and prove some **other** real,
non-trivial theorem still earns `compiles` unfaithfully. That is not free — it is most of the
Lean competence the real task needs — and it is loud. A node whose `compiles` count climbs while
`solved` does not, or whose statements shorten, use fewer Mathlib constants, or lean more on
one-tactic proofs, is flagged `EROSION SUSPECT` on the node, in the evolution log and in the
next round's prompt. Not a rejection; a permanent record on the branch you built.

Statement erosion has been observed in audit on this corpus — files that redefine the objects
until the claim collapses to an existing Mathlib lemma. A measured example inside the
equivalence metric itself: gold `∀ n, n + 0 = n` versus generated `0 + 0 = 0` scored
"equivalent" because both fall to `simp`. The grader's triviality guard exists for that, and
this loop treats a guard that could not run as *not* equivalent.

## 3b. TWO INPUT MODES — the harness must handle both

`problem.txt` holds ONE input containing two tagged blocks, `<informal_theorem>` and
`<informal_proof>`. The proof block **may be empty**. Strip whitespace before deciding: a block
containing only newlines or spaces is empty.

**Mode 1 — theorem AND proof (both blocks non-empty).** Full autoformalization: the statement
must say what `<informal_theorem>` says, and the proof should follow the argument given. A
remaining `sorry` is an unfinished answer.

**Mode 2 — theorem ONLY (the proof block is empty after whitespace removal).** The task is
**statement autoformalization**. There is no proof to follow and none is being asked for. A file
whose statement is faithful and whose proof is `sorry` is a COMPLETE answer in this mode, not a
failure, and the harness must not spend the run trying to discharge it, must not deny delivery
for containing `sorry`, and must not tell the agent that `sorry` scores zero.

This is a REQUIREMENT, not a suggestion: users will send theorem-only inputs to the deployed
harness. A harness that assumes both blocks are present will, on those inputs, burn its whole
budget hunting a proof nobody asked for and then deliver nothing. Whichever mechanisms you add,
keep both modes working, detect the mode from the input rather than from a flag, and say in
`self_critique` how your change behaves in each. Note that any pre-delivery gate, verify hook,
probe or "refuse to deliver" mechanism inherited from an ancestor has to be mode-aware too --
several of them treat `sorry` as a defect unconditionally.

## 4. The noise floor — read this before choosing a mechanism
The standard error on a 300-problem, one-rollout, temperature-1.0 evaluation is around
**0.010–0.013**. The round prompt tells you the measured value and the accept bar in force.
[...]

## 5. Frozen versus mutable
**Frozen — a change here is rejected before any GPU time is spent.**
- `MODEL`, compared against the seed both as text and as RESOLVED values via
[...]

## 6. Environment invariants — each of these has already cost a run
- **`AgentConfig.instructions` is never consumed by vibe 2.24.2.** `$VIBE_HOME/AGENTS.md` is
  the only channel that reaches the model. Prompt text added anywhere else is silently not
[...]

## 7. Known failure modes — priors, superseded by this round's numbers
Measured once on a 220-run batch of a DIFFERENT corpus, at a 60-turn cap, on a host where
most of the Lean tools were broken. Several of these rates have since been observed to be off by
[...]

## 8. The axis map, and its saturation state
Pick exactly one axis and put it in `mutation.json` as `axis`.
- **A. Prompt content and ordering** — heavily worked, nearly exhausted. The seed already does
[...]

## 9. How big a change to make — your call, informed by the trajectory
**There is no limit on how much you change.** A child may be a two-line edit to one symbol, or
a rewrite of the harness's whole control flow, or several unrelated fixes at once. The goal is
[...]

## 10. Anti-overfitting
The line is not "is it about Lean" — general Lean and Mathlib craft is legitimate harness
content. The line is: **did I choose this because of what is in this round's problems?**
[...]

## 11. Workflow — the minimums are not optional
1. Read the selection scoreboard and the evolution log in the round prompt. Note which
   mechanisms and axes have already been tried **on this lineage**.
[...]

## 12. `mutation.json`
```json
{
[...]

## 13. What not to propose
An **unjustified** turn-cap increase — raise it if the evidence supports it, but record the
`turn_budget` block and expect the cost to be reported. A change to the model id, temperature or
[...]

## 14. The harness, structurally
One self-contained Python file. The seed lays it out as:
| section | what lives there |
[...]

## 14b. Mechanisms that already exist in the seed and are switched OFF
The cheapest region of the search space is code that is written, plumbed and unreachable. Verify
each against the file before relying on this list, but it is where to look first:
[...]

## 15. Lean craft that survives the corpus-swap test
General, so allowed under §10. Specific to this round's problems, so forbidden.
- `leanprover/lean4:v4.23.0` against a prebuilt read-only Mathlib. `grind` exists on ≥4.22.
[...]

## 16. Two things deliberately left alone, not closed
Either is a legitimate hypothesis. `bash` use was left alone: compiling runs use it 40.3 times
each, identical to failures, so heavy shell use is not by itself a symptom. The search tools
[...]
\end{lstlisting}

\lsthead{Round prompt (first user message of the mutator agent, each round)}{}
\begin{lstlisting}[style=prompttext,captionpos=t,caption={\textbf{HarnessEvolve round prompt.} Every \texttt{\{\{...\}\}} field is rendered per round from the search state.},label={lst:hevo-round}]
Round {{ROUND}} of {{TOTAL_ROUNDS}} of harness evolution for Lean 4 proof autoformalization.
Follow the guide you were given as a system prompt.

WRITE EXACTLY TWO FILES, at these absolute paths:

  {{CHILD_HARNESS}}
      the child harness: a copy of your parent's harness.py with your change applied.

  {{CHILD_MUTATION}}
      your account of the change, as JSON (schema in the guide §12). Not the diff -- the loop
      derives that from the two harness files. This is the reasoning, which it cannot derive,
      and which later rounds read to know what has already been tried.

End your reply with exactly: MUTATION: <slug>

{{RETRY_BLOCK}}
================================================================================
1. YOUR PARENT
================================================================================

Copy this file and edit it:

    {{PARENT_HARNESS}}

  node            {{PARENT_NAME}}   depth {{PARENT_DEPTH}}, produced in round {{PARENT_ROUND}}
  fitness         {{PARENT_FITNESS}}
  its own diff    {{PARENT_DIFF}}
  its report      {{PARENT_REPORT}}
  its full record {{PARENT_EVAL}}   ({{EVAL_SIZE}}, one JSON object per line)

Selected by: {{SELECTION_REASON}}

{{SCOREBOARD}}

================================================================================
2. THE SEARCH SO FAR
================================================================================

  seed fitness                          {{SEED_FITNESS}}
  best anywhere in memory               {{BEST_FITNESS}}   ({{BEST_NODE}})
  your child is ACCEPTed if it exceeds  {{ACCEPT_BAR}}
  one-sigma noise on this statistic     {{SIGMA}}   (accept margin {{ACCEPT_MARGIN}})
  rounds since the last ACCEPT          {{STAGNATION}}{{WIDENED_NOTE}}

{{MEMORY_TREE}}

Machine-readable index of every node and every path: {{INDEX_PATH}}

--------------------------------------------------------------------------------
Every round so far. `>>` marks your own lineage -- those mechanisms are already IN your
parent, or were tried from it and rejected. `predicted vs actual` is the calibration
record for this loop.

{{FULL_LOG}}

Diffs, most recent first:

{{RECENT_DIFFS}}

Attempts per axis (the axis map is in guide §8):

{{AXIS_TABLE}}

================================================================================
3. WHAT THIS PARENT ACTUALLY DID, ON THESE 300 PROBLEMS
================================================================================

Rung histogram, in counts:

{{RUNG_TABLE}}

{{PROFILE}}

Effort and cost, which is also what bounds the turn budget:

{{TURN_BUDGET_BLOCK}}

Full report: {{PARENT_REPORT}}

================================================================================
4. GOING DEEPER THAN THE AGGREGATES
================================================================================

The full record is {{PARENT_EVAL}} -- every problem, the Lean it produced, the compiler's
reply, and the whole turn-by-turn conversation with tool inputs and outputs. Schema in
DESIGN.md §5. It is {{EVAL_SIZE}}, so slice it rather than reading it in one call:

    grep -c '"rung": "incomplete_faithful"' {{PARENT_EVAL}}
    python3 {{EVOLVE}} show {{PARENT_NAME}} <uuid>        # renders any instance in full

READ THE CERTIFICATES. Every instance carries `score.reward_certificate`: the four checks in
their own words, delimited, saying WHY that instance landed on that rung.

    --- TYPECHECK    ---  Lean's own compiler message, verbatim
    --- COMPLETENESS ---  which `sorry`, or which axiom `#print axioms` found
    --- SEMANTIC     ---  which BEq+ direction failed, what that asymmetry means, and on a miss
                          the REFERENCE statement beside the candidate's (STATEMENTS COMPARED)
    --- BREVITY      ---  the proof's length against the gold's (a PROXY, see the guide)

The rung alone cannot tell you what to change; the certificate can. To pull them out:

    python3 - <<'EOF'
    import json
    for l in open("{{PARENT_EVAL}}"):
        r = json.loads(l)
        if r["score"]["rung"] == "compiles":          # or any rung you are hunting
            print(r["uuid"]); print(r["score"]["reward_certificate"])
    EOF

The SEMANTIC section is the one to read first when `compiles` is high and `solved` is not.
`gold => candidate proved, candidate => gold NOT proved` means the agent stated something
WEAKER than the English asked for -- usually a dropped hypothesis, a fixed value where the
English quantified, or a special case. The reverse means it stated something STRONGER --
usually an added hypothesis. Those two failures need opposite harness changes, and the scalar
cannot distinguish them. When NEITHER direction is proved, compare the two statements shown under
STATEMENTS COMPARED and name the kind of mismatch (re-invented definitions, wrong type or domain,
missing hypothesis, different quantifier structure). Fix the KIND in the harness; never copy a
reference statement, name or lemma into it -- the diff is scanned for reference text.

Some instances are already rendered as markdown, picked mechanically across the three
domains (successes, near-misses, the modal failure, no-answers) -- a starting point, not a
selection that knows what your hypothesis needs:

{{INTERESTING}}

================================================================================
5. DID YOUR PARENT'S OWN CLAIM LAND?
================================================================================

{{PARENT_CLAIM}}

================================================================================
6. WHERE THE FAILURE MASS HAS MOVED, SEED -> THIS PARENT
================================================================================

Both columns are measured on the same 300 problems under the same conditions. The guide's
§7 table lists failure MODES; every magnitude comes from here or from section 3.

{{PRIOR_DELTA}}

================================================================================
7. A STANDING REQUIREMENT: TWO INPUT MODES
================================================================================

`problem.txt` carries one input with two tagged blocks, `<informal_theorem>` and
`<informal_proof>`. THE PROOF BLOCK MAY BE EMPTY -- strip whitespace before deciding.

  * both blocks non-empty  -> full autoformalization; a leftover `sorry` is unfinished.
  * proof block EMPTY      -> STATEMENT autoformalization. A faithful statement with `sorry`
                              as its proof is a COMPLETE answer. Do not spend the run trying
                              to discharge it, do not deny delivery for containing `sorry`,
                              and do not tell the agent that `sorry` scores zero.

Every problem in THIS round's task set has a non-empty proof block, so this changes nothing
you can measure here. It is required anyway: the harness this search produces is what gets
deployed, and users will send theorem-only inputs to it. A harness that assumes both blocks
are present will burn the whole budget on those inputs hunting a proof nobody asked for.

Detect the mode from the input, not from a flag or an environment variable. Any pre-delivery
gate, verify hook, triviality probe or refuse-to-deliver mechanism you inherited has to be
mode-aware too -- some of them treat `sorry` as a defect unconditionally. Say in
`self_critique` how your change behaves in each mode.

================================================================================
8. WHAT WILL BE REJECTED BEFORE ANY GPU TIME IS SPENT
================================================================================

Not advice -- these are the checks that run on your child. Everything else is your call.

  * harness.py must parse, provide `MODEL`, `TASK` and `main`, accept the `agent` CLI mode,
    and exit 0 on `python3 {{CHILD_HARNESS}} config`.
  * `MODEL`'s model id, temperature and reasoning effort must match the seed's, and nothing
    may set `os.environ["AGENT_MODEL"/"AGENT_TEMPERATURE"/"AGENT_THINKING"]`. Those are the
    experiment's control variables: changing one does not make a worse round, it makes every
    fitness in this memory incomparable. `max_turns` and `api_timeout` are exempt.
  * the resolved turn budget must be at or below {{TURN_CEILING}}. That is a wall-clock guard,
    not a rule about the idea: every container is killed at {{TIMEOUT}} s whatever its budget,
    so past some point extra turns buy truncated attempts and an arm that no longer fits in the
    allocation, and an agent that loops without converging burns the wall clock for nothing.
    The measured per-problem distribution is in §3; use it to pick a budget that fits.
  * `mutation.json` must carry the guide's §12 fields.
  * four waste guards, each WAIVABLE with `"override_checks": [...]` and an
    `"override_reason"` if you think the detection is wrong: an AST identical to the parent's,
    a diff confined to the inert symbols (guide §5), a harness byte-identical to a node
    already in the tree, and a 120-character run of gold Lean in the diff.

Two operational facts, not rules: `$VIBE_HOME/AGENTS.md` is the only channel that reaches the
model, and there is no user to ask. The rest of the environment invariants are in guide §6.

Write the two files above. End with `MUTATION: <slug>`.
\end{lstlisting}

\subsection{Semantic-Correctness Judges (Evaluation)}
\label{app:prompts:judge}
The semantic-correctness check of Sec.~\ref{sec:metrics} runs two frontier judges with one system prompt (Listing~\ref{lst:judge-system}) and two user prompts: the FL$\leftrightarrow$FL$'$ comparison of the candidate Lean with the gold formalization (Listing~\ref{lst:judge-flfl}) and the NL$\leftrightarrow$NL$'$ comparison of the candidate's informalization with the NL input (Listing~\ref{lst:judge-nlnl}).

\lsthead{Judge system prompt (both judges)}{}
\begin{lstlisting}[style=prompttext,captionpos=t,caption={\textbf{Judge system prompt.}},label={lst:judge-system}]
You are an expert in formal mathematics, Lean 4 theorem proving, the Mizar proof system, and informal mathematical reasoning. Follow the user's judging instructions precisely.
\end{lstlisting}

\lsthead{FL$\leftrightarrow$FL$'$ judge prompt}{}
\begin{lstlisting}[style=prompttext,captionpos=t,caption={\textbf{FL-to-FL judge.} \texttt{\{gold\_description\}} and \texttt{\{gold\_block\}} depend on whether the gold is Mizar, Lean, or NL.},label={lst:judge-flfl}]
You are an expert in formal mathematics, Lean 4 theorem proving (version v4.23.0), and the Mizar proof system.

You are given a candidate Lean 4 file produced by an automated formalizer and a gold-standard source. Each side may contain multiple theorems, definitions, and proofs.

<candidate_lean_file>
```lean4
{generated_lean_file}
```
</candidate_lean_file>

{gold_block}

{gold_description}

Your tasks are:

A. Semantic equivalence of theorem statements (ignore proof correctness):
   - Determine whether the candidate Lean theorem statement(s) express the same mathematical claims as the gold source, modulo:
     * Notational differences (e.g. `ℕ` vs `Nat`, `→` vs `fun`, operator notation vs function application).
     * Harmless quantifier reordering that does not change the logical content.
     * Renaming of bound variables or hypothesis labels.
     * Equivalent reformulations (e.g. `¬P → Q` vs `¬Q → P`, set-builder vs predicate form).
   - Be rigorous. Minor-seeming differences can be semantically significant:
     * Sign conventions (e.g. `n - 1` in `ℕ` truncates vs `n - 1` in `ℤ`).
     * Missing or extra hypotheses (e.g. `0 < n` omitted).
     * Weaker/stronger conclusions (e.g. `≤` vs `<`, `∃` vs `∃!`).
     * Different algebraic structures (e.g. ring vs field, group vs monoid).
   - Do NOT compare proof correctness or proof strategy.
   - If the gold side has multiple theorems, the candidate should cover the same mathematical content (no missing or extraneous claims).
   - If theorem statements are NOT semantically equivalent, the score MUST be -1 regardless of proofs.

B. Placeholder-proof check (candidate Lean only):
   - After establishing theorem-statement equivalence, inspect each candidate Lean proof body.
   - A proof is a PLACEHOLDER if and only if it is one of: `sorry`, `admit`, `trivial`, an empty `by` block, `by sorry`, or `by admit`.
   - A proof is NOT a placeholder if it uses any genuine tactic (e.g. `by rfl`, `by decide`, `by norm_num`, `by simp`, `by omega`, `by linarith`, `by ring`, `by exact ...`, or any multi-tactic block) — even if the proof is a single line.
   - Do not judge whether a non-placeholder proof is correct or complete; only flag explicit placeholders.

C. Provide your reasoning inside justification tags:

   <justification>
   1. Identify the theorem statement(s) on each side and their mathematical content.
   2. Compare them point-by-point (hypotheses, conclusion, quantifiers, types).
   3. State whether they are semantically equivalent and why/why not.
   4. If equivalent, list each candidate proof and whether it is a placeholder.
   </justification>

D. Scoring — output exactly one of 1, 0.5, or -1 inside score tags:
   - 1: theorem statement(s) semantically equivalent AND no placeholder proofs.
   - 0.5: theorem statement(s) semantically equivalent BUT at least one proof is a placeholder.
   - -1: theorem statement(s) NOT semantically equivalent (regardless of proofs).

   <score>
   your numeric score (1, 0.5, or -1)
   </score>
\end{lstlisting}

\lsthead{NL$\leftrightarrow$NL$'$ judge prompt}{}
\begin{lstlisting}[style=prompttext,captionpos=t,caption={\textbf{NL-to-NL judge.} The judge informalizes the candidate and compares it with the NL specification.},label={lst:judge-nlnl}]
You are an expert in formal mathematics, Lean 4 theorem proving (version v4.23.0), and informal mathematical reasoning.

An automated formalizer was given the natural-language theorem/proof specification below and asked to formalize it in Lean 4. You are given that original specification and the candidate Lean 4 file it produced. Each may contain multiple theorems, definitions, and proofs.

<nl_specification>
{nl_spec}
</nl_specification>

The specification above is the ORIGINAL natural-language (informal) request: an informal theorem statement and possibly its informal proof. Use its statement as the reference claim; the informal proof, if present, only clarifies the intended meaning of the statement.

<candidate_lean_file>
```lean4
{generated_lean_file}
```
</candidate_lean_file>

Your tasks, performed strictly in this order:

A. Informalize the candidate Lean (statement only — ignore all proofs):
   - CANDIDATE informal statement: translate the candidate Lean theorem
     statement(s) into a precise, self-contained natural-language statement, in
     plain mathematical English. Do not reference Lean syntax. Read only what the
     Lean actually states — do NOT assume it matches the specification.

B. Compare the candidate informal statement against the specification's statement:
   - Judge whether they assert the SAME mathematical claim.
   - Acceptable differences: wording, variable names, equivalent reformulations
     (e.g. contrapositive), notational choices.
   - Unacceptable differences: missing/extra hypotheses, altered conclusions, wrong
     quantifier structure, missing cases, sign/structure changes, or different
     mathematical content.
   - Base the comparison on the meaning of the two statements, not on surface form.

C. Provide your work inside the following tags:

   <spec_statement>
   the reference claim, restated from the natural-language specification
   </spec_statement>

   <candidate_informal>
   your natural-language rendering of the candidate Lean theorem statement(s)
   </candidate_informal>

   <justification>
   point-by-point comparison (hypotheses, conclusion, quantifiers), then your
   verdict and why.
   </justification>

D. Verdict — output exactly one of yes or no inside equivalence tags:
   - yes: the candidate informal statement asserts the same claim as the specification.
   - no:  it does not.

   <equivalence>
   yes or no
   </equivalence>
\end{lstlisting}

\subsection{Baseline Agents and LLMs (Inference)}
\label{app:prompts:baselines}
Single-turn models and coding agents receive the system prompt in Listing~\ref{lst:base-system} and the user prompt in Listing~\ref{lst:base-user}. The agentic baselines, Aristotle, Numina-Lean-Agent, and OpenGauss, receive the same task instruction, adapted only to how each takes its input and where it writes its output.

\lsthead{System prompt for single-turn models and coding agents}{}
\begin{lstlisting}[style=prompttext,captionpos=t,caption={\textbf{Baseline system prompt.}},label={lst:base-system}]
You are an expert in mathematics. Your task is to convert an informal, natural-language theorem and proof pair into a correct Lean 4 formalization.
\end{lstlisting}

\lsthead{User prompt for single-turn models, coding agents, and OpenGauss}{}
\begin{lstlisting}[style=prompttext,captionpos=t,caption={\textbf{Baseline user prompt.} \texttt{\{input\_nl\_thm\_proof\_pair\}} carries the NL theorem and proof; \texttt{\{exampleInPrompt\}} is left empty.},label={lst:base-user}]
You task is to take as input an informal theorem and proof in natural language and autoformalize it in Lean 4.

Produce ONLY the final Lean 4 theorem and proof that compiles under Lean 4 (version 4.23.0). DO NOT include explanations, intermediate reasoning, comments, or step-by-step thoughts. Include all necessary imports/header. Use standard Mathlib conventions when needed.

{exampleInPrompt}

Here is the **actual** informal theorem and proof in natural language:
{input_nl_thm_proof_pair}

Return the corresponding Lean 4 theorem and proof strictly in this format, and NOTHING else:

<formal_proof>
```lean4
(Provide the complete Lean 4 code here)
```
</formal_proof>

Formatting rules (follow exactly):
- Open the code block with ```lean4 (not ```lean, not a bare ```), and close it with a single ```.
- Emit exactly one <formal_proof> block and exactly one code fence inside it. Do not add extra ``` fences or nest code blocks.
- Put nothing before <formal_proof> or after </formal_proof>.
\end{lstlisting}

\subsection{Data Pipelines (Data Curation)}
\label{app:prompts:data}
LoCoBench's NL side is produced by two informalization prompts (App.~\ref{app:benchdetails}): Lean to NL for the Mathlib and CSLib training instances (Listing~\ref{lst:inf-lean}) and Mizar to NL for the Mizar instances (Listing~\ref{lst:inf-mizar}). The teacher prompts of data distillation are given in App.~\ref{app:prompts}.

\lsthead{Lean-to-NL informalization prompt (Mathlib and CSLib instances)}{}
\begin{lstlisting}[style=prompttext,captionpos=t,caption={\textbf{Lean-to-NL informalization.} \texttt{\{fl\_proof\}} carries the Lean source. Data construction.},label={lst:inf-lean}]
You are an expert in formal mathematics, Lean 4 theorem proving, and mathematical exposition.

Your task is to translate a Lean 4 theorem and proof into precise natural-language mathematics.

The input Lean code is wrapped inside:
<lean4_theoremproof>
...
</lean4_theoremproof>

Everything after the keyword "theorem" or "lemma" and before the keyword ":= " is the theorem statement. Everything after := is the proof.

### Goal
Produce a natural-language theorem statement and proof that:
- accurately preserve the mathematical meaning,
- faithfully follow the proof structure,
- are precise enough that a human could reconstruct the original formal proof,
- read like textbook-quality mathematical writing.

### Output Format
First output the proof plan between the following tags (plain text is fine here):

<proof_plan>
...
</proof_plan>

Then output the theorem between the following tags. Write in **English** (normal mathematical prose). For symbols, variables, and formulas, use **LaTeX-style notation**: inline math as `\( ... \)`, display math as `\[ ... \]` when needed, and standard LaTeX commands (e.g. `\mathbb`, `\forall`, `\in`, fractions, subscripts). Do **not** emit a full LaTeX document, preamble, or `\begin{document}`; do **not** require `theorem`/`proof` environments unless they read naturally—paragraphs of English with LaTeX math are preferred.

<natural_language_theorem>
...
</natural_language_theorem>

Then output the proof between the following tags, with the same convention (English + LaTeX-style math):

<natural_language_proof>
...
</natural_language_proof>

Do not wrap these sections in markdown code fences. Do not use Markdown `$...$` delimiters; keep math in LaTeX `\( \)` / `\[ \]` form.

### Detailed instructions
1. Proof Plan: Before writing the natural language theorem and proof, analyze the Lean-4 theorem and proof and provide a structured plan.
    - First, state your understanding of the theorem in own words.
    - Next, identify the main proof strategy.
    - List the key tactics and proof-steps used.
    - Highlight intermediate lemmas, subcases, or structural reasoning.
    - Summarize how these pieces fit together to establish the final proof.
    - The proof plan should highlight key ideas, intermediate lemmas, and proof structures that will guide the construction of the final natural language proof.

2. Natural Language Theorem: After the plan, write the theorem accurately in English, with mathematical content in LaTeX-style notation as in the output format above.
    - Write a mathematically precise theorem statement.
    - Use standard mathematical prose, as is typically found in textbooks or written by professional mathematicians.
    - Precisely output a self-contained theorem statement and no extraneous information like name of theorem etc.

3. Natural Language Proof
    - Write a rigorous mathematical proof in English, using LaTeX-style notation for symbols and formulas.
    - Preserve proof order and logical structure,
    - Explicitly explain intermediate deductions,
    - Include case analysis if present,
    - When the proof uses auxiliary lemmas or prior results, **prefer not to state formal or code-like lemma names** (e.g. Lean identifiers). Instead, describe **what the lemma says or does** in plain mathematical language, as a good textbook would—e.g. “by the standard bound on …” or “using the fact that continuous functions on compact sets are uniformly continuous,” not “by lemma `foo_bar`.”
    - **ABSOLUTELY DO NOT** mention or reference ANYTHING about LEAN or the LEAN proof or tactic names such as rw, simp, cases, exact, etc. in the output natural language proof. Prove it in natural language like a mathematician.

4. Style constraints
    - Prefer mathematical prose over code-like wording.
    - When citing lemmas or named results, favor descriptive mathematical phrasing over opaque names, consistent with textbook-quality exposition.
    - Do not invent mathematical facts absent from the formal proof.
    - If the proof relies on direct rewriting, explain exactly what is rewritten.
    - Faithfulness to the formal proof is more important than elegance.

### Actual Input

<lean4_theoremproof>

{fl_proof}

</lean4_theoremproof>
    
\end{lstlisting}

\lsthead{Mizar-to-NL informalization prompt (Mizar instances)}{}
\begin{lstlisting}[style=prompttext,captionpos=t,caption={\textbf{Mizar-to-NL informalization.} \texttt{\{mizar\_source\}} carries the Mizar article excerpt. Data construction.},label={lst:inf-mizar}]
You are an expert in formal mathematics, the Mizar proof system, and mathematical exposition.

Your task is to translate a Mizar source file into precise, **self-contained** natural-language mathematics.

The input Mizar source is wrapped inside:
<mizar_article>
...
</mizar_article>

### Mizar file structure (what to read, what to ignore)

A Mizar source file is laid out as follows:

1. **Header comments** — lines starting with `::`. These contain the article title, authors, license. **Ignore the license/copyright entirely.** The title can be used for context but **must not appear** in the output.

2. **`environ` block** — lists of `vocabularies`, `notations`, `constructors`, `registrations`, `requirements`, `definitions`, `equalities`, `expansions`, `theorems`, `schemes`. This is purely the *import / library* section. **Completely ignore it.** Do not describe, list, or refer to any of those names anywhere in your output.

3. **`begin` section** — the actual mathematics. It typically contains:
   * `reserve x, y for SomeType;` — type reservations (free declarations that fix the types of identifiers used later).
   * `definition ... end;` — local definitions of functors, predicates, modes, attributes, structures.
   * `theorem [LABEL:] STATEMENT proof ... end;` — one or more theorems with proofs. The keyword `proof` separates the statement from the proof; the proof ends at the matching `end;`.
   * Proof-structuring constructs: `now ... end;`, `consider ... such that ...`, `let`, `assume`, `take`, `hence`, `thesis`, labels like `A1:`, `:LblName:`. These are **proof-engine syntax**, not mathematics.

### What to include vs. exclude in the natural-language output

* **Exclude** the entire `environ` block and the header beyond what you need to understand the math.
* **Exclude** any local `definition`, `reserve`, or auxiliary lemma that is **not actually used** by any of the theorems/proofs you are informalizing. Do not waste prose on unused machinery.
* **Include** any local `definition`, `reserve`, or notational convention that **is used** by the theorem statement(s) or their proof(s). When you include one, **state its meaning in plain mathematical English** inside the natural-language theorem (or, when more natural, at the start of the proof). The reader of the natural-language output must NOT need to consult the Mizar file: treat the output as a fully self-contained mathematical excerpt.
* **Multiple theorems in one file:** if the `begin` section contains more than one `theorem ... proof ... end;` block, treat them collectively as the content to informalize. Merge them into **one coherent natural-language theorem statement** — as a conjunction, an enumerated list of claims, or a single result with several parts, whichever reads most naturally. Produce **one matching natural-language proof** that establishes all the claims in the same order they appear in the source.
* **Do not invent** mathematical facts that are not present in the Mizar source. Faithfulness first, elegance second.

### Output format

First a short structured plan (plain prose):

<proof_plan>
- One or two sentences summarising what is proved.
- The main proof strategy and any case analysis.
- Which local definitions / reservations need to be carried over into the natural-language output, and how each will be phrased.
</proof_plan>

Then the theorem — English mathematical prose, with LaTeX-style math `\( ... \)` inline and `\[ ... \]` for display. Do NOT use Markdown `$...$` delimiters and do NOT emit `\begin{document}` or any full LaTeX preamble:

<natural_language_theorem>
...
</natural_language_theorem>

Then the proof, same conventions (English + LaTeX math):

<natural_language_proof>
...
</natural_language_proof>

Do not wrap any of these sections in markdown code fences.

### Style and faithfulness constraints

* Write like a textbook author: rigorous, complete, but readable. Use standard mathematical phrasing.
* **Never** reference Mizar identifiers, MML article filenames (e.g. `MATRIX_3`, `XBOOLE_0`, `NUMBERS`, `XREAL_1`), Mizar label tags (`A1:`, `Th20`, `Lm5`), Mizar scheme names, or Mizar keywords (`thesis`, `consider`, `hence`, `assume`, `take`, `now ... end`, `by`, `from`, `reconsider`, `let ... be`, `func`, `mode`, `attr`, `pred`, `cluster`, `synonym`, `antonym`, `registration`, `theorem`, `definition`, `proof`, `end`) anywhere in the natural-language output. The output must read as pure mathematics, with no trace of the formal system or its file names.
* When the formal proof invokes an external lemma (e.g. `by MATRIX_3:2`, `by FUNCT_1:def 5`), describe **what the lemma says** in plain English (e.g. "by the associativity of matrix addition" or "using the standard identity for the additive inverse"). Never quote the lemma name, theorem number, or article reference.
* Preserve the proof order and the logical structure of the Mizar proof. Make every intermediate step explicit.
* If a `reserve` declaration (e.g. `reserve K for Ring; reserve M1, M2 for Matrix of K;`) introduces type conventions used by the theorem, state those conventions inline in natural language ("Let \( K \) be a ring and \( M_1, M_2 \) matrices over \( K \) ..."), instead of assuming the reader knows them.
* If a local `definition` block introduces a new operation/predicate that is used in the theorem (e.g. `func M1-M2 -> Matrix of K equals M1+(-M2);`), state the definition explicitly inside the natural-language theorem ("Define matrix subtraction by \( M_1 - M_2 := M_1 + (-M_2) \) ...") and then state the theorem.
* If a `definition` block is followed by a `theorem` that depends on it, integrate the definition into the natural-language theorem statement rather than presenting them as two unrelated items.
* The natural-language statement and proof together must be **self-contained**: a competent mathematician reading them, with no access to the Mizar source, should be able to (i) understand the statement unambiguously and (ii) reconstruct a correct mathematical proof.

### Actual input

<mizar_article>
{mizar_source}
</mizar_article>
\end{lstlisting}

\end{document}